\documentclass[11pt,titlepage,a4paper]{article}
\usepackage{bozhomac}  
\usepackage{bozhlogo}  
\usepackage{cite}	
\usepackage{varioref}	

\title{\bfseries	\vspace*{-1.7in}
{\huge Lagrangian quantum field theory\\[1ex] in momentum picture}
 \\[1.3ex]
{\LARGE I.\ Free scalar fields}
}

\vspace{1.7ex}

\author{
Bozhidar Z.\ Iliev
\thanks{Laboratory of Mathematical Modeling in Physics,
Institute for Nuclear Research and \mbox{Nuclear} Energy,
Bulgarian Academy of Sciences,
Boul.\ Tzarigradsko chauss\'ee~72, 1784 Sofia, Bulgaria}
\thanks{E-mail address: bozho@inrne.bas.bg}
\thanks{URL: http://theo.inrne.bas.bg/$\sim$bozho/}
}

\date{	
 \vspace{2.27ex}\ShortTitle{QFT in momentum picture: I}\\[0.27ex]
 \vspace{3.27ex}
\small
	\begin{tabular}{r@{$\colon\to~$}l}
 \vspace{0.09ex} Basic ideas	& June, 2001\\[0.09ex]
 \vspace{0.09ex} Began		& June 2, 2001	\\[0.09ex]
 \vspace{0.09ex} Ended		& July 31, 2001	\\[0.09ex]
 \vspace{0.09ex} Initial typeset& June 8, 2001--August 4, 2001 \\[0.09ex]
 \vspace{0.09ex} Last update	& February 1, 2004	\\[0.09ex]
 \vspace{0.27ex} Produced	& \fbox{\today}	\\[0.27ex]
	\end{tabular} \\[1.27ex]
\normalsize
	\begin{tabular}{r@{$\colon~$}l}
 \vspace{0.27ex} http://www.arXiv.org e-Print archive No. & hep-th/0402006
 							\\[0.27ex]
	\end{tabular} \\[-0.27ex]
 \vspace{4.27ex}{\Huge\BOZHO}	\\[4.27ex]
 \vspace{0.27ex}\Subject{Quantum field theory}
								\\[2.27ex]
	\begin{tabular}{r@{\hspace{0.512em}}|@{\hspace{0.512em}}l}
 \vspace{0.27ex}\MSC[2000]{81Q99, 81T99\\\hspace{0pt}}
&
 \vspace{0.27ex}
	\end{tabular} \\[1.27ex]
 \vspace{0.27ex}\\[0.27ex]
}

\newcommand{\bk}{\boldsymbol{k}}  	

\newcommand{\Hil}{\mathcal{F}}		
	\newcommand{\base}{\mathit{M}}	

\newcommand{\ope}[2][{}]{\lindex[\mathcal{#2}]{}{#1}} 
\newcommand{\tope}[2][{}]{\ope[#1]{\Tilde{#2}}} 
\begin{document}		

\renewcommand{\thepage}{\roman{page}}

\renewcommand{\thefootnote}{\fnsymbol{footnote}} 
\maketitle				
\renewcommand{\thefootnote}{\arabic{footnote}}   

\tableofcontents		


\begin{abstract}
The work contains a detailed investigation of free neutral (Hermitian) or
charged (non\ndash Hermitian) scalar fields and the describing them
(system of) Klein\ndash Gordon equation(s) in momentum picture of motion. A
form of the field equation(s) in terms of creation and annihilation operators
is derived. An analysis of the (anti\ndash)commutation relations on its base
is presented.  The concept of the vacuum and the evolution of state vectors
are discussed.

\end{abstract}

\renewcommand{\thepage}{\arabic{page}}

\section {Introduction}
\label{Introduction}

	This paper pursuits a twofold goal. On one hand, it gives a detailed
illustration of the methods of Lagrangian quantum field theory in momentum
picture, introduced in~\cite{bp-QFT-pictures,bp-QFT-MP}, on the
simplest examples of free Hermitian (neutral, real) or non\ndash Hermitian
(charged, complex) scalar field. On another hand, it contains an in-depth
analysis of the (system of) Klein\ndash Gordon equation(s) in momentum
picture describing such fields. Most of the known fundamental results are
derived in a new way (and in a slightly modified form), but the work contains
and new ones.

	We have to mention, in this paper it is considered only the Lagrangian
(canonical) quantum field theory in which the quantum fields are represented
as operators, called field operators, acting on some Hilbert space, which in
general is unknown if interacting fields are studied. These operators are
supposed to satisfy some equations of motion, from them are constructed
conserved quantities satisfying conservation laws, etc. From the view\ndash
point of present\ndash day quantum field theory, this approach is only a
preliminary stage for more or less rigorous formulation of the theory in
which the fields are represented as operator\ndash valued distributions, a
fact required even for description of free fields. Moreover, in non\ndash
perturbative directions, like constructive and conformal field theories, the
main objects are the vacuum mean (expectation) values of the fields and from
these are reconstructed the Hilbert space of states and the acting in it
fields. Regardless of these facts, the Lagrangian (canonical) quantum field
theory is an inherent component of the most of the ways of presentation of
quantum field theory adopted explicitly or implicitly in books
like~\cite{Bogolyubov&Shirkov,Bjorken&Drell,Roman-QFT,Ryder-QFT,
Akhiezer&Berestetskii,Ramond-FT,Bogolyubov&et_al.-AxQFT,Bogolyubov&et_al.-QFT}.
Besides, the Lagrangian approach is a source of many ideas for other
directions of research, like the axiomatic quantum field
theory~\cite{Roman-QFT,Bogolyubov&et_al.-AxQFT,Bogolyubov&et_al.-QFT}.

	The basic moments of the method, we will follow in this work, are the
next ones:
\\\indent
	(i) In Heisenberg picture is fixed a (second) non\ndash quantized and
non\ndash normally ordered  operator\ndash valued Lagrangian, which is
supposed to be polynomial (or convergent power series) in the field operators
and their first partial derivatives;
\\\indent
	(ii) As conditions additional to the Lagrangian formalism are
postulated the commutativity between the components of the momentum operator
(see~\eref{2.1} below) and the Heisenberg relations between the field
operators and momentum operator (see~\eref{2.28} below);
\\\indent
	(iii) Following the Lagrangian formalism in momentum picture, the
creation and annihilation operators are introduced and the dynamical
variables and field equations are written in their terms;
\\\indent
	(iv) From the last equations, by imposing some additional restrictions
on the creation and annihilation operators, the (anti)commutation relations
for these operators are derived;
\\\indent
	(v) At last, the vacuum and normal ordering procedure are defined, by
means of which the theory can be developed to a more or less complete form.

	The main difference of the above scheme from the standard one is that
we postulate the below\ndash written relations~\eref{2.1} and~\eref{2.28}
and, then, we look for compatible with them and the field equations
(anti)commutation relations. (Recall, ordinary the (anti)commutation
relations are postulated at first and the validity of the
equations~\eref{2.1} and~\eref{2.28} is explored after
that~\cite{Bjorken&Drell-2}.)

\vspace{0.8ex}

	In Sect.~\ref{Sect2} are reviewed the basic moments of the momentum
picture of motion in quantum field theory.

	The rest of the work is divided into two parts.

	Part~\ref{PartA}, involving sections~\ref{Sect3}--\ref{Sect10}, deals
with the case of neutral (Hermitian, real) free scalar field. The contents of
its sections is as follows:

	In Sect.~\ref{Sect3}, the material of Sect.~\ref{Sect2} is specialized
to the case of free Hermitian (neutral, real) scalar field; in particular, the
Klein\ndash Gordon equation in momentum picture is derived.


	Sect.~\ref{Sect5} is devoted to analysis of the Klein-Gordon equation
(in momentum picture) in terms of operators similar to (and, in fact, up to a
phase factor and normalization constant, identical with) the Fourier images
of field operators in Heisenberg picture. From them, in Sect.~\ref{Sect6},
are constructed the creation and annihilation operators which turn to be
identical, up to a phase factor and, possibly, normalization constant, with
the ones known from Heisenberg picture. Their physical meaning is discussed
(or recalled). In Sect.~\ref{Sect7}, the 3\ndash dimensional creation and
annihilation operators (depending on the 3\ndash momentum) are introduced and
the field equation is written in their terms. It happens to be a tri\ndash
linear equation relative to them. This new form of the field equation is
utilized in Sect.~\ref{Sect8} for a detailed analysis of the (additional)
conditions leading to the known commutation relations. In particular, it is
proved that, excluding the vanishing field case, the quantization of a free
Hermitian scalar field by \emph{anti}\ndash commutators is rejected by the
field equation without an appeal to the spin\ndash statistics theorem (or
other equivalent to it additional assertion).

	Sect.~\ref{Sect9} is devoted to the introduction of the concept of
vacuum (state) which requires a modification of the developed theory by
a normal ordering of products of creation and/or annihilation operators. The
vacua of Heisenberg and momentum pictures happen to coincide. The problem of
state vectors, representing in momentum picture a free Hermitian scalar
field, is considered in Sect.~\ref{Sect10}. It turns to be rather trivial due
to the absence of any interaction. However, the construction of Fock base is
recalled  and the basic ideas of scattering theory are illustrated in this
almost trivial case.

	The second part~\ref{PartB} of our work is devoted to the general
case of charged or neural free scalar field. Regardless of some overlap with
part~\ref{PartA}, it concentrates mainly on the case of non-Hermitian field.
Most of the proofs in it refer or are based, at least partially, on similar
ones in the Hermitian case, investigated in part~\ref{PartA}. The problems of
the choice of the initial Lagrangian and the `right' definitions of the
energy\ndash momentum and charge operators are partially discussed. The
layout of part~\ref{PartB}, involving sections~\ref{Sect11}--\ref{Sect19}, is
similar to the one of part~\ref{PartA}.

	The description of arbitrary free scalar field is presented in
Sect.~\ref{Sect11}.

	An analysis of the system of Klein-Gordon equations describing free
scalar fields is presented in Sect.~\ref{Sect13}. A feature of the momentum
picture is that, in a case of non\ndash Hermitian field, the two equations of
this system are not separate equations for the field and its Hermitian
conjugate; they are mixed via the momentum operator. The creation and
annihilation operators are introduced in Sect.~\ref{Sect14} and the field
equations are written in their terms in Sect.~\ref{Sect15}. They turn to be
trilinear equations similar to the ones appearing the parafield theory. The
commutation relations are extracted from them in Sect.~\ref{Sect16}. These
reveal the non-equivalence of the theories build from different initial
Lagrangians. To agree the results, one needs different additional
hypotheses/conditions, depending on the concrete Lagrangian utilized. The
`best' Lagrangian (of three considered) is pointed out. The charge and
orbital angular momentum operators are considered in Sect.~\ref{Sect17}. The
vacuum is defined in Sect.~\ref{Sect18}, where the normal ordering of the
dynamical variables is described too. We point that the last operation is the
final step which leads to identical theories build from different initial
Lagrangians. Some problems concerning the state vectors of free scalar fields
are discussed in Sect.~\ref{Sect19}.

	Sect.~\ref{Conclusion} closes the paper by pointing to its basic
results.

\vspace{1ex}

	The books~\cite{Bogolyubov&Shirkov,Roman-QFT,Bjorken&Drell} will be
used as standard reference works on quantum field theory. Of course, this
is more or less a random selection between the great number of papers on the
theme.%
\footnote{~%
The reader is referred for more details or other points of view, for
instance, to~\cite{Itzykson&Zuber,Ryder-QFT,Schweber} or the literature cited
in~\cite{Bogolyubov&Shirkov,Roman-QFT,Bjorken&Drell,Itzykson&Zuber,Ryder-QFT,
Schweber}.%
}

	Throughout this paper $\hbar$ denotes the Planck's constant (divided
by $2\pi$), $c$ is the velocity of light in vacuum, and $\iu$ stands for the
imaginary unit.

	The Minkowski spacetime is denoted by $\base$. The Greek indices run
from 0 to $\dim\base-1=3$. All Greek indices will be raised and lowered by
means of the standard 4\ndash dimensional Lorentz metric tensor
$\eta^{\mu\nu}$ and its inverse $\eta_{\mu\nu}$ with signature
$(+\,-\,-\,-)$. The Einstein's summation convention over indices repeated on
different levels is assumed over the whole range of their values.

	At the end, a technical remark is in order. The derivatives with
respect to operator\ndash valued (non\ndash commuting) arguments will be
calculated according to the rules of the classical analysis of commuting
variables, which is an everywhere silently accepted
practice~\cite{Bogolyubov&Shirkov,Bjorken&Drell-2}. As it is demonstrated
in~\cite{bp-QFT-action-principle}, this is not quite correct but does not
lead to incorrect results (except the non\ndash uniqueness of the conserved
quantities) when free scalar fields are concerned. We shall pay attention on
that item at the corresponding places in the text.


\section{The momentum picture}
	\label{Sect2}

	The momentum picture in quantum field theory was introduce
in~\cite{bp-QFT-pictures,bp-QFT-MP}. Its essence is in the
following.

	Let us consider a system of quantum fields, represented in Heisenberg
picture of motion by field operators $\tope{\varphi}_i(x)\colon\Hil\to\Hil$,
$i=1,\dots,n\in\field[N]$, in system's Hilbert space $\Hil$ of
states and depending on a point $x$ in Minkowski spacetime $\base$. Here and
henceforth, all quantities in Heisenberg picture will be marked by a tilde
(wave) ``$\tope{\mspace{6mu}}\mspace{3mu}$'' over their kernel symbols. Let
$\tope{P}_\mu$ denotes the system's (canonical) momentum vectorial operator,
defined via the energy\ndash momentum tensorial operator $\tope{T}^{\mu\nu}$
of the system, viz.
	\begin{equation}
			\label{2.0}
\tope{P}_\mu
:=
\frac{1}{c}\int\limits_{x^0=\const} \tope{T}_{0\mu}(x) \Id^3\bs x .
	\end{equation}
Since this operator is Hermitian, $\tope{P}_\mu^\dag=\tope{P}_\mu$, the
operator
	\begin{equation}	\label{12.112}
\ope{U}(x,x_0)
 =
\exp\Bigl( \iih \sum_\mu (x^\mu-x_0^\mu)\tope{P}_{\mu}  \Bigr) ,
	\end{equation}
where $x_0\in\base$ is arbitrarily fixed and $x\in\base$,%
\footnote{~%
The notation $x_0$, for a fixed point in $\base$, should not be confused with
the zeroth covariant coordinate $\eta_{0\mu}x^\mu$ of $x$ which, following
the convention $x_\nu:=\eta_{\nu\mu}x^\mu$, is denoted by the same symbol
$x_0$. From the context, it will always be clear whether $x_0$ refers to a
point in $\base$ or to the zeroth covariant coordinate of a point
$x\in\base$.%
}
is unitary, \ie
\(
\ope{U}^\dag(x_0,x)
:= (\ope{U}(x,x_0))^\dag
 = \ope{U}^{-1}(x,x_0)
 = (\ope{U}(x,x_0))^{-1}
\)
and, via the formulae
	\begin{align}	\label{12.113}
\tope{X}\mapsto \ope{X}(x)
	&= \ope{U}(x,x_0) (\tope{X})
\\			\label{12.114}
\tope{A}(x)\mapsto \ope{A}(x)
	&= \ope{U}(x,x_0)\circ (\tope{A}(x)) \circ \ope{U}^{-1}(x,x_0) ,
	\end{align}
realizes the transition to the \emph{momentum picture}. Here $\tope{X}$ is a
state vector in system's Hilbert space of states $\Hil$ and
$\tope{A}(x)\colon\Hil\to\Hil$ is (observable or not) operator\ndash valued
function of $x\in\base$ which, in particular, can be polynomial or convergent
power series in the field operators $\tope{\varphi}_i(x)$; respectively
$\ope{X}(x)$ and $\ope{A}(x)$ are the corresponding quantities in momentum
picture.
	In particular, the field operators transform as
	\begin{align}	\label{12.115}
\tope{\varphi}_i(x)\mapsto \ope{\varphi}_i(x)
     = \ope{U}(x,x_0)\circ \tope{\varphi}_i(x) \circ \ope{U}^{-1}(x,x_0) .
	\end{align}
	Notice, in~\eref{12.112} the multiplier $(x^\mu-x_0^\mu)$ is regarded
as a real parameter (in which $\tope{P}_\mu$ is linear). Generally,
$\ope{X}(x)$ and $\ope{A}(x)$ depend also on the point $x_0$ and, to be
quite correct, one should write $\ope{X}(x,x_0)$ and $\ope{A}(x,x_0)$ for
$\ope{X}(x)$ and $\ope{A}(x)$, respectively. However, in the most situations
in the present work, this dependence is not essential or, in fact, is not
presented at all. For this reason, we shall \emph{not} indicate it explicitly.

	As it was said above, we consider quantum field theories in which the
components $\tope{P}_\mu$ of the momentum operator commute between themselves
and satisfy the Heisenberg relations/equations with the field operators, \ie
we suppose that $\tope{P}_\mu$ and $\tope{\varphi}_i(x)$ satisfy the
relations:
	\begin{align}	\label{2.1}
& [\tope{P}_\mu, \tope{P}_\nu ]_{\_} = 0
\\			\label{2.28}
& [\tope{\varphi}_i(x), \tope{P}_\mu]_{\_} = \ih\pd_\mu \tope{\varphi}_i(x).
	\end{align}
Here $[A,B]_{\pm}:=A\circ B \pm B\circ A$, $\circ$ being the composition of
mappings sign, is the commutator/anticommutator of operators (or matrices)
$A$ and $B$. The momentum operator $\tope{P}_\mu$ commutes with the
`evolution' operator $\ope{U}(x,x_0)$ (see below~\eref{12.118}) and its
inverse,
	\begin{equation}	\label{2.2}
	[ \tope{P}_\mu, \ope{U}(x,x_0) ]_{\_} =0
\qquad
	[ \tope{P}_\mu, \ope{U}^{-1}(x,x_0) ]_{\_} =0 ,
	\end{equation}
due to ~\eref{2.1} and~\eref{12.112}. So, the momentum operator remains
unchanged in momentum picture, \viz we have (see~\eref{12.114}
and~\eref{2.2})
	\begin{equation}	\label{2.3}
	 \ope{P}_\mu = \tope{P}_\mu.
	\end{equation}

	Since from~\eref{12.112} and~\eref{2.1} follows
	\begin{equation}	\label{12.116}
\ih \frac{\pd\ope{U}(x,x_0)}{\pd x^\mu}
=
\ope{P}_{\mu} \circ \ope{U}(x,x_0)
\qquad
\ope{U}(x_0,x_0) = \id_\Hil ,
	\end{equation}
we see that, due to~\eref{12.113}, a state vector $\ope{X}(x)$ in
momentum picture is a solution of the initial\ndash value problem
	\begin{equation}	\label{12.117}
\ih \frac{\pd\ope{X}(x)}{\pd x^\mu}
=
\ope{P}_{\mu}  (\ope{X}(x))
\qquad
\ope{X}(x)|_{x=x_0}=\ope{X}(x_0) = \tope{X}
	\end{equation}
which is a 4-dimensional analogue of a similar problem for the Schr{\"o}dinger
equation in quantum mechanics~\cite{Messiah-QM,Dirac-PQM,Prugovecki-QMinHS}.

	By virtue of~\eref{12.112}, or in view of the independence of
$\ope{P}_{\mu} $ of $x$, the solution of~\eref{12.117} is
	\begin{equation}	\label{12.118}
\ope{X}(x)
= \ope{U}(x,x_0) (\ope{X}(x_0))
= \e^{\iih(x^\mu-x_0^\mu)\ope{P}_{\mu} } (\ope{X}(x_0)).
	\end{equation}
Thus, if $\ope{X}(x_0)=\tope{X}$  is an eigenvector of
$\ope{P}_{\mu} $ ($=\tope{P}_{\mu} $)
with eigenvalues $p_\mu$,
	\begin{equation}	\label{12.119}
\ope{P}_{\mu}  (\ope{X}(x_0)) = p_\mu \ope{X}(x_0)
\quad
( =p_\mu \tope{X} = \tope{P}_{\mu}  (\tope{X}) ) ,
	\end{equation}
we have the following \emph{explicit} form of the state vectors
	\begin{equation}	\label{12.120}
\ope{X}(x)
=
\e^{ \iih(x^\mu-x_0^\mu)p_\mu } (\ope{X}(x_0)).
	\end{equation}
It should clearly be understood, \emph{this is the general form of all state
vectors} as they are eigenvectors of all (commuting)
observables~\cite[p.~59]{Roman-QFT}, in particular, of the momentum operator.

	In momentum picture, all of the field operators happen to be
constant in spacetime, \ie
	\begin{equation}	\label{2.4}
\varphi_i(x)
= \ope{U}(x,x_0)\circ \tope{\varphi}_i(x) \circ \ope{U}^{-1}(x,x_0)
= \varphi_i(x_0)
= \tope{\varphi}_i(x_0)
=: \varphi_{(0)\, i} .
	\end{equation}
Evidently, a similar result is valid for any (observable or not such)
function of the field operators which is polynomial or convergent power series
in them and/or their first partial derivatives. However, if $\tope{A}(x)$ is an
arbitrary operator or depends on the field operators in a different way, then
the corresponding to it operator $\ope{A}(x)$ according to~\eref{12.114} is,
generally, not spacetime\ndash constant and depends on the both points $x$
and $x_0$. As a rules, if $\ope{A}(x)=\ope{A}(x,x_0)$ is independent of $x$,
we, usually, write $\ope{A}$ for $\ope{A}(x,x_0)$, omitting all arguments.

	It should be noted, the Heisenberg relations~\eref{2.28} in momentum
picture transform into the identities $\pd_\mu\varphi_i=0$ meaning that the
field operators $\varphi_i$ in momentum picture are spacetime constant
operators (see~\eref{2.4}). So, in momentum picture, the Heisenberg
relations~\eref{2.28} are incorporated in the constancy of the field
operators.

	Let $\tope{L}$ be the system's Lagrangian (in Heisenberg picture). It
is supposed to be polynomial or convergent power series in the field
operators and their first partial derivatives, \ie
$\tope{L}=\tope{L}(\varphi_i(x),\pd_\nu\varphi_i(x))$ with $\pd_\nu$ denoting
the partial derivative operator relative to the $\nu^{\mathrm{th}}$
coordinate $x^\nu$. In momentum picture, it transforms into
	\begin{equation}	\label{2.5}
\ope{L} =\tope{L}(\varphi_i(x), y_{j\nu})
\qquad
y_{j\nu}=\iih[\varphi_j,\ope{P}_{\nu} ]_{\_}  ,
	\end{equation}
\ie in momentum picture one has simply to replace the field operators in
Heisenberg picture with their values at a fixed point $x_0$ and the partial
derivatives  $\pd_\nu\tope{\varphi}_j(x)$ in Heisenberg picture with the
above\ndash defined quantities $y_{j\nu}$.
	The (constant) field operators $\varphi_i$ satisfy the following
\emph{algebraic Euler\ndash Lagrange equations in momentum picture}:%
\footnote{~%
In~\eref{12.129} and similar expressions appearing further, the derivatives
of functions of operators with respect to operator arguments are calculated
in the same way as if the operators were ordinary (classical)
fields/functions, only the order of the arguments should not be changed.
This is a silently accepted practice in the
literature~\cite{Roman-QFT,Bjorken&Drell}. In the most cases such a procedure
is harmless, but it leads to the problem of non\ndash unique definitions of
the quantum analogues of the classical conserved quantities, like the
energy\ndash momentum and charge operators. For some details on this range of
problems in quantum field theory, see~\cite{bp-QFT-action-principle}; in
particular, the \emph{loc.\ cit.}\ contains an example of a Lagrangian whose
field equations are \emph{not} the Euler\ndash Lagrange
equations~\eref{12.129} obtained as just described.%
}
	\begin{equation}	\label{12.129}
\Bigl\{
\frac{\pd\tope{L}(\varphi_j,y_{l\nu})} {\pd \varphi_i}
-
\iih
\Bigl[
\frac{\pd\tope{L}(\varphi_j,y_{l\nu})} {y_{i\mu}}
,
\ope{P}_{\mu}
\Bigr]_{\_}
\Bigr\}
\Big|_{ y_{j\nu}=\iih[\varphi_j,\ope{P}_{\nu} ]_{\_} }
= 0 .
	\end{equation}
Since $\ope{L}$ is supposed to be polynomial or convergent power series  in
its arguments, the equations~\eref{12.129}  are \emph{algebraic}, not
differential, ones. This result is a natural one in view of~\eref{2.4}.

	By virtue of~\eref{12.114} and~\eref{2.3}, the momentum
operator~\eref{2.0} can be written as
	\begin{equation}
			\label{2.6}
\ope{P}_\mu = \tope{P}_\mu
=
\frac{1}{c} \int\limits_{x^0=x_0^0}^{}
	\ope{U}^{-1}(x,x_0)\circ \ope{T}_{0\mu} \circ \ope{U}(x,x_0)
	\Id^3\bs{x}  .
	\end{equation}
This expression for $\ope{P}_\mu$ will be employed essentially in the present
paper.


\part{{\Large Free neutral scalar field}}
\label{PartA}
\markright{\itshape\bfseries Bozhidar Z. Iliev:
	\upshape\sffamily\bfseries QFT in momentum picture: I.~Scalar fields}

	The purpose of this part of the present work is a detailed
exploration of a free neutral (Hermitian, real, uncharged) scalar field in
momentum picture.%
\footnote{~\label{HermitianField}%
A classical real field, after quantization, becomes a Hermitian operator
acting on the system's (field's) Hilbert space of states. That is why the
quantum analogue of a classical real scalar field is called Hermitian scalar
field. However, it is an accepted common practice such a field to be called
(also) a real scalar field. In this sense, the terms real and Hermitian
scalar field are equivalent and, hence, interchangeable. Besides, since a
real (classical or quantum) scalar does not carry any charge, it is called
neutral or uncharged scalar field too.%
}

	After fixing the notation and terminology, we write the Klein-Gordon
equation in momentum picture and derive its version in terms of creation and
annihilation operators. The famous commutation relations are extracted from
it. After defining the vacuum for a free Hermitian scalar field, some
problems concerning the state vectors of such a field are investigated.

\section
[Description of free neutral scalar field in momentum picture]
{Description of free neutral scalar field\\ in momentum picture}
\label{Sect3}

	Consider a free neutral (Hermitian) scalar field with mass $m$. The
corresponding to such a field operator will be denoted by $\tope{\varphi}$. It
is Hermitian, \ie
	\begin{gather}
			\label{2.1new}
\tope{\varphi}^\dag(x) = \tope{\varphi}(x) ,
\intertext{where the dagger $\dag$ denotes Hermitian conjugation relative
to the scalar product $\langle\cdot |\cdot \rangle$ of field's Hilbert space
$\Hil$, and it is described in the Heisenberg picture by the Lagrangian}
			\label{3.1}
\tope{L}
= \tope{L} (\tope{\varphi},\pd_\nu\tope{\varphi})
=
 - \frac{1}{2}m^2c^4\tope{\varphi}\circ\tope{\varphi}
 + \frac{1}{2}c^2\hbar^2 (\pd_\mu\tope{\varphi})\circ(\pd^\mu\tope{\varphi}) ,
\\\intertext{which in momentum picture transforms into (see~\eref{2.5})}
			\label{3.2}
\ope{L}
=
\tope{L}(\varphi,y_\nu)
\big|_{ y_{\nu}=\iih[\varphi_0,\ope{P}_{\nu} ]_{\_} }
=
  - \frac{1}{2}m^2c^4\ope{\varphi}_0\circ\ope{\varphi}_0
  - \frac{1}{2}c^2
	[\varphi_0,\ope{P}_{\mu} ]_{\_} \circ
	[\varphi_0,\ope{P}^{\mu} ]_{\_} .
	\end{gather}
Here $\varphi_0$ is the constant value of the field operator in
momentum picture  \ie (see~\eref{2.4})%
\footnote{~%
The index ``0'' in $\varphi_0$ indicates the dependence on $x_0$, according
to~\eref{12.130}.%
}
	\begin{gather}
			\label{12.130}
\varphi(x)
= \ope{U}(x,x_0)\circ \tope{\varphi}(x) \circ \ope{U}^{-1}(x,x_0)
= \varphi(x_0)
= \tope{\varphi}(x_0)
=: \varphi_0 .
\intertext{Since the operator $\ope{U}(x,x_0)$ is unitary (see~\eref{12.112}
and use the Hermiticity of the momentum operator), the (momentum) field
operator $\varphi_0$ is also Hermitian, \ie}
			\label{2.2new}
\varphi^\dag_0 = \varphi_0 .
	\end{gather}
So, we have~%
\footnote{~%
As pointed at the end of the Introduction, the calculation of the derivatives
in~\eref{3.3} and~\eref{3.4} below is not quite correct mathematically.
However, the field equation~\eref{12.131} is correct; for its rigorous
derivation, see~\cite{bp-QFT-action-principle}.%
}
	\begin{gather}
			\label{3.3}
\pi^\mu
: =
\frac{\pd\tope{L}(\varphi_0,y_{\nu})}{\pd y_\mu}
	\Big|_{ y_{\nu}=\iih[\varphi_0,\ope{P}_{\nu} ]_{\_} }
=
 - \ih c^2 [\varphi_0,\ope{P}^{\mu} ]_{\_}
\\			\label{3.4}
\frac{\pd\ope{L}}{\pd\varphi_0}
=
\frac{\pd\tope{L}}{\pd\varphi_0}
= - m^2c^4\varphi_0 .
\\\intertext{Consequently, equation~\eref{12.129}, in this particular
situation, yields}
			\label{12.131}
m^2c^2\varphi_0 - [ [ \varphi_0,\ope{P}_\mu  ]_{\_}, \ope{P}^\mu ]_{\_} = 0 .
	\end{gather}
This is the \emph{Klein-Gordon equation in momentum picture}. It replaces the
usual Klein\ndash Gordon equation
	\begin{equation}	\label{3.4new}
\Bigl( \widetilde{\square}+\frac{m^2c^2}{\hbar^2} \id_\Hil \Bigr)
  						\tope{\varphi}(x) = 0,
	\end{equation}
where $\widetilde{\square}:=\pd_\mu\pd^\mu$, in Heisenberg picture.%
\footnote{~
As a simple exercise, the reader may wish to prove that the D'Alembert
operator (on the space of operator\ndash valued functions) in momentum
picture is
\(
\square(\cdot) =  -\frac{1}{\hbar^2}
[ [\cdot , \ope{P}_\mu ]_{\_}, \ope{P}_\nu ]_{\_} \eta^{\mu\nu} .
\)
(Hint: from the relation
\(
\ih\frac{\pd\ope{A}(x)}{\pd x^\mu}
=
[\ope{A}(x),\tope{P}_\mu^{\mathrm{t}}]_{\_}
\)
with
$ \tope{P}_\mu^{\mathrm{t}} = - \ih\frac{\pd}{\pd x^\mu}+ p_\mu\id_\Hil $
(see~\cite{bp-QFT-momentum-operator}), it follows that
\(
\widetilde{\square}(\tope{\varphi}_i)
=  -\frac{1}{\hbar^2}
[ [
\tope{\varphi}_i ,
\tope{P}_\mu^{\mathrm{t}} ]_{\_}\tope{P}_\nu^{\mathrm{t}}
]_{\_} \eta^{\mu\nu}
\);
now prove that in momentum picture
\(
\ope{P}_\mu^{\mathrm{t}}
	=\tope{P}_\mu^{\mathrm{t}} + \ope{P}_\mu .
\)%
)
Now, the equation~\eref{12.131} follows immediately from here and the `usual'
Klein\ndash Gordon equations~\eref{3.4new}. By the way, the above
representation for $\square(\cdot)$ remains valid in any picture of motion as
it is a polynomial relative to $(\cdot)$ and $\ope{P}_\mu$; in particular,
~\eref{12.131} is equivalent to
\(
m^2c^2\tope{\varphi}_0 -
	[ \tope{\varphi}_0,\tope{P}_\mu  ]_{\_}, \tope{P}^\mu ]_{\_} = 0
\)
which is equivalent to~\eref{3.4new} due to the (Heisenberg) relation
$[ \tope{\varphi}_0,\tope{P}_\mu  ]_{\_}=\ih\pd_\mu\tope{\varphi}$.%
}

	The energy-momentum tensorial operator $\ope{T}_{\mu\nu}$
has two (non\ndash equivalent) forms for free Hermitian scalar field, \viz
	\begin{subequations}	\label{3.8}
	\begin{gather}
			\label{3.8a}
\tope{T}_{\mu\nu}
=
\tope{\pi}_\mu\circ \pd_\nu\tope{\varphi} - \eta_{\mu\nu} \tope{L}
 =
c^2\hbar^2 ( \pd_\mu\tope{\varphi}) \circ (\pd_\nu\tope{\varphi} )
	- \eta_{\mu\nu} \tope{L}
\\
			\label{3.8b}
	\begin{split}
\tope{T}_{\mu\nu}
 =
\frac{1}{2} (\tope{\pi}_\mu\circ \pd_\nu\tope{\varphi}
	+ \pd_\nu\tope{\varphi}\circ \tope{\pi}_\mu )
- \eta_{\mu\nu} \tope{L}
 =
\frac{1}{2} c^2\hbar^2
\bigl( (\pd_\mu\tope{\varphi}) \circ (\pd_\nu\tope{\varphi})
	+ (\pd_\nu\tope{\varphi}) \circ (\pd_\mu\tope{\varphi}) \bigr)
- \eta_{\mu\nu} \tope{L}  ,
	\end{split}
	\end{gather}
	\end{subequations}
which in momentum picture read respectively:
	\begin{subequations}	\label{3.9}
	\begin{gather}
			\label{3.9a}
	\begin{split}
\ope{T}_{\mu\nu}
=
 - c^2 [\varphi_0,\ope{P}_\mu]_{\_} \circ [\varphi_0,\ope{P}_\nu]_{\_}
+ \eta_{\mu\nu} \frac{1}{2} c^2
     \bigl( m^2c^2 \varphi_0\circ\varphi_0
     + [\varphi_0,\ope{P}_\lambda]_{\_} \circ
       [\varphi_0,\ope{P}^\lambda]_{\_} \bigr)
	\end{split}
\\
			\label{3.9b}
	\begin{split}
\ope{T}_{\mu\nu}
=
&- \frac{1}{2} c^2
\bigl(
[\varphi_0,\ope{P}_\mu]_{\_} \circ [\varphi_0,\ope{P}_\nu]_{\_}
+ [\varphi_0,\ope{P}_\nu]_{\_} \circ [\varphi_0,\ope{P}_\mu]_{\_}
\bigr)
\\&+ \eta_{\mu\nu} \frac{1}{2} c^2
     \bigl( m^2c^2 \varphi_0\circ\varphi_0
     + [\varphi_0,\ope{P}_\lambda]_{\_} \circ
       [\varphi_0,\ope{P}^\lambda]_{\_} \bigr).
	\end{split}
	\end{gather}
	\end{subequations}
	Let us say a few words on the \emph{two} versions,~\eref{3.8a}
and~\eref{3.8b}, of the energy\ndash momentum operator. The former one is a
direct analogue of the classical expression for the energy\ndash momentum
tensor, while the latter variant is obtained from~\eref{3.8a} by a `Hermitian
symmetrization'. The second expression is symmetric and Hermitian, \ie
$\tope{T}_{\mu\nu}=\tope{T}_{\nu\mu}$ and
$\tope{T}_{\mu\nu}^\dag=\tope{T}_{\mu\nu}$, while the first one is such if
$\pd_\mu\tope{\varphi}$ and $\pd_\nu\tope{\varphi}$ commute for all
subscripts $\mu$ and $\nu$. However, as we shall see, both forms of
$\tope{T}_{\mu\nu}$  lead to one and the same (Hermitian) momentum operator.
In this sense, the both forms of $\tope{T}_{\mu\nu}$ are equivalent in
quantum field theory. More details on this problem will be given in
Sect.~\ref{Sect11}.

	Before going on, let us make a technical remark. The derivatives
in~\eref{3.3} and~\eref{3.4} are calculated according to the rules of
classical analysis of commuting variables, which is not correct, as explained
in~\cite{bp-QFT-action-principle}. However, the field equation~\eref{12.131}
or~\eref{3.4new} is completely correct for the reasons given in \emph{loc.\
cit.} Besides, the two forms~\eref{3.8} of the energy\ndash momentum tensor
operator are also due to an incorrect applications of the rules mentioned to
the region of analysis of non\ndash commuting variables; the correct rigorous
expression turns to be~\eref{3.8b}. The reader is referred for more detail on
that item to~\cite{bp-QFT-action-principle}, in particular to section~5.1 in
it. The only reason why we use a non\ndash rigorous formalism is our
intention to stay closer to the standard books on Lagrangian quantum field
theory. This approach will turn to be harmless for the range of problems
considered in the present paper.

	A free neutral or charged scalar field has a vanishing spin angular
momentum and possesses an orbital angular momentum, which coincides with its
total angular momentum. The orbital angular momentum density operator is
	\begin{align}	\label{3.9-1}
\tope{L}_{\mu\nu}^{\lambda} (x)
& =
-\tope{L}_{\nu\mu}^{\lambda} (x)
=
  x_\mu \Sprindex[\tope{T}]{\nu}{\lambda} (x)
- x_\nu \Sprindex[\tope{T}]{\mu}{\lambda} (x)
\\			\label{3.9-2}
\ope{L}_{\mu\nu}^{\lambda}
& =
-\ope{L}_{\nu\mu}^{\lambda}
=
  x_\mu \Sprindex[\ope{T}]{\nu}{\lambda}
- x_\nu \Sprindex[\ope{T}]{\mu}{\lambda}
	\end{align}
in Heisenberg and momentum pictures, respectively. The angular momentum
operator in Heisenberg and momentum picture, respectively, is
	\begin{align}	\notag
& \tope{L}_{\mu\nu}
:=
\frac{1}{c} \int\limits_{x^0=\const}
\tope{L}_{\mu\nu}^0(x) \Id^3\bs x
=
\frac{1}{c} \int\limits_{x^0=\const}
\{
  x_\mu \Sprindex[\tope{T}]{\nu}{0}(x)
- x_\nu \Sprindex[\tope{T}]{\mu}{0}(x)
\} \Id^3\bs x
\\			\label{3.9-3}
& \hphantom{\tope{L}_{\mu\nu}}
=
\frac{1}{c} \int\limits_{x^0=\const}
	\ope{U}^{-1}(x,x_0)\circ \{
  x_\mu \Sprindex[\ope{T}]{\nu}{0} - x_\nu \Sprindex[\ope{T}]{\mu}{0}
\} \circ \ope{U}(x,x_0)
	\Id^3\bs{x}
\\			\label{3.9-4}
& \ope{L}_{\mu\nu}(x,x_0)
=
  \ope{U}(x,x_0)\circ \tope{L}_{\mu\nu} \circ \ope{U}^{-1}(x,x_0) .
	\end{align}
Since the spin angular momentum of a free scalar field is zero, the
equations
	\begin{equation}	\label{3.9-5}
\pd_\lambda \tope{L}_{\mu\nu}^{} = 0 \quad
\frac{\od\tope{L}_{\mu\nu}}{\od x^0} = 0 \quad
\pd_\lambda \tope{L}_{\mu\nu}^{\lambda} = 0
	\end{equation}
express equivalent forms of the conservation law of angular momentum. As the
treatment of the orbital angular momentum for a neutral and charged scalar
fields is quite similar, we shall present a unified consideration of the both
cases in Sect.~\ref{Sect17}.

%

	A simple, but important, conclusion from~\eref{12.131} is that the
operator
	\begin{equation}	\label{3.5}
\ope{M}^2\colon\varphi_0
\mapsto
\frac{1}{c^2} [ [ \varphi_0,\ope{P}_\mu  ]_{\_}, \ope{P}^\mu ]_{\_}
	\end{equation}
should be interpreted as square-of-mass operator of the field in momentum
(and hence in any) picture. This does not contradict to the accepted opinion
that the square-of-mass operator  is equal to the Lorentz square of the
(divided by $c$) momentum operator. The problem here is in what is called a
momentum operator and in what picture the considerations are done. Indeed, in
Heisenberg picture, we can write
	\begin{multline}	\label{3.6}
\tope{M}^2(\tope{\varphi})
=
\frac{1}{c^2} [ [ \tope{\varphi},\tope{P}_\mu  ]_{\_}, \tope{P}^\mu ]_{\_}
=
\frac{1}{c^2} [ \ih \pd_\mu\tope{\varphi} , \tope{P}^\mu ]_{\_}
=
\frac{(\ih)^2}{c^2} (\pd_\mu\pd^\mu) (\tope{\varphi}(x))
\\
   \Bigl( = - \frac{(\hbar)^2}{c^2} \tope{\square} \tope{\varphi}(x) \Bigr)
=
\frac{1}{c^2} \ope{P}_\mu^{\mathrm{QM}} \circ \ope{P}^{\mathrm{QM}\,\mu}
	(\tope{\varphi}(x))
= \cdots =
\frac{1}{c^2}
      [ [ \tope{\varphi},
      \tope{P}_\mu^{\mathrm{t}}  ]_{\_}, \tope{P}^{\mathrm{t}\,\mu} ]_{\_} ,
	\end{multline}
where~\cite{bp-QFT-momentum-operator} $\ope{P}_\mu^{\mathrm{QM}}=\ih\pd_\mu$
and
\(
\tope{P}_\mu^{\mathrm{t}}
=
\tope{P}_\mu^{\mathrm{QM}} + p_\mu\id_\Hil ,
\)
with $p_\mu=\const$ and $\id_\Hil$ being the identity mapping of $\Hil$, are
respectively the quantum mechanical and translational momentum operators. So,
we see that the conventional identification of $\tope{M}^2$ with the square
of (divided by $c$) momentum operator  corresponds to the identification of
the last operator with $\tope{P}_\mu^{\mathrm{QM}}$ (or of its square with
$-\hbar^2\tope{\square}$). For details on the last item the reader is referred
to~\cite{bp-QFT-momentum-operator}. It should be noted, the eigenvalues of
the operator~\eref{3.5} on the solutions of~\eref{12.131} characterize the
field (or its particles), while the eigenvalues of
$\frac{1}{c^2}\ope{P}_\mu\circ\ope{P}^\mu$ on the same solutions are
characteristics of the particular states these solutions represent.


\section {Analysis of the Klein-Gordon equation}
\label{Sect5}

	Our next aim is to find, if possible the (general) explicit form of
the (constant) field operator $\varphi_0$. The Klein\ndash Gordon
equation~\eref{12.131} is not enough for the purpose due to the simple fact
that the (canonical) momentum operator $\ope{P}_\mu$ depends on $\varphi_0$.
To show this, recall the definition~\eref{2.0} of $\ope{P}_\mu$ and the
expression~\eref{2.6} for it through the energy\ndash momentum operators
$\ope{T}_{\mu\nu}$ and $\tope{T}_{\mu\nu}$, given in Heisenberg and momentum
pictures, respectively, by~\eref{3.8} and~\eref{3.9}.
	Therefore~\eref{12.131},~\eref{2.6},~\eref{3.9}, and the explicit
relation~\eref{12.112} form a closed algebraic\ndash functional system of
equations for determination of $\varphi_0$ (and $\ope{P}_\mu$).

	At the beginning of our analysis of the equations defining $\varphi$,
we notice the evident solution
	\begin{gather}
			\label{3.11}
[\varphi_0,\ope{P}_\mu]_{\_} = 0 \qquad\text{for $m=0$}
\intertext{of~\eref{12.131} which, by virtue of the Heisenberg
equations/relations}
			\label{3.12}
[\tope{\varphi}(x),\tope{P}_\mu]_{\_}
			= \ih\frac{\pd\tope{\varphi}(x)}{\pd x^\mu}
\intertext{leads to
$ \tope{\varphi}(x) = \tope{\varphi}(x_0) = \const$ for $m=0$.
Consequently (in the zero mass case) our system of equations admits a,
generally non\ndash vanishing, solution (see also~\eref{12.130})}
			\label{3.13}
\tope{\varphi}(x) = \tope{\varphi}(x_0) = \const =\varphi_0
\qquad
\tope{P}_\mu=\ope{P}_\mu=0
\qquad\text{for $m=0$} .
	\end{gather}
Evidently,~\eref{3.11} implies
	\begin{equation}	\label{3.13-1}
\ope{T}_{\mu\nu} = 0 \qquad \ope{L}_{\mu\nu}^\lambda = 0,
	\end{equation}
due to~\eref{3.9} and~\eref{3.9-2}, and, consequently, the dynamical
variables for the solutions~\eref{3.11} vanish, \ie
	\begin{equation}	\label{3.13-2}
\ope{P}_{\mu} = 0 \qquad \ope{L}_{\mu\nu} = 0.
	\end{equation}

	For the general structure of the solutions of~\eref{12.131} is valid
the following result.

	\begin{Prop}	\label{Prop3.1}
Every solution $\varphi_0$ of~\eref{12.131} and~\eref{2.28} is of the form
	\begin{gather}
			\label{3.14}
\varphi_0
=
\int\Id^3\bs k
	\bigl\{
	  f_+(\bs k) \varphi_0(k)\big|_{k_0=+\sqrt{m^2c^2+{\bs k}^2} }
	+ f_-(\bs k) \varphi_0(k)\big|_{k_0=-\sqrt{m^2c^2+{\bs k}^2} }
	\bigr\}
\intertext{which can, equivalently, be rewritten as}
			\label{3.15}
\varphi_0
=
\int\Id^4k \delta(k^2-m^2c^2) f(k) \varphi_0(k) .
	\end{gather}
Here: $k=(k^0,k^1,k^2,k^3)$ is a 4-vector with dimension of 4-momentum,
$k^2=k_\mu k^\mu=k_0^2 -k_1^2 -k_2 -k_3^2=k_0^2 -{\bs k}^2$ with $k_\mu$ being
the components of $k$ and $\bs k:=(k^1,k^2,k^3)=-(k_1,k_2,k_3)$ being the
3\ndash dimensional part of $k$,
$\delta(\cdot)$ is the (1\ndash dimensional) Dirac delta function,
$\varphi_0(k)\colon\Hil\to\Hil$ is a solution of the equation
	\begin{equation}	\label{12.132}
[ \varphi_0(k), \ope{P}_\mu ]_{\_} = - k_\mu \varphi_0(k) ,
	\end{equation}
$f_{\pm}$ are complex\ndash valued functions (resp.\ distributions
(generalized functions)) of $\bs k$ for solutions different from~\eref{3.13}
(resp.\ for the solutions~\eref{3.13}),
and $f$ is a complex\ndash valued function (resp.\ distribution) of $k$
for solutions different from~\eref{3.13} (resp.\ for the
solutions~\eref{3.13}). Besides
\(
f(k)|_{k_0=\pm\sqrt{m^2c^2+{\bs k}^2} }
	= 2\sqrt{m^2c^2+{\bs k}^2} f_{\pm}(\bs k)
\)
for solutions different from~\eref{3.13}.
	\end{Prop}

	\begin{Rem}	\label{Rem3.1}
	In fact, in~\eref{3.14} enter only those solutions of~\eref{12.132}
for which
	\begin{align}	\label{12.133}
k^2 := k_\mu k^\mu =k_0^2 - {\bs k}^2 = m^2c^2 .
	\end{align}
Besides, a  non-vanishing solution of~\eref{12.132} is a solution
of~\eref{12.131} iff the condition~\eref{12.133} holds. (Proof:
write~\eref{12.131} with $\varphi_0(k)$ for $\varphi_0$ and use~\eref{12.132}
twice.)

	\end{Rem}

	\begin{Rem}	\label{Rem3.2}
	Evidently, the l.h.s. of~\eref{12.132} vanishes for the
solutions~\eref{3.13}. Therefore, we have
	\begin{gather}	\label{3.16}
\tope{\varphi}_0(x,0) = \tope{\varphi}_0(x_0,0) = \const =\varphi_0(0)
\qquad
\ope{P}_\mu = \tope{P}_\mu = 0
\intertext{where}	\label{3.17}
\tope{\varphi}_0(x,k)
	:= \ope{U}^{-1}(x,x_0)\circ \varphi_0(k)\circ \ope{U}(x,x_0) .
	\end{gather}
In terms of~\eref{3.14}, this solution is described by  $m=0$ and, for
example, $f_\pm(\bs k)=(\frac{1}{2}\pm a)\delta^3(\bs k)$, $a\in\field[C]$ or
$f(k)$ such that
$ f(k)|_{k_0=\pm|\bs k|} = (1\pm 2a) |\bs k| \delta^3(\bs k)$,
$a\in\field[C]$.
	\end{Rem}

	\begin{Rem}	\label{Rem3.3}
	Since $\varphi_0$ is Hermitian (see~\eref{2.2new}), we have
	\begin{gather}	\label{3.17new1}
\bigl(
	f_\pm(k)\varphi_0(k) \big|_{k_0=\pm\sqrt{m^2c^2+{\bs k}^2} }
\bigr)^\dag
=
 - f_\mp(-k)\varphi_0(-k) \big|_{k_0=\mp\sqrt{m^2c^2+{\bs k}^2} } .
\intertext{due to~\eref{3.14}. Also the similar relation}
			\label{3.17new}
(f(\bs k)\varphi_0(k))^\dag = f(-\bs k)\varphi_0(-k),
	\end{gather}
holds as a corollary of~\eref{3.15}.%
\footnote{~%
The same result follows also from the below-written equations~\eref{3.22}
and~\eref{3.24}. Indeed,~\eref{3.22} and
$\tope{\varphi}^\dag=\tope{\varphi}$ imply
$\underline{\tope{\varphi}}_0(k)^\dag = \underline{\tope{\varphi}}_0(-k)$,
which, in view of~\eref{3.24}, entails~\eref{3.17new}.%
}
	\end{Rem}

	\begin{Proof}
	Since the proposition was proved for the `degenerate'
solutions~\eref{3.13} in remark~\ref{Rem3.2}, below we shall suppose that
$(m,k)\not=(0,0)$ (and, hence $\ope{P}_\mu\not=0$).

	The equality~\eref{3.15} is equivalent to~\eref{3.14} (for solutions
different from~\eref{3.13}) due to
 $\delta(y^2-a^2)=\frac{1}{2a}(\delta(y-a)+\delta(y+a))$ for $a>0$.

	Let  $\varphi_0$ be given by~\eref{3.15}. Using~\eref{12.132}, we get
	\begin{gather}
			\label{3.18}
[\varphi_0,\ope{P}_\mu]_{\_}
=
-\int\Id^4k k_\mu\delta(k^2-m^2c^2) f(k) \varphi_0(k) .
\intertext{Inserting this into~\eref{12.131} and, again,
applying~\eref{12.131}, we see that~\eref{3.15} is a solution
of~\eref{12.131}. So, it remains to be proved that any solution
of~\eref{12.131} is of the form~\eref{3.15}.
\newline\indent
	The equation~\eref{12.132} in Heisenberg picture reads
(see~\eref{12.115})}
			\label{3.19new}
[\tope{\varphi}(x,k),\tope{P}_\mu]_{\_} = - k_\mu \tope{\varphi}(x,k) .
\intertext{Combining this with~\eref{3.12} (with $\tope{\varphi}(x,k)$ for
$\tope{\varphi}(x)$), we find}
			\label{3.19}
\tope{\varphi}(x,k)
=
\e^{-\iih(x^\mu-x_0^\mu)k_\mu} \tope{\varphi}(x_0,k)
=
\e^{-\iih(x^\mu-x_0^\mu)k_\mu} \ope{\varphi}_0(k)
\intertext{as}
			\label{3.20}
\ope{\varphi}_0(k) = \tope{\varphi}(x_0,k) .
\intertext{Hence, due to~\eref{12.115}, the equation~\eref{3.15} in
Heisenberg picture reads}
			\label{3.21}
\tope{\varphi}(x)
=\int\Id^4k \delta(k^2-m^2c^2)
	\e^{+\iu \frac{1}{\hbar}k_\mu x^\mu}
	\e^{\iih k_\mu x_0^\mu}
	f(k) \varphi_0(k) .
	\end{gather}

	At the end, recalling that any solution (different from~\eref{3.13})
of the Klein\ndash Gordon equation~\eref{3.4new}, satisfying the Heisenberg
relation~\eref{3.28}, admits a Fourier expansion of the
form~\cite{Bogolyubov&Shirkov,Bjorken&Drell,Roman-QFT}
	\begin{gather}
			\label{3.22}
\tope{\varphi}(x)
=\int\Id^4k \delta(k^2-m^2c^2)
	\e^{\iu \frac{1}{\hbar}k_\mu x^\mu}
	\underline{\tope{\varphi}}(k)
	\end{gather}
for some operator-valued function $\underline{\tope{\varphi}}_0(k)$
satisfying~\eref{12.132} under the condition~\eref{12.133}, viz.%
\footnote{~%
The first equality in~\eref{3.23} is the Fourier image of~\eref{2.28}.%
}
	\begin{gather}
			\label{3.23}
[\underline{\tope{\varphi}}(k),\ope{P}_\mu]_{\_}
		= -k_\mu \underline{\tope{\varphi}}(k)
\qquad
k^2=m^2c^2 ,
\intertext{the proof is completed by the identification}
			\label{3.24}
\underline{\tope{\varphi}}(k)
=
e^{\iih k_\mu x_0^\mu}f(k) \varphi_0(k)
=
e^{\iih k_\mu x_0^\mu}f(k) \tope{\varphi}(x_0,k)
	\end{gather}
and subsequent return from Heisenberg to momentum picture.
	\end{Proof}

	Meanwhile, we have proved two quite important results. On one hand,
by virtue of~\eref{3.24} and the homogeneous character of~\eref{12.132}, any
solution of~\eref{12.131} can be written as
	\begin{equation}	\label{3.25}
\varphi_0=\int \delta(k^2-m^2c^2) \underline{\varphi}_0(k) \Id^4k
	\end{equation}
where $\varphi_0(k)$ are \emph{appropriately normalized (scaled) solutions
of}~\eref{12.132} which solutions can be identified, up to the phase factor
$\e^{\iih x_0^\mu k_\mu}$, with the Fourier coefficients of
$\tope{\varphi}(x)$, \viz (see~\eref{3.22} and~\eref{3.24})
	\begin{equation}	\label{3.25new}
\underline{{\tope{\varphi}}}(k)
=
\e^{\iih x_0^\mu k_\mu} \underline{\varphi}_0(k) .
	\end{equation}
	Therefore, by virtue of~\eref{3.17new} and~\eref{3.24}, the
operators $\varphi_0(k)$ appearing in~\eref{3.25} satisfy the relation
(cf.~\eref{3.17new})
	\begin{equation}	\label{3.25new1}
(\varphi_0(k))^\dag = \varphi_0(-k) .
	\end{equation}
On the other hand, this implies that the operator $\varphi_0(k)$, entering
in~\eref{3.25}, is (up to a constant) identical with the usual momentum
representation of a scaler field  $\tope{\varphi}(x)$ in Heisenberg
picture~\cite{Bogolyubov&Shirkov,Bjorken&Drell,Roman-QFT}.
\emph{
Consequently, the operators $\underline{\tope{\varphi}}(k)$ with
$k^2=m^2c^2$, representing a free Hermitian scalar field in momentum
representation of Heisenberg picture of motion in quantum field theory, are
nothing else but (suitably normalized) solutions of~\eref{12.132}  under the
condition~\eref{12.133}, \ie of the Klein\ndash Gordon
equation~\eref{12.131}, which form a basis in the operator space of all
solutions of~\eref{12.131}.
}
In short, the momentum representation of a scalar field in Heisenberg picture
is a suitably chosen base for the solutions of the Klein\ndash Gordon
equation in momentum picture. This result is quite important from two aspects.
On one hand, it reveals the real meaning of the momentum representation in
Heisenberg picture of quantum field theory. On another hand, it allows us to
apply freely in momentum picture  all of the already established results
concerning the Fourier images of the field operators, observables and other
operators in Heisenberg picture. In particular, this concerns the frequency
decompositions and creation and annihilation operators theory.


\section
[Frequency decompositions and their physical meaning]
{Frequency decompositions and their physical meaning}
	\label{Sect6}

	Since the decomposition~\eref{3.14} (or~\eref{3.15} if the
solutions~\eref{3.13} are excluded)  is similar to the one leading to the
frequency decompositions (in Heisenberg picture for free fields) in quantum
field theory, we shall introduce similar (in fact identical) notation.
Defining%
\footnote{~%
Notice, we do not exclude the case $k_0=0$ as it is done in the literature.%
}
	\begin{gather}
			\label{3.26}
\varphi_0^\pm(k)=
	\begin{cases}
f_\pm(\pm \bs k) \varphi_0(\pm k)	& \text{for $k_0\ge0$} \\
0					& \text{for $k_0<0$}
	\end{cases}
\quad ,
\intertext{we see that}
			\label{3.27}
\varphi_0 = \varphi_0^+ + \varphi_0^-
\\			\label{3.28}
\varphi_0^\pm
=
\int\Id^3\bs k \varphi_0^\pm (k) |_{k_0=\sqrt{m^2c^2+{\bs k}^2}}
\\			\label{3.29}
[\varphi_0^\pm(k),\ope{P}_\mu]_{\_} = \mp k_\mu \varphi_0^\pm(k)
\qquad
k_0=\sqrt{m^2c^2+{\bs k}^2}
	\end{gather}
due to~\eref{3.14} and~\eref{12.132}. Notice, since $\varphi_0$ and
$\varphi_0(k)$ satisfy~\eref{2.2new}, \eref{3.17new} and~\eref{3.17new1}, we
have
	\begin{equation}	\label{3.29new}
( \varphi_0^\pm(k) )^\dag = \varphi_0^\mp(k) .
	\end{equation}

	The equation~\eref{3.29} implies the interpretation of a free scalar
field in terms of particles. Indeed, if $\ope{X}_p$ is a state vector of a
state with 4\ndash momentum $p$, \ie
	\begin{gather}	\label{3.30}
\ope{P}_\mu(\ope{X}_p) = p_\mu \ope{X}_p ,
\intertext{then, applying~\eref{3.29}, we obtain}
			\label{3.31}
\ope{P}_\mu (\varphi_0^\pm(k) (\ope{X}_p))
=
( p_\mu\pm k_\mu ) \varphi_0^\pm(k)(\ope{X}_p)
\qquad
k_0=\sqrt{m^2c^2+{\bs k}^2}
	\end{gather}
So,  $\varphi_0^+(k)$ (resp.\ $\varphi_0^-(k)$) can be interpreted as an
operator creating (resp.\ annihilating) a particle (quant of the field)
with mass $m$ and 4\ndash momentum $k$ with $k_0=\sqrt{m^2c^2+{\bs k}^2}$,
\ie a particle with energy $\sqrt{m^2c^2+{\bs k}^2}$ and 3\ndash momentum
$\bs k$. In this situation, the vacuum for a scalar field should be a state
$\ope{X}_0$ with vanishing 4\ndash momentum and such that
	\begin{equation}	\label{3.32}
\varphi_0^-(k)(\ope{X}_0) = 0   \qquad k_0=\sqrt{m^2c^2+{\bs k}^2} .
	\end{equation}
The action of the operators $\varphi_0^+(k)$ on $\ope{X}_0$ produces one- or
multiple\ndash particle states. More details concerning the definition of a
vacuum and construction of state vectors from it will be given in
sections~\ref{Sect9} and~\ref{Sect10}.

	For a comparison with existing literature (see,
e.g.,~\cite{Bogolyubov&Shirkov,Bjorken&Drell,Roman-QFT,Itzykson&Zuber}), we
notice that the creation/annihilation operators $\tope{\varphi}^\pm$,
introduced in momentum representation of Heisenberg picture of motion, are
defined by
	\begin{equation}
			\label{3.32-1}
\tope{\varphi}^\pm(k)=
	\begin{cases}
\underline{{\tope{\varphi}}} (\pm k)	& \text{for $k_0\ge0$} \\
0					& \text{for $k_0<0$}
	\end{cases}
	\end{equation}
where $\underline{{\tope{\varphi}}} (k)$ is the Fourier image of
$\tope{\varphi}(x)$ (see~\eref{3.22}). Therefore, by virtue of~\eref{3.24}
and proposition~\ref{Prop3.1}, we have the following connection between the
creation/annihilation operators in Heisenberg and momentum picture:
	\begin{equation}	\label{3.32-2}
\tope{\varphi}^\pm(k)
=
\e^{\pm\iih x_0^\mu k_\mu} \varphi_0^\pm(k)
\qquad
k_0=\sqrt{m^2c^2+{\bs k}^2} .
	\end{equation}
We leave to the reader to prove as an exercise that the general
formula~\eref{12.115} implies
	\begin{equation}	\label{3.32-3}
\varphi^\pm(k)
=
\e^{\mp\iih x^\mu k_\mu}
\ope{U}(x,x_0)\circ \tope{\varphi}_0^\pm(k) \circ \ope{U}^{-1}(x,x_0)
\qquad
k_0=\sqrt{m^2c^2+{\bs k}^2} ,
	\end{equation}
which, in view of~\eref{2.28} and~\eref{3.29}, is equivalent
to~\eref{3.32-2}. (Hint: apply the Fourier decomposition from
Sect.~\ref{Sect5}, then show that $\frac{\pd}{\pd x^\mu}\varphi^\pm(k)=0$
and, at last, set $x=x_0$ in~\eref{3.32-3}.)


\section
[The field equations in terms of creation and annihilation operators]
{The field equations in terms of creation and \\annihilation operators}
\label{Sect7}

	Let us return now to our main problem: to be found, if possible, the
explicit form of the field operator $\varphi_0$ (and momentum operator
$\ope{P}_\mu$).

	At first, we shall express $\ope{P}_\mu$ in terms of the creation and
annihilation operators $\varphi_0^\pm$. Regardless of the fact that the
result is known (see, e.g.,~\cite[eq.~(3.26)]{Bogolyubov&Shirkov}
or~\cite[eq.~(12.11)]{Bjorken&Drell}), we shall reestablish it in a
completely new way.

	Since~\eref{3.27}--\eref{3.29} entail
	\begin{gather}	\notag
[\varphi_0,\ope{P}_\mu]_{\_}
=
\int\Id^3\bs k \bigl\{ k_\mu (-\varphi_0^+(k) + \varphi_0^-(k)) \bigr\}
	|_{ k_0=\sqrt{m^2c^2+{\bs k}^2} } ,
	\end{gather}
from~\eref{3.9}, after some algebra (involving~\eref{3.27}
and~\eref{3.28}), we get the energy\ndash momentum operator as
	\begin{subequations}	\label{3.33}
	\begin{multline}	\label{3.33a}
\ope{T}_{\nu\mu}
=
c^2\int\Id^3\bs k\Id^3\bs k'
	\bigl\{
( - k_\nu k'_\mu + \frac{1}{2}\eta_{\nu\mu} k_\lambda k^{\prime\,\lambda} )
( - \varphi_0^+(k) + \varphi_0^-(k) )
\\ \times
( - \varphi_0^+(k') + \varphi_0^-(k') )
+ \frac{1}{2}\eta_{\nu\mu} m^2c^2
( \varphi_0^+(k) + \varphi_0^-(k) )
( \varphi_0^+(k') + \varphi_0^-(k') )
	\bigr\}
	\end{multline}
\vspace{-3ex}
	\begin{multline}	\label{3.33b}
\ope{T}_{\nu\mu}
=
 \frac{1}{2} c^2\int\Id^3\bs k\Id^3\bs k'
	\bigl\{
( - k_\nu k'_\mu - k_\mu k'_\nu
+ \eta_{\nu\mu} k_\lambda k^{\prime\,\lambda} )
( - \varphi_0^+(k) + \varphi_0^-(k) )
\\ \times
( - \varphi_0^+(k') + \varphi_0^-(k') )
+ \eta_{\nu\mu} m^2c^2
( \varphi_0^+(k) + \varphi_0^-(k) )
( \varphi_0^+(k') + \varphi_0^-(k') )
	\bigr\} ,
	\end{multline}
	\end{subequations}
where
 $k_0=\sqrt{m^2c^2+{\bs k}^2}$,
 $k'_0=\sqrt{m^2c^2+{\bs k'}^2}$
and the two expressions for $\ope{T}_{\mu\nu}$ correspond to the two its
versions in~\eref{3.9}.

The idea is now the last result to be inserted into~\eref{2.6} and to
commute $\varphi_0^\pm(k)$ and $\varphi_0^\pm(k')$ with $\ope{U}(x,x_0)$ in
order to `move' $\varphi_0^\pm(k)$ and $\varphi_0^\pm(k')$  to the right of
$\ope{U}(x,x_0)$. Rewriting~\eref{12.132} as
$\varphi_0(k)\circ\ope{P}_\mu =(\ope{P}_\mu-k_\mu\id_\Hil)\circ\varphi_0(k)$,
by induction, we derive
	\begin{gather}
			\label{3.34}
\varphi_0(k)\circ( \ope{P}_{\mu_1}\circ \dots \circ\ope{P}_{\mu_a} )
=
\{
 (\ope{P}_{\mu_1} - k_{\mu_1} \id_\Hil)
 	\circ\dots\circ (\ope{P}_{\mu_a} - k_{\mu_a} \id_\Hil)
\} \circ \varphi_0(k)
\intertext{for any $a\in\field[N]$, which, in view of the expansion of the
r.h.s.\ of~\eref{12.112} into a power series, implies}
			\label{3.35}
\varphi_0(k)\circ\ope{U}(x,x_0)
=
\e^{-\iih(x^\mu-x_0^\mu)k_\mu} \ope{U}(x,x_0) \circ \varphi_0(k) .
\\\intertext{So, we have (see~\eref{3.26})}
			\label{3.35new}
	\begin{split}
\varphi_0^\pm(k)\circ\ope{U}(x,x_0)
&=
\e^{-\iih(x^\mu-x_0^\mu)(\pm k_\mu)} \ope{U}(x,x_0) \circ \varphi_0^\pm(k)
\\
\varphi_0^+(k)\circ \varphi_0^\pm(k')\circ\ope{U}(x,x_0)
&=
\e^{-\iih(x^\mu-x_0^\mu)(k_\mu\pm k'_\mu)}
		\ope{U}(x,x_0) \varphi_0^+(k) \circ \varphi_0^\pm(k')
\\
\varphi_0^-(k)\circ \varphi_0^\pm(k')\circ\ope{U}(x,x_0)
&=
\e^{-\iih(x^\mu-x_0^\mu)(-k_\mu\pm k'_\mu)}
		\ope{U}(x,x_0) \varphi_0^-(k) \circ \varphi_0^\pm(k') .
	\end{split}
	\end{gather}
At the end, substituting~\eref{3.33} into~\eref{2.6}, applying the derived
commutation rules~\eref{3.35new}, performing the integration over $x$, which
yields $\delta$\ndash functions like $\delta(\bs k\pm\bs k')$, and the
integration over $\bs k'$, we finally get after, a simple, but lengthy and
tedious, calculation, the following result
	\begin{gather}	\label{3.36}
\ope{P}_\mu
=
\frac{1}{2}\int
	k_\mu |_{ k_0=\sqrt{m^2c^2+{\bs k}^2} }
\{
\varphi_0^+(\bs k)\circ\varphi_0^-(\bs k)
+
\varphi_0^-(\bs k)\circ\varphi_0^+(\bs k)
\}
\Id^3\bs k ,
\intertext{where we have introduced the 3-dimensional renormalized creation
and annihilation operators}
			\label{3.37}
\varphi_0^\pm(\bs k)
:=
\bigl\{
(2c(2\pi\hbar)^3k_0)^{1/2}
\varphi_0^\pm(k)
\bigr\}	\big|_{ k_0=\sqrt{m^2c^2+{\bs k}^2} } .
	\end{gather}
Notice, the result~\eref{3.36} is independent of from what form of
$\ope{T}_{\mu\nu}$, \eref{3.9a} or~\eref{3.9b}, we have started.
The only difference of our derivation of~\eref{3.36} from similar one in the
literature (see, e.g.,~\cite[sec.~3.2]{Bogolyubov&Shirkov}) is that we have
\emph{not} exclude the massless case and the degenerate solution~\eref{3.13}
from our considerations. The operator $\varphi_0^+(\bs k)$ (resp.\
$\varphi_0^-(\bs k)$) is called the \emph{creation} (resp.\
\emph{annihilation}) operator (of the field).

	We would like to emphasize on the relations
	\begin{equation}	\label{3.37new}
(\varphi_0^\pm(\bs k))^\dag = \varphi_0^\mp(\bs k)
	\end{equation}
which follow from~\eref{3.37} and~\eref{3.29new}. Actually these equalities
are equivalent to the supposition that $\varphi_0$ is Hermitian field
operator.

	As a result of~\eref{3.32-2}, the operators
$\tope{\varphi}^\pm(\bk)$, corresponding to~\eref{3.36} in (the momentum
representation of) Heisenberg picture of motion,
are~\cite[Sec.~3.2]{Bogolyubov&Shirkov}
	\begin{equation}	\label{3.37new1}
	\begin{split}
\tope{\varphi}^\pm(\bk)
:& =
\bigl\{
\bigl( c(2\pi\hbar)^3 \bigr)^{1/2}  (2k_0)^{-1/2} \tope{\varphi}^\pm(k)
\bigr\} \big|_{k_0=\sqrt{m^2c^2+\bk^2}}
\\
& =
\e^{\pm\iih x_0^\mu k_\mu} \big|_{k_0=\sqrt{m^2c^2+\bk^2}}
\varphi_0^\pm(\bk) .
	\end{split}
	\end{equation}
Similarly, we have also the connection (see~\eref{3.32-3})
	\begin{equation}	\label{3.37new2}
\tope{\varphi}^\pm(\bk)
=
\e^{\pm\iih x^\mu k_\mu} \big|_{k_0=\sqrt{m^2c^2+\bk^2}}
\ope{U}^{-1}(x,x_0)\circ \varphi_0^\pm(\bk) \circ \ope{U}(x,x_0) .
	\end{equation}
Consequently, the integrand in~\eref{3.36} and similar ones which will be met
further in this work, look identically in terms of $\varphi_0^\pm(\bk)$ and
$\tope{\varphi}^\pm(\bk)$.

	Summarizing the above results, we are ready to state and analyze the
following problem.

	\begin{Prob}	\label{Prob3.1}
Let
	\begin{gather}	\label{3.38}
	\begin{split}
\varphi_0
& = \varphi_0^+ + \varphi_0^-
= \int (\varphi_0^+(k) + \varphi_0^-(k))
				|_{k_0=\sqrt{m^2c^2+{\bs k}^2}} \Id^3\bs k
\\
& = \int
\Bigl( 2c(2\pi\hbar)^3 \sqrt{m^2c^2+{\bs k}^2} \, \Bigr)^{-1/2}
			(\varphi_0^+(\bs k) + \varphi_0^-(\bs k)) \Id^3\bs k
	\end{split}
\intertext{be a solution of the Klein-Gordon equation~\eref{12.131}. Find the
general explicit form of the operators $\varphi_0^\pm(\bs k)$ which are
solutions of the equations (see~\eref{3.29})}
			\label{3.39}
[ \varphi_0^\pm(\bs k),\ope{P}_\mu]_{\_} = \mp k_\mu \varphi_0^\pm(\bs k)
\qquad
k_0=\sqrt{m^2c^2+{\bs k}^2}
	\end{gather}
where $\ope{P}_\mu$ is given by~\eref{3.36}.
	\end{Prob}

	Inserting the representation~\eref{3.36} into~\eref{3.39}
and writing the expression $\mp k_\mu\varphi_0^\pm(\bs k)$ as
$\int \mp q_\mu\varphi_0^\pm(\bs k)\delta^3(\bs k-\bs q)\Id^3\bs q$
with $q_0=\sqrt{m^2c^2+{\bs q}^2}$, we, after a simple algebraic
manipulation, obtain
	\begin{multline}	\label{3.40}
\int
q_\mu |_{q_0=\sqrt{m^2c^2+{\bs q}^2}}
\bigl\{
[ \varphi_0^\pm(\bs k), \varphi_0^+(\bs q)\circ\varphi_0^-(\bs q)
		      + \varphi_0^-(\bs q)\circ\varphi_0^+(\bs q) ]_{\_}
\\
\pm 2 \varphi_0^\pm(\bs k) \delta^3(\bs k-\bs q)
\bigr\}
\Id^3\bs q
= 0
	\end{multline}
Consequently $\varphi_0^\pm(\bs k)$ must be solutions of
	\begin{equation}	\label{3.41}
[ \varphi_0^\pm(\bs k), [ \varphi_0^+(\bs q), \varphi_0^-(\bs q) ]_{+} ]_{\_}
\pm2 \varphi_0^\pm(\bs k) \delta^3(\bs k-\bs q)
= f^\pm(\bs k, \bs q)
	\end{equation}
where $[A,B]_{+}:=A\circ B+B\circ A$ is the anticommutator of operators
(or matrices) $A$ and $B$ and $f^\pm(\bs k, \bs q)$ are operator\ndash
valued (generalized) functions such that
	\begin{equation}	\label{3.42}
\int q_\mu |_{q_0=\sqrt{m^2c^2+{\bs q}^2}} f^\pm(\bs k, \bs q) \Id^3\bs q = 0.
	\end{equation}

	Looking over the derivation of~\eref{3.41}, we see its equivalence
with the initial system of Klein\ndash Gordon equation~\eref{12.131} and
Heisenberg relations~\eref{2.28} in momentum picture.
Consequently,~\eref{3.41} is the \emph{system of field equations in terms
of creation and annihilation operators in momentum picture}.

	As a simple test of our calculations, one can prove that the
commutativity of the components $\ope{P}_\mu$ of the momentum operator,
expressed by~\eref{2.1} (see also~\eref{2.3}), is a consequence
of~\eref{3.36}, \eref{3.41}, and~\eref{3.42}.


\section {The commutation relations}
\label{Sect8}

	The equations~\eref{3.41} are the corner stone of the famous
(anti)commutation relations in quantum field theory for the considered here
free Hermitian scalar field.%
\footnote{~%
The idea for arbitrary (Hermitian/real or non\ndash Hermitian/complex, free
or interacting) fields remains the same: one should derive the Euler\ndash
Lagrange equations in momentum picture  and to solve them relative to the
field operators by using the explicit expression of the (canonical) momentum
operator through the field operators.%
}
The equations~\eref{3.41} form a system of two homogeneous algebraic equations
of third order relative to the functions $\varphi_0^\pm(\bs k)$. Generally,
it has infinitely may solutions, but, at present, only a selected class of
them has a suitable physical meaning and interpretation. This class will be
described a little below.

	Since for the physics is essential only the restriction of
$\varphi_0^\pm(\bs k)$ on the \emph{physically realizable states}, not on the
whole system's Hilbert space $\Hil$, we shall analyze~\eref{3.41} in this
case, following the leading idea in similar cases in,
e.g.,~\cite[subsec.~10.1]{Bogolyubov&Shirkov},
or~\cite[\S~70]{Bjorken&Drell}, or~\cite[p.~65]{Roman-QFT}. It consists in
\emph{admitting} that the commutator or anticommutator of the creation and/or
annihilation operators (for free fields) is a  $c$\ndash number, \ie it is
proportional to the identity mapping $\id_\Hil$ of the system's Hilbert space
$\Hil$ of states. In particular, in our case, this (additional hypothesis)
results in
	\begin{equation}	\label{3.43}
[ \varphi_0^\pm(\bs k) , \varphi_0^\pm(\bs q) ]_\varepsilon
=
\alpha_\varepsilon^\pm(\bs k,\bs q) \id_\Hil
\qquad
[ \varphi_0^\mp(\bs k) , \varphi_0^\pm(\bs q) ]_\varepsilon
=
\beta_\varepsilon^\pm(\bs k,\bs q) \id_\Hil
	\end{equation}
where $\varepsilon=\pm$, $[A,B]_\pm:=A\circ B\pm B\circ A$, and
$\alpha_\varepsilon^\pm$ and $\beta_\varepsilon^\pm$ are complex\ndash valued
(generalized) functions. It should clearly be understood, the
conditions~\eref{3.43} are \emph{additional} to the Lagrangian formalism and
do not follow from it. In fact, they or the below written commutation
relations~\eref{3.57} (which under some conditions follow from~\eref{3.43})
have to be \emph{postulated}.

	Defining
	\begin{gather}
			\label{3.43-1}
[A,B]_{\eta} = A\circ B + \eta B\circ A \qquad \eta\in\field[C]
\\\intertext{and applying the identity}
			\label{3.43-2}
[A,B\circ C]_{\eta}
=
[A,B]_{-\eta\varepsilon} \circ C
	+\eta\varepsilon B\circ [A,C]_\varepsilon
\qquad \varepsilon=\pm 1
\intertext{for $\eta=-1$, \viz}
			\label{3.43-3}
[A,B\circ C]_{\_}
= [A,B]_\varepsilon \circ C -\varepsilon B\circ [A,C]_\varepsilon ,
	\end{gather}
due to $\varepsilon=\pm1$ and $[A,B]_{\pm1}=[A,B]_{\pm}$, we
rewrite~\eref{3.41} as
	\begin{equation}	\label{3.44}
	\begin{split}
[ \varphi_0^\pm(\bs k), \varphi_0^+(\bs q) ]_\varepsilon
	\circ\varphi_0^-(\bs q)
	- \varepsilon \varphi_0^+(\bs q)\circ
& [ \varphi_0^\pm(\bs k),\varphi_0^-(\bs q) ]_\varepsilon
\\
+ [ \varphi_0^\pm(\bs k),\varphi_0^-(\bs q) ]_\varepsilon
	\circ\varphi_0^+(\bs q)
	- \varepsilon \varphi_0^-(\bs q)\circ
& [ \varphi_0^\pm(\bs k),\varphi_0^+(\bs q) ]_\varepsilon
\\
\pm 2 & \varphi_0^\pm(\bs k) \delta^3(\bs k-\bs q)
=
f^\pm(\bs k,\bs q) .
	\end{split}
	\end{equation}
Now, substituting the \emph{additional conditions}~\eref{3.43} into~\eref{3.44},
we get
	\begin{multline}	\label{3.45}
(1-\varepsilon) \varphi_0^-(\bs q)\circ
		[ \varphi_0^\pm(\bs k),\varphi_0^+(\bs q) ]_\varepsilon
\\+
(1-\varepsilon) \varphi_0^+(\bs q)\circ
		[ \varphi_0^\pm(\bs k),\varphi_0^-(\bs q) ]_\varepsilon
\pm 2 \varphi_0^\pm(\bs k) \delta^3(\bs k-\bs q)
=f^\pm(\bs k,\bs q) .
	\end{multline}

	For $\varepsilon=+1$, this equality reduces to
$ \pm 2 \varphi_0^\pm(\bs k) \delta^3(\bs k-\bs q) =f^\pm(\bs k,\bs q) $
which, when inserted into~\eref{3.42}, entails
	\begin{gather}
			\label{3.45-1}
k_\mu |_{k_0=\sqrt{m^2c^2+{\bs k}^2}} \varphi_0^\pm(\bs k) = 0
\\\intertext{for any $\mu=0,1,2,3$ and $k$. Consequently, the
choice $\varepsilon=+1$, \ie}
			\label{3.46}
[ \varphi_0^\pm(\bs k) , \varphi_0^\pm(\bs q) ]_+
=
\alpha_+^\pm(\bs k,\bs q) \id_\Hil
\qquad
[ \varphi_0^\mp(\bs k) , \varphi_0^\pm(\bs q) ]_+
=
\beta_+^\pm(\bs k,\bs q) \id_\Hil ,
\\\intertext{leads, in view of~\eref{3.36} and~\eref{3.45-1}, to}
			\label{3.45-2}
\ope{P}_\mu = 0
\\\intertext{which, by virtue of Klein-Gordon equation~\eref{12.131}, implies}
			\label{3.45-3}
m^2c^2\varphi_0 = 0.
\\\intertext{So, if $m=0$, we get the solution~\eref{3.13} and, if $m\not=0$,
we derive}
			\label{3.47}
\varphi_0 = 0 \qquad\text{for $m\not=0$}
\\\intertext{and, consequently,}
			\label{3.46new}
\varphi_0^\pm(\bs k) \equiv 0 \qquad\text{for $m\not=0$}
	\end{gather}
We interpret the obtained solution~\eref{3.47} (or~\eref{3.46new}) of the
Klein\ndash Gordon equation~\eref{12.131} as a complete absence of the
physical scalar field.%
\footnote{~%
Since, for a scalar field, the solutions~\eref{3.13} do not lead to any
physically predictable results, they, in the massless case, should also be
interpreted as absence of the field.%
}

	Consider now~\eref{3.45} with $\varepsilon=-1$. Writing it explicitly
for the upper, ``$+$'', and lower, ``$-$'', signs, we see that~\eref{3.45} is
equivalent to
	\begin{equation}	\label{3.48}
\varphi_0^\mp(\bs q)\circ [ \varphi_0^\pm(\bs k),\varphi_0^\pm(\bs q) ]_{\_}
+
[ \varphi_0^\pm(\bs k),\varphi_0^\mp(\bs q) ]_{\_} \circ \varphi_0^\pm(\bs q)
\pm  \varphi_0^\pm(\bs q) \delta^3(\bs k-\bs q)
=
\frac{1}{2}f^\pm(\bs k,\bs q)
	\end{equation}
where, in accord with~\eref{3.43}, the commutators are
	\begin{equation}	\label{3.49}
[ \varphi_0^\pm(\bs k),\varphi_0^\pm(\bs q) ]_{\_}
				= \alpha^\pm(\bs k,\bs q) \id_\Hil
\qquad
[ \varphi_0^\mp(\bs k),\varphi_0^\pm(\bs q) ]_{\_}
				=  \beta^\pm(\bs k,\bs q) \id_\Hil .
	\end{equation}
Here, for brevity, we have omit the subscript ``$-$'', \ie we write
$\alpha^\pm$ and $\beta^\pm$ for
$\alpha_-^\pm$ and $\beta_-^\pm$ respectively.

	The properties of $\alpha^\pm$ and $\beta^\pm$  can be specified as
follows.%
\footnote{~%
For the initial idea, see~\cite[subsec.~10.1]{Bogolyubov&Shirkov}, where the
authors (premeditated or not?) make a number of implicit assumptions which
reduce the generality of the possible (anti)commutation relations.%
}

	Let $\ope{X}_p$  be a state vector with fixed 4-momentum $p$
(see~\eref{3.30}). Defining
	\begin{gather}	\label{3.50}
\ope{X}_{k,q}^\pm
	:= (\varphi_0^\pm(\bs k) \circ \varphi_0^\pm(\bs q) ) (\ope{X}_p)
\qquad
\ope{Y}_{k,q}^\pm
	:= (\varphi_0^\pm(\bs k) \circ \varphi_0^\mp(\bs q) ) (\ope{X}_p) ,
\\\intertext{we, applying~\eref{3.31}, get}
			\label{3.51}
\ope{P}_\mu (\ope{X}_{k,q}^\pm )
	= (p_\mu \pm k_\mu \pm q_\mu) \ope{X}_{k,q}^\pm
\qquad
\ope{P}_\mu (\ope{Y}_{k,q}^\pm )
	= (p_\mu \pm k_\mu \mp q_\mu) \ope{Y}_{k,q}^\pm  .
\\\intertext{Noticing that}
			\notag
\ope{X}_{k,q}^\pm  + \varepsilon \ope{X}_{q,k}^\pm
	= [ \varphi_0^\pm(\bs k),\varphi_0^\pm(\bs q) ]_\varepsilon (\ope{X}_p)
\qquad
\ope{Y}_{k,q}^\mp  + \varepsilon \ope{Y}_{q,k}^\pm
	= [ \varphi_0^\mp(\bs k),\varphi_0^\pm(\bs q) ]_\varepsilon (\ope{X}_p)
\intertext{for $\varepsilon=\pm 1$, we see that}
			\label{3.52}
	\begin{split}
\bigl( \ope{P}_\mu\circ
	[ \varphi_0^\pm(\bs k),\varphi_0^\pm(\bs q) ]_\varepsilon \bigr)
	(\ope{X}_p)
& =
(p_\mu \pm k_\mu \pm q_\mu)
	[ \varphi_0^\pm(\bs k),\varphi_0^\pm(\bs q) ]_\varepsilon
	(\ope{X}_p)
\\
\bigl( \ope{P}_\mu\circ
	[ \varphi_0^\mp(\bs k),\varphi_0^\pm(\bs q) ]_\varepsilon \bigr)
	(\ope{X}_p)
& =
(p_\mu \mp k_\mu \pm q_\mu)
	[ \varphi_0^\mp(\bs k),\varphi_0^\pm(\bs q) ]_\varepsilon
	(\ope{X}_p) .
	\end{split}
	\end{gather}
It is worth to mention, in the derivation of~\eref{3.52} no additional
hypothesis, like~\eref{3.43}, have been used.

	Applying~\eref{3.52} for $\varepsilon=-1$ and imposing the
\emph{additional conditions}~\eref{3.49}, we, due to~\eref{3.30}, find
	\begin{gather}	\label{3.53}
(\bs k +\bs q) \alpha^\pm(\bs k,\bs q) = 0
\qquad
(\bs k -\bs q) \beta^\pm(\bs k,\bs q) = 0
\\			\label{3.53-1}
	\begin{split}
\bigl( \sqrt{m^2c^2+{\bs k}^2} + \sqrt{m^2c^2+{\bs q}^2} \bigr)
	\alpha^\pm(\bs k,\bs q) &= 0
\\
\bigl( \sqrt{m^2c^2+{\bs k}^2} - \sqrt{m^2c^2+{\bs q}^2} \bigr)
	\beta^\pm(\bs k,\bs q)  &= 0 .
	\end{split}
	\ \Bigg\}
	\end{gather}
Here, with necessity, $\alpha^\pm$ and $\beta^\pm$ must be regarded as
distributions (generalized functions) as otherwise the
equations~\eref{3.53} have only the trivial solutions
	\begin{equation}	\label{3.53new}
\alpha^\pm(\bs k,\bs q) = 0 \quad\text{for $\bs k+\bs q\not=0$}
\qquad
\beta^\pm(\bs k,\bs q) = 0 \quad\text{for $\bs k-\bs q\not=0$}
	\end{equation}
which, as it can easily be seen, reproduce the trivial solution~\eref{3.47}
of the Klein\ndash Gordon equation. Rewriting~\eref{3.53} as
$\bs q \alpha^\pm(\bs k,\bs q) = - \bs k \alpha^\pm(\bs k,\bs q)$ and
$\bs q  \beta^\pm(\bs k,\bs q) =   \bs k \beta^\pm(\bs k,\bs q)$,
we see that
	\begin{equation}	\label{3.54}
f(\bs q) \alpha^\pm(\bs k,\bs q) = f(-\bs k) \alpha^\pm(\bs k,\bs q)
\qquad
f(\bs q) \beta^\pm(\bs k,\bs q)  = f(\bs k) \beta^\pm(\bs k,\bs q)
	\end{equation}
for a function $f(\bs k)$ which is supposed to be polynomial or convergent
power series.%
\footnote{~%
We cannot write, e.g.,
$\alpha^\pm(\bs k,\bs q)=\const\times\delta^3(\bs k+\bs q)$
as the equation $yg(y)=0$, $y\in\field[R]$, has a solution
$g(y)=\const\times\delta(y)$, but this is \emph{not} its general solution;
e.g., its solutions are
\(
g(y)
= \alpha_0\delta(y) + \alpha_1 y \frac{\od\delta(y)}{\od y}
  + \alpha_2 y^2 \frac{\od^2\delta(y)}{\od y^2}
  + \cdots
\)
with $\alpha_0,\alpha_1,\dots$ being constant numbers.%
}
As a result of~\eref{3.54}, the second equation in~\eref{3.53-1} is
equivalent to the identity $0=0$, while the first one reduces to
	\begin{equation}	\label{3.53-2}
\sqrt{m^2c^2+{\bs k}^2} \alpha^\pm(\bs k,\bs q) = 0 .
	\end{equation}

	Inserting~\eref{3.49} into~\eref{3.48}, we find
	\begin{equation}	\label{3.54new}
\frac{1}{2} f^\pm(\bs k,\bs q)
=
 \varphi_0^\mp(\bs q) \alpha^\pm(\bs k,\bs q)
+
 \varphi_0^\pm(\bs q) (\beta^\mp(\bs k,\bs q) \pm \delta^3(\bs k-\bs q) ) .
	\end{equation}

	At the end, besides~\eref{3.53-2}, the condition~\eref{3.42} is the
only one remaining to be satisfied. Substituting~\eref{3.54new} into it and
using~\eref{3.54} for $f(\bs q)=\sqrt{m^2c^2+{\bs q}^2}$ in the case $\mu=0$
and for $f(\bs q)=q_a$ in the one with $\mu=a=1,2,3$, we obtain

	\begin{subequations}	\label{3.54-1}
	\begin{align}	\label{3.54-1a}
0 &=
k_a \int \bigl\{
- \alpha^\pm(\bs k,\bs q) \varphi_0^\mp(\bs q)
+ (\beta^\mp(\bs k,\bs q) \pm \delta^3(\bs k-\bs q) ) \varphi_0^\pm(\bs q)
\bigr\} \Id^3\bs q
\\			\label{3.54-1b}
0 &=
\sqrt{m^2c^2+{\bs k}^2} \int \bigl\{
+ \alpha^\pm(\bs k,\bs q) \varphi_0^\mp(\bs q)
+ (\beta^\mp(\bs k,\bs q) \pm \delta^3(\bs k-\bs q) ) \varphi_0^\pm(\bs q)
\bigr\} \Id^3\bs q ,
	\end{align}
	\end{subequations}
where, in the second equation, the term containing $\alpha^\pm$ vanishes due
to~\eref{3.53-2}. Since these equations must be valid for arbitrary $\bs k$,
the integrals in them should vanish if $\bs k\not=0$. Forming the sum and
difference of these integrals in that case, we get
	\begin{subequations}	\label{3.55}
	\begin{align}	\label{3.55a}
\int \alpha^\pm(\bs k,\bs q) \varphi_0^\mp(\bs q )  \Id^3\bs q &=0
\\			\label{3.55b}
\int (\beta^\mp(\bs k,\bs q) \pm \delta^3(\bs k-\bs q) )
				\varphi_0^\pm(\bs q )  \Id^3\bs q &=0 .
	\end{align}
	\end{subequations}

	If $(m,\bs k)\not=(0,\bs0)$, the standard (Bose-Einstein) commutation
relations are extracted from equations~\eref{3.54-1} (or~\eref{3.55} if
$\bs k\not=0$) by imposing a \emph{second, after~\eref{3.43}, additional
condition}. If we demand~\eref{3.54-1} to be valid for \emph{arbitrary}
$\varphi_0^\pm(\bs q)$ whose commutators are $c$\ndash numbers,
then~\eref{3.54-1} entail
	\begin{equation}	\label{3.56}
\alpha^\pm(\bs k,\bs q) = 0
\qquad
\beta^\mp(\bs k,\bs q) = \pm \delta^3(\bs k-\bs q)
	\end{equation}
which convert~\eref{3.53-2} into identity.

	Let us say also a few words on the special case
when $m=0$ and $\bs k=\bs0$. In it one cannot extract any information
from~\eref{3.54-1} and~\eref{3.53-2}. Recall, if $m=0$, the operators
$\varphi_0^\pm(\bs0)$ describe creation/annihilation of massless particles
with zero 4\ndash momentum. Since, in the Hermitian scalar case, the 4\ndash
momentum and mass are the only characteristics of the field's quanta, the
particles corresponding to
	\begin{equation}	\label{3.56-1}
\varphi_0^\pm(\bs k) \qquad\text{with $m=0$ and $\bs k=\bs0$}
	\end{equation}
are unphysical as they cannot lead to some observable consequences.  The
operators~\eref{3.56-1} are special kind of the solutions~\eref{3.13} of the
Klein\ndash Gordon equation and, consequently, can be interpreted as absence
of the field under consideration. Besides,
as we proved above under the hypothesis~\eref{3.49}, the only restrictions to
which $\varphi_0^\pm(\bs 0)$ with $m=0$ should be subjected are the
conditions~\eref{3.49} with $\bs k=\bs 0$, arbitrary $\bs q$, and any
(generalized) functions $\alpha^\pm$ and $\beta^\pm$. Thus, to ensure
a continuous limit when $(m,\bs k)\to(0,\bs0)$, we, \emph{by convention},
choose $\alpha^\pm(\bs 0,\bs q)$ and $\beta^\pm(\bs 0,\bs q)$ to be given
by~\eref{3.56} with $\bs k=\bs0$.%
\footnote{~%
In this way we exclude from the theory a special kind of `absent'
(unphysical) field described by $m=0$ and $\varphi_0=0$ or
$\varphi_0^{\pm}(\bs k)=0$.%
}

	So, we have obtained the next solution of the problem~\ref{Prob3.1}.
In momentum picture, the Klein\ndash Gordon equation~\eref{12.131} admits a
solution~\eref{3.38}, \ie
\[
\varphi_0
=
\int \Bigl( 2c(2\pi\hbar)^3 \sqrt{m^2c^2+{\bs k}^2} \Bigr)^{-1/2}
	(\varphi_0^+(\bs k) + \varphi_0^-(\bs k)) \Id^3\bs k ,
\]
in which the creation/annihilation operators $\varphi_0^\pm(\bs k)$
satisfy only the commutation relations
	\begin{equation}	\label{3.57}
[ \varphi_0^\pm(\bs k),\varphi_0^\pm(\bs q) ]_{\_} = 0
\qquad
[ \varphi_0^\mp(\bs k),\varphi_0^\pm(\bs q) ]_{\_}
				= \pm  \delta^3(\bs k-\bs q) \id_\Hil .
	\end{equation}

	Now, it is a trivial exercise to be verified that, in view
of~\eref{3.36} and~\eref{3.57}, the equality~\eref{3.39} is identically
valid.

	We would like to remark, in the above considerations the massless
case, \ie $m=0$, is not neglected. However, obviously, the commutation
relations~\eref{3.57} exclude the degenerate solution~\eref{3.47}
or, equivalently,~\eref{3.53new} or~\eref{3.46new}.

	More generally, one can look for solutions of the tri-linear
(para?)`com\-mu\-tation' relation~\eref{3.41}, under the
condition~\eref{3.42} which do \emph{not} satisfy the additional
conditions~\eref{3.43}. But this is out of the aims of this work.%
\footnote{~%
Tri-linear relations, like~\eref{3.41}, are known as paracommutation
relations and were discovered in~\cite{Green-1953} (See
also~\cite{Volkov-1959,Volkov-1960}). However, it seems that at present are
not indications that solutions of~\eref{3.41}, which do not
satisfy~\eref{3.57}, may describe actually existing physical objects or
phenomena~\cite{Greenberg&Messiah-1964,Greenberg&Messiah-1965,
Bogolyubov&et_al.-QFT}. This is one of the reasons the quantum field theory
to deal with~\eref{3.57} instead of~\eref{3.41} (or the equivalent to it
Klein\ndash Gordon equation~\eref{12.131} (in momentum picture)
or~\eref{3.4new} (in Heisenberg picture). Elsewhere we shall demonstrate how
the parabose\ndash commutation relations for a free scalar field can be
derived from~\eref{3.41}.%
}

	As we have noted several times above, the concepts of a distribution
(generalized function) and operator-valued distribution appear during the
derivation of the commutation relations~\eref{3.57}. We first met them in the
tri\ndash linear relations~\eref{3.41}. In particular, the canonical
commutation relations~\eref{3.57} have a sense iff
 $[\varphi_0^\mp(\bs{k}),\varphi_0^\pm(\bs{k})]_{\_}$
is an operator-valued distribution (proportional to $\id_\Hil$), which is
\emph{not} the case if the field $\varphi_0$ (or $\tope{\varphi}_0$) is an
ordinary operator acting on $\Hil$. These facts point to inherent
contradiction of quantum field theory if the field variables are considered
as operators acting on a Hilbert space. The rigorous mathematical setting
requires the fields variables to be regarded as operator\ndash valued
distributions. However, such a setting is out of the scope of the present
work and the reader is referred to books
like~\cite{Streater&Wightman,Jost,
Bogolyubov&et_al.-AxQFT,Bogolyubov&et_al.-QFT}
for more details and realization of that program. In what follows, the
distribution character of the quantum fields will be encoded in the Dirac's
delta function, which will appear in relations like~\eref{3.41}
and~\eref{3.57}.

	Ending the discussion of the commutation relations for a free
Hermitian scalar field, we would like to note that the commutation, not
anticommutation, relations for it were derived directly from the
Klein\ndash Gordon equation~\eref{12.131} without involving the spin\ndash
statistics theorem, as it is done everywhere in the
literature~\cite{Roman-QFT,Bogolyubov&Shirkov}. In fact, this theorem is
practically derived here in the special case under consideration. Besides, we
saw that the commutation relations can be regarded as additional restrictions,
postulated for the field operators as stated, e.g.,
in~\cite[pp.~59--60]{Roman-QFT}, which must be compatible with the equations
of motion. In fact, as we proved, these relations are, under some
assumptions, equivalent to the equations of motion, \ie to the Klein-Gordon
equation in our case. Said differently, the commutation relations convert the
field equation(s) into identity (identities) and, in this sense are their
solutions. An alternative viewpoint is the commutation relations~\eref{3.57}
to be considered as field equations (under the conditions specified above)
with respect to the creation and annihilation operators as field operators
(variables).

	To close this section, we have to make the general remark that the
tri\ndash linear relations~\eref{3.41} (together with~\eref{3.42}) are
equivalent to the initial Klein\ndash Gordon equation (in terms of
creation/annihilation operators) and all efforts for the establishment of the
commutation relations~\eref{3.57} reflect, first of all, the fact of
extraction of physically essential solutions of these equations.%
\footnote{~%
One may recognize in~\eref{3.41} a kind of paracommutation relations which
are typical for the so\ndash called
parastatistics~\cite{Green-1953,Volkov-1959,Volkov-1960,Govorkov}.%
}
In this sense, we can say that the commutation relations~\eref{3.57} are a
reduction of the initial Klein\ndash Gordon equation~\eref{12.131} (in
momentum picture) or~\eref{3.4new} (in Heisenberg picture), under the
conditions~\eref{3.43} and the assumption that~\eref{3.54-1} hold for any
$\varphi_0^\pm(\bs q)$.


\section {The vacuum and normal ordering}
\label{Sect9}

	The vacuum of a free Hermitian scalar field $\varphi_0$ is a
particular its state, described by a state vector $\ope{X}_0$ which carries
no energy\ndash momentum and, correspondingly, it is characterized by a
constant (in spacetime) state vector, \ie
	\begin{gather}	\label{8.1}
\ope{P}_\mu(\tope{X}_0(x)) = 0
\\			\label{8.2}
\tope{X}_0(x) = \tope{X}_0(x_0) = \ope{X}_0
	\end{gather}
where $\tope{X}_0$ and $\ope{X}_0$ are the vacua in Heisenberg and momentum
pictures respectively.
	Equation~\eref{8.1} can be taken as a `macroscopic' definition of the
vacuum state vector $\ope{X}_0$. Since, `microscopically', the field is
considered as a collection of particles (see Sect.~\ref{Sect6}), the vacuum
should be considered as a vector characterizing a state with no particles in
it. Recalling the interpretation of creation, $\varphi_0^+(\bs k)$, and
annihilation, $\varphi_0^-(\bs k)$, operators from Sect.~\ref{Sect6}, we can
make the definition~\eref{8.1} more precise by demanding
	\begin{equation}	\label{8.3}
\varphi_0^-(\bs k) (\ope{X}_0) = 0
\qquad
\ope{X}_0 \not= 0 .
	\end{equation}
This is the everywhere accepted definition of a vacuum for a free Hermitian
scalar field. However, it does not agree with the expression~\eref{3.36} for
the momentum operator and the commutation relations~\eref{3.57}. In fact,
commuting $\varphi_0^+(\bs k)$ and $\varphi_0^-(\bs k)$, according
to~\eref{3.57}, in the second term in the integrand in~\eref{3.36} and
using~\eref{8.3}, we get
	\begin{equation}	\label{8.4}
\ope{P}_\mu (\ope{X}_0)
=
\frac{1}{2}\int k_\mu \delta^3(\bs 0) \Id^3\bs k  \ope{X}_0
=
\ope{X}_0 \times \delta^3(\bs 0) \times
	\begin{cases}
\infty^4			&\text{for $\mu=0$}	\\
\infty^2 (\infty^2 - \infty^2)	&\text{for $\mu=1,2,3$}
	\end{cases}
\mspace{6.5mu} .
	\end{equation}
Of course, this is a nonsense which must be corrected. The problem can be
solved by `repairing' the r.h.s.\ of~\eref{3.36}, by replacing the commutation
relations~\eref{3.57} with other relations (compatible with~\eref{3.41}
and~\eref{3.42}), or by some combination of these possibilities. At this
point, we agree with the established procedure for removing~\eref{8.4} from
the theory. If one accepts \emph{not to change the logical structure of the
theory}, the only possibility is a change in the Lagrangian from which all
follows. Since in~\eref{8.4} the infinities come from the term
$\varphi_0^-(\bs k)\circ\varphi_0^+(\bs k)$ in~\eref{3.36}, it should be
eliminated somehow. The known and, it seems, well working procedure for doing
this, which we accept, is the following one. At first the Lagrangian and the
dynamical variables, obtained from it and containing the field $\varphi_0$,
should be written, by means of~\eref{3.26}--\eref{3.28} in terms of creation,
$\varphi_0^+(\bs k)$, and annihilation, $\varphi_0^-(\bs k)$ operators. Then,
any composition (product) of these operators, possibly appearing under some
integral sign, must be change so that all creation operators to stand to the
left relative to all annihilation operators.%
\footnote{~%
The so-formulated rule is valid only for integer spin particles/fields.
By virtue of~\eref{3.57}, the order of the different creation/annihilation
operators relative to each other is insignificant, \ie it does not influence
the result of described procedure.%
}
The described procedure for transforming compositions of creation and
annihilation operators is known as \emph{normal ordering} and the result of it
is called \emph{normal product} of the corresponding operators. This item
is discussed at length in the literature to which the reader is referred for
details~\cite{Bogolyubov&Shirkov,Bjorken&Drell,Roman-QFT,Wick}.

	The mapping assigning to a composition of creation/annihilation
operators their normal product will be denoted by $\ope{N}$ and called normal
ordering operator. The action of $\ope{N}$ on polynomials or convergent power
series of creation/annihilation operators is defined by extending it by
linearity; the resulting mapping being denoted by $\ope{N}$ too.

	Since, evidently,
\(
\ope{N}\bigl( \varphi_0^\pm(\bs k)\circ\varphi_0^\mp(\bs k) \bigr)
=
\varphi_0^+(\bs k)\circ\varphi_0^-(\bs k) ,
\)
the representation~\eref{3.36} of the momentum operator changes, after normal
ordering, into
	\begin{equation}	\label{8.5}
\ope{P}_\mu
=
\int\Id^s\bs k
k_\mu|_{k_0=\sqrt{m^2c^2+{\bs k}^2}}
\varphi_0^+(\bs k)\circ\varphi_0^-(\bs k) .
	\end{equation}
(Notice, after normal ordering, we retain the notation $\ope{P}_\mu$ for the
object resulting from~\eref{3.36}. This is an everywhere accepted system of
notation in the literature and it is applied to all similar situations, \eg
for the Lagrangian $\ope{L}$ or energy\ndash momentum operator
$\ope{T}^{\mu\nu}$. After some experience with such a system of doubling the
meaning of the symbols, one finds it useful and harmless.) Now the
equality~\eref{8.1} is a trivial corollary of~\eref{8.5} and~\eref{8.3}.

	As $\ope{X}_0\not=0$, we shall assume that the vacuum $\ope{X}_0$ can
be normalized to unity, \viz
	\begin{equation}	\label{8.6}
\langle\ope{X}_0|\ope{X}_0\rangle = 1 ,
	\end{equation}
where $\langle\cdot | \cdot\rangle\colon\Hil\times\Hil\to\field[C]$ is the
Hermitian scalar product of the Hilbert space $\Hil$ of states.
In fact, the value $\langle\ope{X}_0|\ope{X}_0\rangle$ is insignificant and
its choice as the number 1 is of technical character. In this way, in many
calculations, disappears the coefficient $\langle\ope{X}_0|\ope{X}_0\rangle$.
\emph{Prima facie} one can loosen~\eref{8.6} by demanding $\ope{X}_0$ to have
a finite norm, but this only adds to the theory the insignificant constant
$\langle\ope{X}_0|\ope{X}_0\rangle$ which can be eliminated by a rescaling of
$\ope{X}_0$.%
\footnote{~%
If $\ope{X}_0$ has an infinite norm, so is the situation with any other state
vector obtained from $\ope{X}_0$ via action with creation operators, which
makes the theory almost useless.%
}

	Let us summarize the above discussion. The vacuum of a free scalar
field is its physical state which does not contains any particles and has zero
energy and 3\ndash momentum. It is described by a state vector, denoted by
$\ope{X}_0$ (in momentum picture) and also called the vacuum of the field,
such that:
	\begin{subequations}	\label{8.7}
	\begin{gather}
				\label{8.7a}
\ope{X}_0 \not= 0
\\				\label{8.7b}
\ope{X}_0 = \tope{X}_0
\\				\label{8.7c}
\varphi_0^-(\bs k) (\ope{X}_0) = 0
\\				\label{8.7d}
\langle\ope{X}_0|\ope{X}_0\rangle = 1
	\end{gather}
	\end{subequations}
for any 3-momentum $\bs k$. Since the existence of the vacuum $\tope{X}_0$ in
the Heisenberg picture is a known theorem, the condition~\eref{8.7b},
expressing the coincidence of `Heisenberg vacuum' and `momentum vacuum',
ensures the existence of the vacuum $\ope{X}_0$ in momentum picture. Besides,
to make the theory sensible, we have assumed normal ordering of the
creation/annihilation operators in the Lagrangian and all observables derived
from it.

	The normal ordering of products changes the field
equations~\eref{3.41} into
	\begin{equation}	\label{8.8}
[ \varphi_0^\pm(\bs k) ,
  \varphi_0^+(\bs q)\circ \varphi_0^-(\bs q) ]_{\_}
\pm \varphi_0^\pm(\bs k) \delta^3(\bs k-\bs q)
= \frac{1}{2} f^\pm(\bs k, \bs q)
	\end{equation}
as the anticommutator in~\eref{3.41} originates from the
expression~\eref{3.36} for the momentum operator (before normal ordering). The
conditions~\eref{3.42} remains unchanged. However, by means of~\eref{3.43-3}
with $\varepsilon=-1$ and the commutation relations~\eref{3.57}, one can
verify that~\eref{8.8} and~\eref{3.42} are identically valid. This means that,
in fact, the commutation relations~\eref{3.57} play a role of field equations
under the hypotheses made.


\section {State vectors and transitions between them}
\label{Sect10}

	According to the general theory of Sect.~\ref{Sect2}, the general form
of a state vector $\ope{X}(x)$ of a free Hermitian scalar field in momentum
picture is
	\begin{equation}	\label{9.1}
\ope{X}(x)
=
\ope{U}(x,x_0) (\ope{X}(x_0))
=
\e^{ \iih(x^\mu-x_0^\mu)\ope{P}_\mu } (\ope{X}(x_0)) ,
	\end{equation}
where $x_0\in\base$ is an arbitrarily fixed point in the Minkowski spacetime
$\base$, $\ope{P}_\mu$ is given via~\eref{8.5}, and the initial value
$\ope{X}(x_0)$ of $\ope{X}$ at $x_0$ is identical with the (constant) state
vector $\tope{X}$ representing the same state as $\ope{X}(x)$ but in
Heisenberg picture, \ie

	\begin{gather}	\label{9.2}
\ope{X}(x_0) = \tope{X} .
\intertext{In particular, if $\ope{X}_p$ is a state vector at $x_0$ with
fixed 4-momentum $p_\mu$, \ie}
			\label{9.3}
\ope{P}_\mu (\ope{X}_p) = p_\mu \ope{X}_p ,
\intertext{then its general form  at a point $x$ is}
			\label{9.4}
\ope{X}_p(x)
=
\e^{ \iih(x^\mu-x_0^\mu)p_\mu } \ope{X}_p
	\end{gather}
and  $\ope{P}_\mu (\ope{X}_p(x)) = p_\mu \ope{X}_p(x)$

	From the vacuum $\ope{X}_0$ can be constructed a base for the
system's Hilbert space $\Hil$, called the Fock base.
	A general $s$\ndash particle state containing
$s$ particles with momenta $k_1\dots,k_s$ (some of these vectors may be
identical) of a scalar field $\varphi_0$ has the form
	\begin{equation}	\label{12.141}
\ope{X}(k_1,\dots,k_s)
=
f_s(k_1,\dots,k_s)
\bigl( \varphi^+_{0}(k_1) \circ\dots\circ \varphi^+_{0}(k_s) \bigr)
								(\ope{X}_0)
	\end{equation}
for some  function $f_s$ characterizing the distribution of the particles.%
\footnote{~%
In~\eref{12.141} we have omit a spacetime dependent factor which the reader
may recover, using~\eref{9.4}; the vector~\eref{12.141} corresponds to the
vector $\ope{X}_p$ in~\eref{9.4} in a case of $s$\ndash particle state.%
}
Notice, by virtue~\eref{12.133}, the 4\ndash momenta $k_1,\dots,k_s$ are
subjected to the conditions $k_{a}^2=m^2c^2$, $a=1,\dots,s$, with $m$
being the mass of the quanta of the field $\varphi_0$. An arbitrary
state, described via the Klein\ndash Gordon
equation~\eref{12.131} or its version expressed by~\eref{12.132}
and~\eref{12.133}, can be presented as a superposition of all possible states
like~\eref{12.141}, \viz
	\begin{equation}	\label{12.142}
\ope{X}=
\sum_{s\ge0} \int\Id k_1\dots\int\Id k_s
	f_s(k_1,\dots,k_s)
\bigl( \varphi^+_{0}(k_1) \circ\dots\circ \varphi^+_{0}(k_s) \bigr)
								(\ope{X}_0).
	\end{equation}
	The above results, concerning free Hermitian scalar field, are
identical with similar ones in the momentum representation in ordinary
quantum field theory (in Heisenberg picture); the difference being that now
$\varphi_0(k)$ is the field operator in the \emph{momentum picture}, which,
as we proved, are identical with the Fourier images of the field operators in
Heisenberg picture. So, the Fock base goes without changes from Heisenberg
into momentum picture.

	One of the main problems in quantum field theory is to find the
amplitude for a transition from some initial state $\ope{X}_i(x_i)$ into a
final state $\ope{X}_f(x_f)$, \ie the quantities
	\begin{equation}	\label{9.5}
S_{fi}(x_f,x_i) := \langle\ope{X}_f(x_f) | \ope{X}_i(x_i)\rangle
	\end{equation}
called elements of the so-called $S$-matrix (scattering matrix). Ordinary one
considers the limits of~\eref{9.5} with
$x_f^0\to+\infty$  and $x_i^0\to-\infty$ or, more generally,
$x_f^\mu\to+\infty$  and $x_i^\mu\to-\infty$. These cases are important in
the scattering theory but not for the general theory, described here, and,
respectively, will not be discussed in our work. If we know $\ope{X}_f$ and
$\ope{X}_i$ at some points $x_f^{(0)}$ and $x_i^{(0)}$, respectively, then,
combining~\eref{9.1} and~\eref{9.5}, we get
	\begin{multline}	\label{9.6}
S_{fi}(x_f,x_i)
= \langle\ope{X}_f(x_f^{(0)}) |
	\e^{\iih(x_i^\mu-x_f^\mu)\ope{P}_\mu}
	(\ope{X}_i(x_i^{(0)}))\rangle
=
\langle\ope{X}_f(x_f^{(0)}) | \ope{U}(x_i,x_f)
	(\ope{X}_i(x_i^{(0)}))\rangle .
	\end{multline}
Consequently (cf.~\cite[\S~107]{Bjorken&Drell-2}), the operator
$(\ope{U}(x_f,x_i))^\dag=\ope{U}^{-1}(x_f,x_i)=\ope{U}(x_i,x_f)$, where
$\dag$ means (Hermitian) conjugation, has to be identified with the $S$\ndash
operator (often called also  $S$\ndash matrix). To continue the analogy with
the $S$\ndash matrix theory, we can expand the exponent in~\eref{9.6} into a
power series. This yields (see~\eref{8.5})
	\begin{equation}	\label{9.7}
\ope{U}(x_i,x_f) = \id_\Hil + \sum_{n=1}^{\infty} \ope{U}^{(n)}(x_i,x_f)
	\end{equation}
\vspace{-2.4ex}
	\begin{multline}	\label{9.8}
\ope{U}^{(n)}(x_i,x_f)
:=
\frac{1}{n!} \frac{1}{(\ih)^n}
(x_i^{\mu_1}-x_f^{\mu_1}) \dots (x_i^{\mu_n}-x_f^{\mu_n})
\\
\times \int \mspace{-7mu} \Id^3\bs k^{(1)}\dots\Id^3\bs k^{(n)}
k_{\mu_1}^{(1)}\dotsb k_{\mu_n}^{(n)}
\varphi_0^+(\bs k^{(1)})\circ\varphi_0^-(\bs k^{(1)})
\circ\dotsb\circ
\varphi_0^+(\bs k^{(n)})\circ\varphi_0^-(\bs k^{(n)})
	\end{multline}
where $k_0^{(a)}=\sqrt{m^2c^2+ ({\bs k}^{(a)})^2}$, $a=1,\dots,n$.
This expression is extremely useful in `scattering theory' when one deals
with states having a fixed number of particles.%
\footnote{~%
Since we are dealing with a free field, there is no interaction between its
quanta (particles) and, hence, there is no real scattering. However, the
method, we present below, is of quite general nature and can be applied in
real scattering problems. This will be illustrated in a separate paper.%
}

	The first thing one notices, is that the vacuum, defined
via~\eref{8.7}, cannot be changed, \viz
	\begin{gather}	\label{9.8new}
\ope{U}(x_i,x_f) (\ope{X}_0) \equiv \ope{X}_0 ,
\intertext{and if $\ope{X}_{\ge1}(x)$ is a state vector of a state
containing at least one particle, then}
			\label{9.9}
\langle \ope{X}_{\ge1}(x)|\ope{X}_0 \rangle \equiv 0
\intertext{This simple result means that the only non-forbidden transition
from the vacuum is into itself, \ie}
			\label{9.9new}
\langle \ope{X}_0|\ope{X}_0 \rangle = 1 \not= 0 .
	\end{gather}
The results just obtained are known as the \emph{stability of the
vacuum}.

	In accord with~\eref{9.4},~\eref{8.5} and~\eref{3.57}, the vector
	\begin{equation}	\label{9.10}
\ope{X}(x,p)
=
\e^{\iih(x^\mu-x_0^\mu)p_\mu} \varphi_0^+(\bs p) (\ope{X}_0)
	\end{equation}
describes a particle at a point $x\in\base$ with 4-momentum $p$. Similarly,
an $n$\ndash particle state, $n\ge1$, in which the $i^{\text{th}}$ particle
is at a point $x_i$ with 4\ndash momentum $p_i$, $i=1,\dots,n$, is
represented by%
\footnote{~%
Since the vacuum and creation/annihilation operators in Heisenberg and
momentum pictures coincide, we use the usual Fock base to expand the state
vectors. The spacetime depending factor comes from~\eref{9.4}.%
}
	\begin{equation}	\label{9.11}
\ope{X}(x_1,p_1;\ldots;x_n,p_n)
=
\frac{1}{\sqrt{n!}}
\exp\Bigl\{ \iih \sum_{i=1}^{n} (x_i^\mu-x_0^\mu) (p_i)_\mu \Bigr\}
(\varphi_0^+(\bs p_1)\circ\dots\circ\varphi_0^+(\bs p_n) )
(\ope{X}_0) .
	\end{equation}
After some simple algebra with creation and annihilation operators, which
uses~\eref{3.57}, ~\eref{3.37new} and~\eref{8.7c}, one finds the following
transition amplitude from $m$\ndash particle state into $n$\ndash particle
state, $m,n\in\field[N]$:
	\begin{multline}	\label{9.12}
\langle\ope{X}(y_1,q_1;\ldots;y_n,q_n)|\ope{X}(x_1,p_1;\ldots;x_m,p_m)\rangle
=
\frac{1}{n!} \delta_{mn}
\exp\Bigl\{ \iih \sum_{i=1}^{n} (x_i^\mu-y_i^\mu) (p_i)_\mu \Bigr\}
\\ \times
\sum_{(i_1,\dots,i_n)}
\delta^3(\bs p_{n} - \bs q_{i_1}) \delta^3(\bs p_{n-1} - \bs q_{i_2})
\dots
\delta^3(\bs p_{1} - \bs q_{i_n}) ,
	\end{multline}
where $\delta_{mn}$ is the Kronecker $\delta$-symbol, \ie $\delta_{mn}=1$ for
$m=n$ and $\delta_{mn}=0$ for $m\not=n$, and the summation is over all
permutations $(i_1,\dots,i_n)$ of $(1,\dots,n)$. The presence of
$\delta_{mn}$ in~\eref{9.12} means that an $n$\ndash particle state can be
transformed only into an $n$\ndash particle state; all other transitions are
forbidden. Besides, the $\delta$\ndash functions in~\eref{9.12}  say that if
the 4\ndash momentum of a particle changes, the transition is also forbidden.
So, the only change an $n$\ndash particle state can experience is the change
in the coordinates of the particles it contains. All these results are quite
understandable (and trivial too) since we are dealing with a \emph{free} field
whose quanta move completely independent of each other, without any
interactions between them.

	Notice, if in~\eref{9.12} we set $m=n$ and integrate over all
momenta, we get a pure phase factor, equal to
$\exp\Bigl\{ \iih \sum_{i=1}^{n} (x_i^\mu-y_i^\mu) (p_i)_\mu \Bigr\}$.
Since, in this case, the module of the square of~\eref{9.12} is interpreted
as a probability for the transition between the corresponding states, this
means that the transition between two $n$\ndash particle states is completely
sure, \ie with 100\%
probability.

	At the end of this section, we remark that the states~\eref{9.11}
are normalized to unity,
	\begin{equation}	\label{9.13}
\int \Id^3\bs{q}_1\ldots \Id^3\bs{q}_n
\langle\ope{X}(x_1,q_1;\ldots;x_n,q_n)|\ope{X}(x_1,p_1;\ldots;x_m,p_m)\rangle
=
1 ,
	\end{equation}
due to~\eref{8.7d}. However, the norm
\(
\langle\ope{X}(x_1,p_1;\ldots;x_n,p_n)|\ope{X}(x_1,p_1;\ldots;x_m,p_m)\rangle
\)
is infinity as it is proportional to $(\delta^3(0))^n$, due to~\eref{9.12}.
If one works with a vacuum not normalized to unity, in~\eref{9.11} the factor
 $\langle\ope{X}_0 | \ope{X}_0 \rangle^{-1/2}$
will appear. This will change~\eref{9.12} with the factor
 $\langle\ope{X}_0 | \ope{X}_0 \rangle^{-1}$.


\part{{\Large Free arbitrary scalar field}}
\label{PartB}
\markright{\itshape\bfseries Bozhidar Z. Iliev:
	\upshape\sffamily\bfseries QFT in momentum picture: I.~Scalar fields}

	Until now the case of free Hermitian (neutral) scalar field was
explored. In the present, second, part of the present investigation, the
results obtained for such a field will be transferred to the general case of
free arbitrary, Hermitian (real, neutral, uncharged) or non\ndash Hermitian
(complex, charged), scalar field.%
\footnote{~%
A classical complex field, after quantization, becomes a non\ndash Hermitian
operator acting on the system's (field's) Hilbert space of states. That is
why the quantum analogue of a classical complex scalar field is called
non\ndash Hermitian scalar field. However, it is an accepted common practice
such a field to be called (also) a complex scalar field. In this sense, the
terms complex scalar field and non\ndash Hermitian scalar field are
equivalent and, hence, interchangeable. Besides, since a complex (classical
or quantum) scalar field carries a charge, it is called also charged scalar
field. Cf.~footnote~\ref{HermitianField}.%
}

	As we shall see, there are two essential peculiarities in the
non\ndash Hermitian case. On one hand, the field operator and its Hermitian
conjugate are so `mixed' in the momentum operator that one cannot write (in
momentum picture) separate field equations for them. On the other hand, a
non\ndash Hermitian field carries a charge. These facts will later be
reflected in the corresponding commutation relations.

\section{Description of free scalar field}
	\label{Sect11}

	In Heisenberg picture a free arbitrary, Hermitian or non\ndash
Hermitian, scalar field is described via a field operator
$\tope{\varphi}(x)$, which
may be Hermitian,
	\begin{subequations}	\label{11.1}
	\begin{gather}	\label{11.1a}
\tope{\varphi}^\dag(x) = \tope{\varphi}(x)
\\\intertext{or non-Hermitian,}
			\label{11.1b}
\tope{\varphi}^\dag(x) \not= \tope{\varphi}(x) .
	\end{gather}
	\end{subequations}
The properties of a free non-Hermitian scalar field are,
usually~\cite{Bogolyubov&Shirkov,Bjorken&Drell,Roman-QFT,Pauli&Weisskopf},
encoded in the Lagrangian
	\begin{gather}
			\label{11.2}
\tope{L} :=
- m^2c^4\tope{\varphi}^\dag(x)\circ \tope{\varphi}(x)
+ c^2\hbar^2	(\pd_\mu\tope{\varphi}^\dag(x)) \circ
		(\pd^\mu\tope{\varphi}(x)),
\intertext{which in momentum picture, according to the general rules of
Sect.~\ref{Sect2} (see equation~\eref{2.5}), reads}
			\label{11.3}
\ope{L} =
- m^2c^4\varphi_0^\dag \circ \varphi_0
- c^2
    [ \varphi_0^\dag,\ope{P}_\mu ]_{\_}\circ [ \varphi_0,\ope{P}^\mu ]_{\_} ,
	\end{gather}
where (see~\eref{2.4} and cf.~\eref{12.130})
	\begin{subequations}	\label{11.4}
	\begin{gather}
			\label{11.4a}
\varphi(x)
= \ope{U}(x,x_0)\circ \tope{\varphi}(x) \circ \ope{U}^{-1}(x,x_0)
= \varphi(x_0)
= \tope{\varphi}(x_0)
=: \varphi_0
\\			\label{11.4b}
\varphi^\dag(x)
= \ope{U}(x,x_0)\circ \tope{\varphi}^\dag(x) \circ \ope{U}^{-1}(x,x_0)
= \varphi^\dag(x_0)
= \tope{\varphi}^\dag(x_0)
=: \varphi_0^\dag
	\end{gather}
	\end{subequations}
are the corresponding to $\tope{\varphi}(x)$ and $\tope{\varphi}^\dag(x)$,
respectively, constant field operators in momentum picture. In
general, the operator $\varphi_0$ can be Hermitian,
	\begin{subequations}	\label{11.5}
	\begin{gather}	\label{11.5a}
\varphi_0^\dag(x) = \varphi_0
\\\intertext{or non-Hermitian,}
			\label{11.5b}
\varphi_0^\dag \not= \varphi_0 .
	\end{gather}
	\end{subequations}
as a result of~\eref{11.1} and the unitarity of $\ope{U}(x,x_0)$.

	Defining
	\begin{equation}	\label{11.5new}
\tau(\tope{\varphi}) :=
	\begin{cases}
1  &\text{for $\tope{\varphi}^\dag=\tope{\varphi}$
					(Hermitian (neutral) field)}
\\
0  &\text{for $\tope{\varphi}^\dag\not=\tope{\varphi}$
					(non-Hermitian (charged) field)}
	\end{cases}
\ ,
	\end{equation}
we can unify the Lagrangians~\eref{11.2} and~\eref{3.1} by writing
	\begin{equation}
			\label{11.2-1}
\lindex[\mspace{-5mu}\tope{L}]{}{\prime} :=
- \frac{1}{1+\tau(\tope{\varphi})} m^2c^4
	\tope{\varphi}^\dag(x)\circ \tope{\varphi}(x)
+ \frac{1}{1+\tau(\tope{\varphi})} c^2\hbar^2
	(\pd_\mu\tope{\varphi}^\dag(x)) \circ (\pd^\mu\tope{\varphi}(x)) .
	\end{equation}

	As an alternative to~\eref{11.2-1}, one may also put
	\begin{equation}
			\label{11.2-2}
\lindex[\mspace{-5mu}\tope{L}]{}{\prime\prime} :=
- \frac{1}{1+\tau(\tope{\varphi})} m^2c^4
	\tope{\varphi}(x)\circ \tope{\varphi}^\dag(x)
+ \frac{1}{1+\tau(\tope{\varphi})} c^2\hbar^2
	(\pd_\mu\tope{\varphi}(x)) \circ (\pd^\mu\tope{\varphi}^\dag(x))
	\end{equation}

	There is also one more candidate for a Lagrangian for a free
arbitrary scalar field. Since such a field $\tope{\varphi}$ is equivalent to
two independent free Hermitian scalar fields
$\tope{\varphi}_1=\tope{\varphi}_1^\dag$ and
$\tope{\varphi}_2=\tope{\varphi}_2^\dag$ with masses equal to the one of
$\tope{\varphi}$ and such that
$\tope{\varphi} = \tope{\varphi}_1 + \iu \tope{\varphi}_2$,
 $\iu$ being the imaginary unit,
we can set
\(
\lindex[\mspace{-5mu}\tope{L}]{}{\prime\prime\prime}
=
\ope{L}_0(\tope{\varphi}_1) + \ope{L}_0(\tope{\varphi}_2)
\)
with $\ope{L}_0$ defined by~\eref{3.1}. Taking into account that
$\tope{\varphi}^\dag = \tope{\varphi}_1 - \iu \tope{\varphi}_2$,
one can transform the last Lagrangian into the form
	\begin{equation}
			\label{11.2-3}
	\begin{split}
\lindex[\mspace{-5mu}\tope{L}]{}{\prime\prime\prime} :=
& - \frac{1}{2(1+\tau(\tope{\varphi}))} m^2c^4
	\bigl( \tope{\varphi}^\dag(x)\circ \tope{\varphi}(x)
	+ \tope{\varphi}(x)\circ \tope{\varphi}^\dag(x) \bigr)
\\
& + \frac{1}{2(1+\tau(\tope{\varphi}))} c^2\hbar^2
\bigl(
       (\pd_\mu\tope{\varphi}^\dag(x)) \circ (\pd^\mu\tope{\varphi}(x))
+      (\pd_\mu\tope{\varphi}(x)) \circ (\pd^\mu\tope{\varphi}^\dag(x))
\bigr) .
	\end{split}
	\end{equation}
This Lagrangian, which is the half of the sum of~\eref{11.2-1}
and~\eref{11.2-2}, also reduces to~\eref{3.1} in the Hermitian case
$\tope{\varphi}^\dag=\tope{\varphi}$ (or $\tope{\varphi}_2=0$ in terms of
the Hermitian fields $\tope{\varphi}_1$ and $\tope{\varphi}_2$). Evidently,
the Lagrangian~\eref{11.2-3} is a `symmetrization' of the r.h.s.\
of~\eref{11.2-1} or~\eref{11.2-2} relative to $\tope{\varphi}$ and
$\tope{\varphi}^\dag$ with coefficient $\frac{1}{2}$. The advantage
of~\eref{11.2-3} is that in it the field $\tope{\varphi}$ and its Hermitian
conjugate $\tope{\varphi}^\dag$ enter in a symmetric way, which cannot be
said relative to~\eref{11.2}, \eref{11.2-1} and~\eref{11.2-2}.

	Going some steps ahead, the consequences of the Lagrangians
$\lindex[\mspace{-5mu}\tope{L}]{}{}$,
$\lindex[\mspace{-5mu}\tope{L}]{}{\prime}$ ,
$\lindex[\mspace{-5mu}\tope{L}]{}{\prime\prime}$, and
$\lindex[\mspace{-5mu}\tope{L}]{}{\prime\prime\prime}$
can be summarized as follows:
	(i) All of these Lagrangians lead to identical (Klein\ndash
Gordon) field equations for $\tope{\varphi}$ and $\tope{\varphi}^\dag$;
	(ii) The energy\ndash momentum, momentum and charge operators
generated by these Lagrangians are, generally, different;
	(iii) After the establishment of the commutation relations and a
normal ordering of products (compositions), the momentum and charge operators
generated by
$\lindex[\mspace{-5mu}\tope{L}]{}{\prime}$ ,
$\lindex[\mspace{-5mu}\tope{L}]{}{\prime\prime}$, and
$\lindex[\mspace{-5mu}\tope{L}]{}{\prime\prime\prime}$
become identical and equal to one half of the ones induced by $\ope{L}$.
	Therefore, in view of these assertions, all of the Lagrangians
given above can be considered as equivalent.%
\footnote{~%
However, in Sect.~\ref{Sect16}, we shall see that the
Lagrangian~\eref{11.2-3} carries more information than~\eref{11.2-1},
\eref{11.2-2} and~\eref{11.2}. In this sense, it is the `best' one.%
}

	In momentum picture, by virtue of~\eref{2.5}, the
Lagrangians~\eref{11.2-1}, \eref{11.2-2} and~\eref{11.2-3} are:
	\begin{gather}
			\label{11.3-1}
\lindex[\mspace{-5mu}\tope{L}]{}{\prime} =
- \frac{1}{1+\tau(\ope{\varphi})} m^2c^4
	\varphi_0^\dag \circ \varphi_0
- \frac{1}{1+\tau(\ope{\varphi})} c^2
   [ \varphi_0^\dag,\ope{P}_\mu ]_{\_}\circ [ \varphi_0,\ope{P}^\mu ]_{\_}
\\
			\label{11.3-2}
\lindex[\mspace{-5mu}\tope{L}]{}{\prime\prime} =
- \frac{1}{1+\tau(\ope{\varphi})} m^2c^4
	\varphi_0 \circ \varphi_0^\dag
- \frac{1}{1+\tau(\ope{\varphi})} c^2
  [ \varphi_0,\ope{P}_\mu ]_{\_}\circ  [ \varphi_0^\dag,\ope{P}^\mu ]_{\_}
\\
			\label{11.3-3}
	\begin{split}
\lindex[\mspace{-5mu}\tope{L}]{}{\prime\prime\prime} =
& - \frac{1}{2(1+\tau(\ope{\varphi}))} m^2c^4
	\bigl( \varphi_0^\dag \circ \varphi_0 +
	\varphi_0 \circ \varphi_0^\dag 	\bigr)
\\
& - \frac{1}{2(1+\tau(\ope{\varphi}))} c^2
  \bigl(
   [ \varphi_0^\dag,\ope{P}_\mu ]_{\_}\circ [ \varphi_0,\ope{P}^\mu ]_{\_}
  +
   [ \varphi_0,\ope{P}_\mu ]_{\_}\circ  [ \varphi_0^\dag,\ope{P}^\mu ]_{\_}
  \bigr) ,
	\end{split}
	\end{gather}
where (cf.~\eref{11.5new})
	\begin{equation}	\label{11.37}
\tau(\varphi_0) :=
	\begin{cases}
1	&\text{for $\varphi_0^\dag=\varphi_0$ (Hermitian (neutral) field)}
\\
0	&\text{for $\varphi_0^\dag\not=\varphi_0$
					  (non-Hermitian (charged) field)}
	\end{cases}
=
\tau(\tope{\varphi}) .
	\end{equation}
From here, we derive:
	\begin{gather}
			\label{11.6}
\frac{\pd\lindex[\mspace{-6mu}\ope{L}]{}{a}}{\pd\varphi_0}
=
\frac{\pd\tope{L}}{\pd\varphi_0}
= - m^2c^4\varphi_0^\dag
\qquad
\frac{\pd\lindex[\mspace{-6mu}\ope{L}]{}{a}}{\pd\varphi_0^\dag}
=
\frac{\pd\tope{L}}{\pd\varphi_0^\dag}
= - m^2c^4\varphi_0
\\			\label{11.7}
\pi^\mu
 : =
\frac{ \pd\lindex[\mspace{-6mu}\ope{L}]{}{a} }{\pd y_\mu}
 =
 - \ih c^2 [\varphi_0^\dag,\ope{P}^{\mu} ]_{\_}
\quad
\pi^{\dag\,\mu}
: =
\frac{\pd\lindex[\mspace{-6mu}\ope{L}]{}{a} }{\pd y_\mu^\dag}
 =
 - \ih c^2 [\varphi_0,\ope{P}^{\mu} ]_{\_}
	\end{gather}
where $a=\prime,\prime\prime,\prime\prime\prime$,
$ y_\mu:=\iih [\varphi_0,\ope{P}_{\mu} ]_{\_} $,
$ y_\mu^\dag:=\iih [\varphi_0^\dag,\ope{P}_{\mu} ]_{\_} $
and we have followed the differentiation rules of classical analysis of
commuting variables. At this point, all remarks, made in Sect.~\ref{Sect3} in
a similar situation, are completely valid too. For a rigorous derivation of
the field equations~\eref{11.10} and energy\ndash momentum
tensors~\eref{11.8} below, the reader is referred
to~\cite{bp-QFT-action-principle}.

	By means of the above equalities, from~\eref{12.129}, we get the
field equations for $\varphi_0$ and $\varphi_0^\dag$:
	\begin{subequations}	\label{11.10}
	\begin{gather}
				\label{11.10a}
m^2c^2 \varphi_0 - [[\varphi_0,\ope{P}_\mu]_{\_} ,\ope{P}^\mu]_{\_} = 0
\\				\label{11.10b}
m^2c^2 \varphi_0^\dag -
		[[\varphi_0^\dag,\ope{P}_\mu]_{\_} ,\ope{P}^\mu]_{\_} = 0 .
	\end{gather}
	\end{subequations}
So, regardless of the Lagrangians one starts, the fields $\varphi_0$ and
$\varphi_0^\dag$ satisfy one and the same Klein\ndash Gordon equation.
However, in contrast to the Heisenberg picture, in momentum picture these
equations are not independent as the momentum operator $\ope{P}_\mu$,
appearing in~\eref{11.10} and given via~\eref{2.0} or~\eref{2.6}, also
depends on $\varphi_0$ and $\varphi_0^\dag$ through the energy\ndash momentum
operator $\ope{T}_{\mu\nu}$. Hence, to determine $\varphi_0$,
$\varphi_0^\dag$ and $\ope{P}_\mu$, we need an explicit expression for
$\ope{T}_{\mu\nu}$  as a function of $\varphi_0$ and $\varphi_0^\dag$.

	If $\tope{\varphi}$ was a classical complex/real field, we would have
	\begin{equation}	\notag
\tope{T}_{\mu\nu}
=
\frac{1}{1+\tau(\tope{\varphi})}
\{ \tope{\pi}_\mu (\pd_\nu\tope{\varphi})
+
\tope{\pi}_\mu^* (\pd_\nu\tope{\varphi}^*) \}
-
\eta_{\mu\nu} \tope{L}
	\end{equation}
where the $*$ means complex conjugation. The straightforward transferring of
this expression in the quantum case results in
	\begin{equation}	\label{11.7a}
\tope{T}_{\mu\nu}^{(1)}
=
\frac{1}{1+\tau(\tope{\varphi})}
\{ \tope{\pi}_\mu\circ (\pd_\nu\tope{\varphi})
+
\tope{\pi}_\mu^\dag\circ (\pd_\nu\tope{\varphi}^\dag) \}
-
\eta_{\mu\nu} \tope{L} .
	\end{equation}
However, if
$\tope{\pi}_\mu$ and $\pd_\nu\tope{\varphi}$
and
$\tope{\pi}_\mu^\dag$ and $\pd_\nu\tope{\varphi}^\dag$
do not commute, this $\tope{T}_{\mu\nu}$ is non\ndash Hermitian,
$\tope{T}_{\mu\nu}^\dag\not=\tope{T}_{\mu\nu}$ . This situation can be
corrected by a `Hermitian symmetrization' of the first two terms
in~\eref{11.7a}, which gives
	\begin{equation}	\label{11.7b}
\tope{T}_{\mu\nu}^{(2)}
=
\frac{1}{2(1+\tau(\tope{\varphi}))} \bigl\{
\tope{\pi}_\mu\circ (\pd_\nu\tope{\varphi})
+
(\pd_\nu\tope{\varphi}^\dag)\circ \tope{\pi}_\mu^\dag
+
\tope{\pi}_\mu^\dag\circ (\pd_\nu\tope{\varphi}^\dag)
+
(\pd_\nu\tope{\varphi})\circ \tope{\pi}_\mu
\bigr\}
-
\eta_{\mu\nu} \tope{L} .
	\end{equation}
Evidently, if $\tope{\varphi}^\dag=\tope{\varphi}$, and the
Lagrangians~\eref{11.2-1} and~\eref{11.2-2} are employed, the
equations~\eref{11.7a} and~\eref{11.7b} reduce to~\eref{3.8a}
and~\eref{3.8b}, respectively. But these are not the only possibilities for
the energy\ndash momentum operator.  Often (see,
e.g.,~\cite[eq.~(3.34)]{Bogolyubov&Shirkov},
or~\cite[eq.~(2-151)]{Roman-QFT}, or~\cite[eq.~(6)]{Pauli&Weisskopf}), one
writes it in the form
	\begin{equation}	\label{11.7c}
\tope{T}_{\mu\nu}^{(3)}
=
\frac{1}{1+\tau(\tope{\varphi})}
\{ \tope{\pi}_\mu\circ (\pd_\nu\tope{\varphi})
+
\tope{\pi}_\nu\circ (\pd_\mu\tope{\varphi}) \}
-
\eta_{\mu\nu} \tope{L} .
	\end{equation}
As an alternative, we may also choose
	\begin{equation}	\label{11.7d}
\tope{T}_{\mu\nu}^{(4)}
=
\frac{1}{1+\tau(\tope{\varphi})}
\{ \tope{\pi}_\mu^\dag\circ (\pd_\nu\tope{\varphi}^\dag)
+
\tope{\pi}_\nu^\dag\circ (\pd_\mu\tope{\varphi}^\dag) \}
-
\eta_{\mu\nu} \tope{L} .
	\end{equation}
The last two expressions are suitable as $\tope{\pi}_\mu$ and
$\tope{\pi}_\mu^\dag$ are proportional to $\pd_\mu\tope{\varphi}^\dag$ and
$\pd_\mu\tope{\varphi}$ respectively.

	Partial discussion of the problem how should be defined the quantum
energy\ndash momentum tensorial operator $\ope{T}_{\mu\nu}$ and a similar one
for the current (see below), the reader can find
in~\cite[\S~2]{Pauli&Weisskopf}, where a reference to an early work of
B.~W.~Gordon in Z.~Phys.\ (vol.~40, p.~117. 1926) is given.

	From the view-point of symmetry, $\ope{T}_{\mu\nu}=\ope{T}_{\nu\mu}$,
and Hermiticity $\ope{T}_{\mu\nu}^\dag=\ope{T}_{\mu\nu}$, the energy\ndash
momentum operators~\eref{11.7b}--\eref{11.7d} are indistinguishable (if one
and the same Lagrangian is used in them).

	So, if we want to explore all possibilities, we have to look for the
consequences of the four energy\ndash momentum
operators~\eref{11.7a}--\eref{11.7d} for any one of the three
Lagrangians~\eref{11.2-1}--\eref{11.2-3}. However, instead of investigating
these~12 cases, we shall study only~3 of them. The reason is that
in~\cite{bp-QFT-action-principle} we have proved that for a given Lagrangian,
from the Schwinger's action principle~\cite{Roman-QFT,Schwinger-QFT-1},
follow, between other things, also \emph{unique expressions for all conserved
quantities in terms of the field operators}. In particular, it implies unique
energy\ndash momentum and current operators. These operators, for the
Lagrangians~\eref{11.2-1}--\eref{11.2-3} will be pointed below. Since the
discussion of the way of their selection is out of the range of the present
work, the reader may:
(i) take this by faith;
(ii) consider it as a lucky choice;
(iii) look on it as an additional hypothesis/postulate;
(iv) explore the consequences of the other~8 cases to see that they lead to
contradictions in the theory;%
\footnote{\label{Contradictions}~%
For instance, for them, generally, there are not one particle states with
fixed energy, \ie there are not one particle eigenvectors of the zeroth
component of the momentum operator.%
}
(v) apply the results obtained in~\cite{bp-QFT-action-principle}.

	The correct energy-momentum operators for the
Lagrangians~\eref{11.2-1}--\eref{11.2-3} are
respectively~\eref{11.7c}, \eref{11.7d} and~\eref{11.7b}.%
\footnote{~%
Since~\eref{11.2} and~\eref{11.2-1} are proportional, all results
for~\eref{11.2-1} can trivially be formulated for~\eref{11.2}.%
}
Explicitly, in view of~\eref{11.7}, we have
	\begin{subequations}	\label{11.8}
	\begin{gather}
			\label{11.8a}
{ \lindex[\mspace{-3mu}\tope{T}]{}{\prime} }_{\mu\nu}^{(3)}
=
\frac{1}{1+\tau(\tope{\varphi})} c^2\hbar^2
\{
     (\pd_\mu\tope{\varphi}^\dag) \circ (\pd_\nu\tope{\varphi})
   + (\pd_\nu\tope{\varphi}^\dag) \circ (\pd_\mu\tope{\varphi})
\}
-\eta_{\mu\nu} \lindex[\mspace{-5mu}\tope{L}]{}{\prime}
\\
			\label{11.8b}
{ \lindex[\mspace{-3mu}\tope{T}]{}{\prime\prime} }_{\mu\nu}^{(4)}
=
\frac{1}{1+\tau(\tope{\varphi})} c^2\hbar^2
\{
     (\pd_\mu\tope{\varphi}) \circ (\pd_\nu\tope{\varphi}^\dag)
   + (\pd_\nu\tope{\varphi})  \circ (\pd_\mu\tope{\varphi}^\dag)
\}
-\eta_{\mu\nu} \lindex[\mspace{-5mu}\tope{L}]{}{\prime\prime}
\\
			\label{11.8c}
	\begin{split}
{ \lindex[\mspace{-3mu}\tope{T}]{}{\prime\prime\prime} }_{\mu\nu}^{(2)}
=
\frac{1}{2(1+\tau(\tope{\varphi}))} c^2\hbar^2
& \{
     (\pd_\mu\tope{\varphi}^\dag) \circ (\pd_\nu\tope{\varphi})
   + (\pd_\nu\tope{\varphi}^\dag)  \circ (\pd_\mu\tope{\varphi})
\\
&   + (\pd_\mu\tope{\varphi}) \circ (\pd_\nu\tope{\varphi}^\dag)
    + (\pd_\nu\tope{\varphi}) \circ (\pd_\mu\tope{\varphi}^\dag)
\}
-\eta_{\mu\nu} \lindex[\mspace{-5mu}\tope{L}]{}{\prime\prime\prime} .
	\end{split}
	\end{gather}
	\end{subequations}
In momentum picture, these operators read (see~\eref{11.3-1}--\eref{11.7}):
	\begin{subequations}	\label{11.9}
	\begin{gather}
			\label{11.9a}
	\begin{split}
{ \lindex[\mspace{-3mu}\ope{T}]{}{\prime} }_{\mu\nu}^{(3)}
=
& - \frac{1}{1+\tau(\varphi_0)} c^2
\{
  [\varphi_0^\dag,\ope{P}_{\mu} ]_{\_} \circ
  [\varphi_0     ,\ope{P}_{\nu} ]_{\_}
+
  [\varphi_0^\dag,\ope{P}_{\nu} ]_{\_} \circ
  [\varphi_0,      \ope{P}_{\mu} ]_{\_}
\}
\\
& + \eta_{\mu\nu}\frac{1}{1+\tau(\varphi_0)}c^2
\{
  m^2c^2\varphi_0^\dag \circ \varphi_0
+
  [\varphi_0^\dag,\ope{P}_\lambda ]_{\_} \circ
  [ \varphi_0,    \ope{P}^\lambda ]_{\_}
\}
	\end{split}
\\
			\label{11.9b}
	\begin{split}
{ \lindex[\mspace{-3mu}\ope{T}]{}{\prime\prime} }_{\mu\nu}^{(4)}
=
& - \frac{1}{1+\tau(\varphi_0)} c^2
\{
  [\varphi_0,     \ope{P}_{\mu} ]_{\_} \circ
  [\varphi_0^\dag,\ope{P}_{\nu} ]_{\_}
+
  [\varphi_0,     \ope{P}_{\nu} ]_{\_} \circ
  [\varphi_0^\dag,\ope{P}_{\mu} ]_{\_}
\}
\\
& + \eta_{\mu\nu}\frac{1}{1+\tau(\varphi_0)}c^2
\{
  m^2c^2\varphi_0      \circ \varphi_0^\dag
+
  [\varphi_0,      \ope{P}_\lambda ]_{\_} \circ
  [\varphi_0^\dag,\ope{P}^\lambda ]_{\_}
\}
	\end{split}
\\
			\label{11.9c}
	\begin{split}
{ \lindex[\mspace{-3mu}\ope{T}]{}{\prime\prime\prime} }_{\mu\nu}^{(2)}
=
& - \frac{1}{2(1+\tau(\varphi_0))} c^2
\{
  [\varphi_0^\dag,\ope{P}_{\mu} ]_{\_} \circ
  [\varphi_0     ,\ope{P}_{\nu} ]_{\_}
+
  [\varphi_0^\dag,\ope{P}_{\nu} ]_{\_} \circ
  [\varphi_0,     \ope{P}_{\mu} ]_{\_}
\\ &+
  [\varphi_0,     \ope{P}_{\mu} ]_{\_} \circ
  [\varphi_0^\dag,\ope{P}_{\nu} ]_{\_}
+
  [\varphi_0,     \ope{P}_{\nu} ]_{\_} \circ
  [\varphi_0^\dag,\ope{P}_{\mu} ]_{\_}
\}
\\
& + \eta_{\mu\nu}\frac{1}{2(1+\tau(\varphi_0))}c^2
\{
  m^2c^2\varphi_0^\dag \circ \varphi_0
+
  m^2c^2\varphi_0      \circ \varphi_0^\dag
\\& +
  [\varphi_0^\dag,\ope{P}_\lambda ]_{\_} \circ
  [ \varphi_0,    \ope{P}^\lambda ]_{\_}
+
  [\varphi_0,      \ope{P}_\lambda ]_{\_} \circ
  [\varphi_0^\dag,\ope{P}^\lambda ]_{\_}
\} .
	\end{split}
	\end{gather}
	\end{subequations}

	Any one of the equations~\eref{11.9}, together
with~\eref{11.10},~\eref{2.6} and~\eref{12.112} form a complete system of
equations for explicit determination of $\varphi$, $\varphi^\dag$ and
$\ope{P}_\mu$. It will be analyzed in the subsequent sections.

	Since the Lagrangians of a free general scalar field are invariant
under (constant) phase transformations, such a field carries a, possibly
vanishing, charge (see, e.g.,~\cite{Bogolyubov&Shirkov,Roman-QFT}. The
(total) charge operator $\tope{Q}$ is defined by
	\begin{equation}	\label{11.11}
\tope{Q} := \frac{1}{c} \int\limits_{x^0=\const}^{} \tope{J}_0(x) \Id^3\bs x
	\end{equation}
where $\tope{J}_\mu(x)$ is a Hermitian operator,
	\begin{equation}	\label{11.11-2}
\tope{J}_\mu^\dag(x) = \tope{J}_\mu(x),
	\end{equation}
describing the field's current considered a little below. Since $\tope{Q}$
and $\tope{J}_\mu$ are conserved quantities, \viz they satisfy the equivalent
conservation laws
	\begin{gather}	\label{11.12}
\frac{\od\tope{Q}}{\od x^0} = 0
\qquad
\pd^\mu\tope{J}_\mu = 0,
\intertext{and $\pd_a\tope{Q}\equiv0$ for $a=1,2,3$, due to~\eref{11.11}, we
can write}
			\label{11.13}
\pd_\mu\tope{Q} = 0 .
	\end{gather}
The consideration of $\tope{Q}$ as a generator of (constant) phase
transformations~\cite{Bogolyubov&Shirkov,Roman-QFT} leads to the following
(Heisenberg) equations/relations%
\footnote{~%
The equations~\eref{11.14} follow from the transformation properties of
$\tope{\varphi}$ and $\tope{\varphi}^\dag$ too.%
}
	\begin{equation}	\label{11.14}
[\tope{\varphi}     ,\tope{Q}]_{\_} = q \tope{\varphi}
\qquad
[\tope{\varphi}^\dag,\tope{Q}]_{\_} = -q \tope{\varphi}^\dag
	\end{equation}
where $q$ is a constant, equal to the opposite charge of the particles
(quanta) of $\tope{\varphi}$ (see below Sect.~\ref{Sect14}), such that
	\begin{equation}	\label{11.14new}
q=0  \qquad\text{for }     \tope{\varphi}^\dag=\tope{\varphi} .
	\end{equation}
The charge operator $\tope{Q}$ is Hermitian, \ie
	\begin{gather}	\label{11.15}
\tope{Q}^\dag = \tope{Q}
\intertext{and commutes with the momentum operator $\tope{P}_\mu$,}
			\label{11.16}
[\tope{Q},\tope{P}_\mu]_{\_} = 0 .
	\end{gather}

	As a consequence of~\eref{11.16} and~\eref{12.112}, the charge
operator commutes with the (evolution) operator $\ope{U}(x,x_0)$ responsible
to the transition from Heisenberg picture to momentum one,
	\begin{gather}	\label{11.17}
[\tope{Q},\ope{U}(x,x_0)]_{\_} = 0 .
\intertext{Combining this with~\eref{12.114}, we get}
			\label{11.18}
\ope{Q}(x) = \tope{Q} =: \ope{Q} ,
	\end{gather}
that is the charge operators in Heisenberg and momentum pictures coincide.%
\footnote{~%
Obviously,~\eref{11.17} and~\eref{11.18} are valid for any operator
$\tope{Q}$ commuting with the momentum operator.%
}

	In momentum picture, the equations~\eref{11.14} and~\eref{11.15},
evidently, read:
	\begin{gather}
			\label{11.19}
[\ope{\varphi}     ,\ope{Q}]_{\_} =  q \ope{\varphi}
\qquad
[\ope{\varphi}^\dag,\ope{Q}]_{\_} = -q \ope{\varphi}^\dag
\\			\label{11.15-1}
\ope{Q}^\dag = \ope{Q} .
	\end{gather}
In this picture the equality~\eref{11.11} is convenient to be rewritten as
(see~\eref{12.114})
	\begin{equation}	\label{11.11-1}
\ope{Q}
=
\frac{1}{c} \int\limits_{x^0=x_0^0}^{}
	\ope{U}^{-1}(x,x_0)\circ \ope{J}_0(x)\circ \ope{U}(x,x_0) \Id^3\bs x .
	\end{equation}

	The only thing, we need for a complete determination of $\ope{Q}$, is
the explicit definition of the (quantum) current $\ope{J}_\mu$. If
$\tope{\varphi}$ was a free classical arbitrary, real or complex, scalar
field, we would have
	\begin{equation}
			\label{11.20-1}
\tope{J}_\mu (x)
=
\frac{q}{\ih} ( \tope{\pi}_\mu(x) \tope{\varphi}(x)
		- \tope{\pi}_\mu^*(x) \tope{\varphi}^*(x) ) .
	\end{equation}
The straightforward transferring of this result into the quantum case gives
	\begin{subequations}
			\label{11.20}
	\begin{equation}
	    		\label{11.20a}
\tope{J}_\mu^{(1)} (x)
=
\frac{q}{\ih} \{ \tope{\pi}_\mu(x)\circ \tope{\varphi}(x)
		- \tope{\pi}_\mu^\dag(x)\circ \tope{\varphi}^\dag(x) \} .
	\end{equation}
But, since the current operator must satisfy~\eref{11.11-2}, the quantities
~\eref{11.20a} are not suitable for components of a current operator if
 $[\tope{\pi}_\mu,\tope{\varphi}]_{\_}\not=0$ and/or
 $[\tope{\pi}_\mu^\dag,\tope{\varphi}^\dag]_{\_}\not=0$.
Evidently, here the situation is quite similar to the one with the definition
of the energy\ndash momentum operator considered above. So, without going
into details, we shall write here is a list of three admissible candidates for
a current operator:%
\footnote{~%
For a partial discussion of the problem, see~\cite[\S~2]{Pauli&Weisskopf}.
The expression~\eref{11.20c} is the one most often used in the
literature~\cite{Bogolyubov&Shirkov,Bjorken&Drell-2,Pauli&Weisskopf};
however, the definition~\eref{11.20a} is utilized too, for instance,
in~\cite[p.~99]{Roman-QFT}.%
}
	\begin{align}
	    		\label{11.20b}
\tope{J}_\mu^{(2)}
& =
\frac{q}{2\ih}
\bigl(
\tope{\pi}_\mu\circ \tope{\varphi}
- \tope{\pi}_\mu^\dag\circ \tope{\varphi}^\dag
- \tope{\varphi}^\dag\circ  \tope{\pi}_\mu^\dag
+ \tope{\varphi} \circ  \tope{\pi}_\mu
\bigr)
\\	    		\label{11.20c}
\tope{J}_\mu^{(3)}
& =
\frac{q}{\ih}
\bigl(
\tope{\pi}_\mu\circ \tope{\varphi}
- \tope{\varphi}^\dag\circ  \tope{\pi}_\mu^\dag
\bigr)
\\	    		\label{11.20d}
\tope{J}_\mu^{(4)}
& =
\frac{q}{\ih}
\bigl(
  \tope{\varphi} \circ  \tope{\pi}_\mu
- \tope{\pi}_\mu^\dag\circ \tope{\varphi}^\dag
\bigr) .
	\end{align}
	\end{subequations}

	Similarly to the case of energy-momentum operator, to any one of the
Lagrangians~\eref{11.2-1}--\eref{11.2-3}, there corresponds a unique current
operator. These operators are as follows
(see~\cite{bp-QFT-action-principle}):
	\begin{subequations}	\label{11.21}
	\begin{align}
	    		\label{11.21a}
{ \lindex[\mspace{-6mu}\tope{J}]{}{\prime} }_\mu^{(3)}
& =
\frac{q}{\ih}
\bigl(
\lindex{}{\prime}\mspace{-3mu}\tope{\pi}_\mu\circ \tope{\varphi}
- \tope{\varphi}^\dag\circ  \lindex{}{\prime}\mspace{-3mu}\tope{\pi}_\mu^\dag
\bigr)
\\	    		\label{11.21b}
{ \lindex[\mspace{-6mu}\tope{J}]{}{\prime\prime} }_\mu^{(4)}
& =
\frac{q}{\ih}
\bigl(
  \tope{\varphi} \circ  \lindex{}{\prime\prime}\mspace{-3mu}\tope{\pi}_\mu
- \lindex{}{\prime\prime}\mspace{-3mu}\tope{\pi}_\mu^\dag\circ \tope{\varphi}^\dag
\bigr)
\\	    		\label{11.21c}
{ \lindex[\mspace{-6mu}\tope{J}]{}{\prime\prime\prime} }_\mu^{(2)}
& =
\frac{q}{2\ih}
\bigl(
\lindex{}{\prime\prime\prime}\mspace{-3mu}\tope{\pi}_\mu\circ \tope{\varphi}
+ \tope{\varphi} \circ  \lindex{}{\prime\prime\prime}\mspace{-3mu}\tope{\pi}_\mu
- \lindex{}{\prime\prime\prime}\mspace{-3mu}\tope{\pi}_\mu^\dag\circ \tope{\varphi}^\dag
- \tope{\varphi}^\dag\circ  \lindex{}{\prime\prime\prime}\mspace{-3mu}\tope{\pi}_\mu^\dag
\bigr) .
	\end{align}
	\end{subequations}
As a consequence of~\eref{11.7}, these current operators in momentum picture
read respectively:
	\begin{subequations}	\label{11.22}
	\begin{align}
	    		\label{11.22a}
{ \lindex[\mspace{-6mu}\ope{J}]{}{\prime} }_\mu^{(3)}
& =
\frac{1}{2} q c^2
\bigl(
  \varphi_0^\dag\circ [\varphi_0,\ope{P}_\mu]_{\_}
- [\varphi_0^\dag,\ope{P}_\mu]_{\_}\circ \varphi_0
\bigr)
\\	    		\label{11.22b}
{ \lindex[\mspace{-6mu}\ope{J}]{}{\prime\prime} }_\mu^{(4)}
& =
\frac{1}{2} q c^2
\bigl(
  [\varphi_0,\ope{P}_\mu]_{\_}\circ \varphi_0^\dag
- \varphi_0\circ [\varphi_0^\dag,\ope{P}_\mu]_{\_}
\bigr)
\\	    		\label{11.22c}
{ \lindex[\mspace{-6mu}\ope{J}]{}{\prime\prime\prime} }_\mu^{(2)}
& =
\frac{1}{4} q c^2
\bigl(
  \varphi_0^\dag\circ [\varphi_0,\ope{P}_\mu]_{\_}
+  [\varphi_0,\ope{P}_\mu]_{\_}\circ \varphi_0^\dag
- [\varphi_0^\dag,\ope{P}_\mu]_{\_}\circ \varphi_0
- \varphi_0\circ [\varphi_0^\dag,\ope{P}_\mu]_{\_}
\bigr) .
	\end{align}
	\end{subequations}

	A free scalar field has no spin angular momentum and possesses a,
generally, non\ndash vanishing orbital angular momentum, as described in
Sect.~\ref{Sect3} (in particular, see equations~\eref{3.9-1}--\eref{3.9-5}).
It will be explored in Sect.~\ref{Sect17} directly in terms of creation and
annihilation operators.

%

	According to the Klein-Gordon equations~\eref{11.10}, the field
operators $\varphi_0$ and $\varphi_0^\dag$ are eigen\ndash operators for the
mapping~\eref{3.5} with eigenvalues equal to the square $m^2$ of the mass
(parameter) $m$ of the field (more precisely, of its quanta). Therefore the
interpretation of the operator~\eref{3.5} as a square\ndash of\ndash mass
operator of the field is preserved also in the case of free arbitrary,
Hermitian of non\ndash Hermitian, scalar field. At the same time, the square
of the momentum operator, $\frac{1}{c^2}\ope{P}_\mu\circ\ope{P}^\mu$, has an
interpretation of a square of mass operator for the solutions of the field
equations, \ie for the \emph{field's states}.


\section{Analysis of the field equations}
	\label{Sect13}

	As we know, the field operator $\varphi_0$ and its Hermitian conjugate
$\varphi_0^\dag$ satisfy the Klein\ndash Gordon equations~\eref{11.10} which
are `mixed' through the momentum operator $\ope{P}_\mu$, due to the
simultaneous presentation of $\varphi_0$ and $\varphi_0^\dag$ in the
energy\ndash momentum operator(s)~\eref{11.9}. However, for $\varphi_0$ and
$\varphi_0^\dag$ are completely valid all of the results of Sect.~\ref{Sect5}
as in it is used only the Klein\ndash Gordon equation (in momentum picture
for $\varphi_0$) and it does not utilize any hypotheses about the concrete
form of the momentum operator $\ope{P}_\mu$. Let us formulate the main of
them.
	\begin{Prop}	\label{Prop11.1}
	The solutions $\varphi_0$ and $\varphi_0^\dag$ of~\eref{11.10} can be
written as
	\begin{subequations}	\label{11.23}
	\begin{gather}
			\label{11.23a}
\varphi_0
=
\int\Id^3\bs k
	\bigl\{ f_+(\bs k) \varphi_0(k) \big|_{k_0=+\sqrt{m^2c^2+{\bs k}^2} }
	+ f_-(\bs k) \varphi_0(k) \big|_{k_0=-\sqrt{m^2c^2+{\bs k}^2} }
	\bigr\}
\\			\label{11.23b}
\varphi_0^\dag
=
\int\Id^3\bs k
	\bigl\{
f_+^\dag(\bs k) \varphi_0^\dag(k) \big|_{k_0=+\sqrt{m^2c^2+{\bs k}^2} }
     + f_-^\dag(\bs k) \varphi_0^\dag(k) \big|_{k_0=-\sqrt{m^2c^2+{\bs k}^2} }
	\bigr\} ,
	\end{gather}
	\end{subequations}
where $\varphi_0(k), \varphi_0^\dag(k) \colon\Hil\to\Hil$ are solutions of
	\begin{equation}	\label{11.24}
[ \varphi_0(k), \ope{P}_\mu ]_{\_} = - k_\mu \varphi_0 (k)
\qquad
[ \varphi_0^\dag(k), \ope{P}_\mu ]_{\_} = - k_\mu \varphi_0^\dag (k)
	\end{equation}
and, for solutions different from the `degenerate' solutions
	\begin{gather}
			\label{11.26}
[\varphi_0,\ope{P}_\mu]_{\_} = 0
\quad
[\varphi_0^\dag,\ope{P}_\mu]_{\_} = 0
\qquad\text{\upshape for $m=0$}
\intertext{or, in Heisenberg picture,}
			\tag{\ref{11.26}$^\prime$}
			\label{11.26'}
\tope{\varphi}(x) = \tope{\varphi}(x_0) =\varphi_0
\quad
\tope{\varphi}^\dag(x) = \tope{\varphi}^\dag(x_0) =\varphi_0^\dag
\quad
\tope{P}_\mu = \ope{P}_\mu = 0
\qquad\text{\upshape for $m=0$},
	\end{gather}
the symbols $f_\pm$ and $f_\pm^\dag$ denote complex-valued functions of
$\bs k$ and for the solutions~\eref{11.26} they stand for some distributions
of $\bs k$.
	\end{Prop}

	Notice, as a result of the restriction
$k_0=\pm\sqrt{m^2c^2+{\bs k}^2}$ in~\eref{11.23}, only the solutions
of~\eref{11.24} for which
	\begin{equation}	\label{11.25}
k^2=k_0^2 - {\bs k}^2 = m^2c^2
	\end{equation}
are significant.

	It should be emphasized, the operator $\varphi_0^\dag(k)$
in~\eref{11.23} is \emph{not} the Hermitian conjugate of $\varphi_0(k)$. In
fact, the reader can verify that~\eref{11.23} imply the equalities
(cf.~\eref{3.17new1})
	\begin{gather}
			\label{11.25new}
	\begin{split}
\bigl(
  f_\pm(\bs k) \varphi_0(k) \big|_{k_0=\pm\sqrt{m^2c^2+{\bs k}^2}}
\bigr)^\dag
=
-f_\mp^\dag(-\bs k) \varphi_0^\dag(-k) \big|_{k_0=\mp\sqrt{m^2c^2+{\bs k}^2}}
\\
\bigl(
  f_\pm^\dag(\bs k) \varphi_0^\dag(k) \big|_{k_0=\pm\sqrt{m^2c^2+{\bs k}^2}}
\bigr)^\dag
=
- f_\mp(-\bs k) \varphi_0(-k) \big|_{k_0=\mp\sqrt{m^2c^2+{\bs k}^2}} .
	\end{split}
\\\intertext{However, in the Hermitian case, \ie for
$\varphi_0^\dag=\varphi_0$, the equations~\eref{11.23a} and~\eref{11.23b}
must be identical and, consequently, we have}
			\label{11.25new1}
f_\pm^\dag(\bs k) = f_\pm(\bs k)
\quad
\varphi_0^\dag(k) = \varphi_0(k)
\qquad\text{for $\varphi_0^\dag=\varphi_0$}.
	\end{gather}
In this case, the equations~\eref{11.25new} reduce to~\eref{3.17new1}.

	\begin{Prop}	\label{Prop11.2}
The solutions of~\eref{11.10} have the representations
	\begin{gather}
			\label{11.27}
\varphi_0 = \int\delta(k^2-m^2c^2) \underline{\varphi}_0(k) \Id^4k
\qquad
\varphi_0^\dag = \int\delta(k^2-m^2c^2) \underline{\varphi}_0^\dag(k) \Id^4k
	\end{gather}
where $\underline{\varphi}_0(k)$ and $\underline{\varphi}_0^\dag(k)$ are
suitably normalized solutions of~\eref{11.24} which, up to a phase factor
equal to $\e^{\iih x_0^\mu k^\mu}$, coincide with the Fourier coefficients of
$\tope{\varphi}_0(x)$ and $\tope{\varphi}_0^\dag(x)$ (in Heisenberg picture
for solutions different from~\eref{11.26}).  \end{Prop}

	It should be emphasized, the solutions~\eref{11.26} are completely
`unphysical' as they have zero (energy\ndash)momentum operator
(see~\eref{11.9} and~\eref{2.6}), zero total charge (see~\eref{11.22})
and zero orbital angular momentum (see~\eref{3.9-3}) and, consequently,
they cannot lead to some physically predictable consequences.


\section{Frequency decompositions and their physical meaning}
	\label{Sect14}

	The presented in Sect.~\ref{Sect6} frequency decompositions of a free
Hermitian scalar field are based on the Klein\ndash Gordon equation, or, more
precisely, on~\eref{3.14} and~\eref{12.132}, and do not rely on a concrete
representation of the energy\ndash momentum operator. Hence they can
\emph{mutatis mutandis} be transferred in the general case of Hermitian or
non\ndash Hermitian scalar field. The basic moments of that procedure are as
follows.

	Let us put
	\begin{gather}
			\label{11.28}
	\begin{split}
\varphi_0^\pm(k) & :=
	\begin{cases}
f_\pm(\pm \bs k) \varphi_0(\pm k)	& \text{for $k_0\ge0$} \\
0					& \text{for $k_0<0$}
	\end{cases}
\\
\varphi_0^{\dag\,\pm}(k) &:=
	\begin{cases}
f_\pm^\dag((\pm \bs k) \varphi_0^\dag(\pm k)& \text{for $k_0\ge0$} \\
0						& \text{for $k_0<0$}
	\end{cases}
\quad .
	\end{split}
\intertext{As an evident consequence of~\eref{11.25new}, we have the
equalities}
			\label{11.28new}
\bigl( \varphi_0^{\pm}( k) \bigr)^\dag =  \varphi_0^{\dag\,\mp}( k)
\quad
\bigl( \varphi_0^{\dag\,\pm}(k) \bigr)^\dag =  \varphi_0^{\mp}(k)
\intertext{which mean that the operators $\varphi_0^{\dag\,\pm}(k) $ are
\emph{not} the Hermitian conjugate of $\varphi_0^{\pm}(k) $. In the
Hermitian case, $\varphi_0^\dag=\varphi_0$,~\eref{11.28} reduce
to~\eref{3.26} due to~\eref{11.25new1}.
\newline\indent
In view of~\eref{11.23} and~\eref{11.24}, we have:}
			\label{11.29}
\varphi_0 = \varphi_0^+ + \varphi_0^-
\qquad
\varphi_0^\dag = \varphi_0^{\dag\,+} + \varphi_0^{\dag\,-}
\\			\label{11.30}
\varphi_0^\pm
=
\int\Id^3\bs k \varphi_0^\pm (k) |_{k_0=\sqrt{m^2c^2+{\bs k}^2}}
\qquad
\varphi_0^{\dag\,\pm}
=
\int\Id^3\bs k \varphi_0^{\dag\,\pm} (k) |_{k_0=\sqrt{m^2c^2+{\bs k}^2}}
\displaybreak[1]\\			\label{11.31}
	\begin{split}
[\varphi_0^\pm(k),\ope{P}_\mu]_{\_}  = \mp k_\mu \varphi_0^\pm(k)
\qquad
[\varphi_0^{\dag\,\pm}(k),\ope{P}_\mu]_{\_}
				 = \mp k_\mu \varphi_0^{\dag\,\pm}(k)
\qquad
k_0=\sqrt{m^2c^2+{\bs k}^2}
	\end{split}
\displaybreak[1]\\			\label{11.31-1}
	\begin{split}
[\varphi_0^\pm(k),\ope{Q}]_{\_} & = q \varphi_0^\pm(k)
\quad
[\varphi_0^{\dag\,\pm}(k),\ope{Q}]_{\_} = - q \varphi_0^{\dag\,\pm}(k)
\\
[\varphi_0^\pm,\ope{Q}]_{\_} & = q \varphi_0^\pm
\quad
[\varphi_0^{\dag\,\pm} , \ope{Q}]_{\_} = - q \varphi_0^{\dag\,\pm} .
	\end{split}
	\end{gather}

	If $\ope{X}_p$ is a state vector characterizing a state with
4-momentum $p$ (see equation~\eref{3.30}), then~\eref{11.31} entail
	\begin{equation}	\label{11.32}
	\begin{split}
\ope{P}_\mu (\varphi_0^\pm(k) (\ope{X}_p))
=
( p_\mu\pm k_\mu ) \varphi_0^\pm(k)(\ope{X}_p)
\qquad
k_0=\sqrt{m^2c^2+{\bs k}^2}
\\
\ope{P}_\mu (\varphi_0^{\dag\,\pm}(k) (\ope{X}_p))
=
( p_\mu\pm k_\mu ) \varphi_0^{\dag\,\pm}(k)(\ope{X}_p)
\qquad
k_0=\sqrt{m^2c^2+{\bs k}^2} .
	\end{split}
	\end{equation}
So, $\varphi_{0}^{+}$ and $\varphi_{0}^{\dag\,+}$ create particles with
4-momentum $k$, while $\varphi_{0}^{-}$ and $\varphi_{0}^{\dag\,-}$ annihilate
such particles. If $\varphi^\dag=\varphi$, the operators
$\varphi_{0}^{\pm}$ and $\varphi_{0}^{\dag\,\pm}$ coincide, while for
$\varphi^\dag\not=\varphi$ they are different. In the last case the
difference comes from the existence of non\ndash zero charge
operator~\eref{11.11} for which the (Heisenberg
equations/)relations~\eref{11.14} hold. If $\ope{X}_e$ is a state vector
corresponding to a state with total charge $e$, \ie
	\begin{equation}	\label{11.33}
\ope{Q}(\ope{X}_e) = e \ope{X}_e,
	\end{equation}
then, from~\eref{11.19} and~\eref{11.31-1}, we get:%
\footnote{~%
These considerations do not use concrete forms, like~\eref{11.22}, of the
current operator $\ope{J}_\mu$.%
}
	\begin{equation}	\label{11.34}
	\begin{split}
\ope{Q}(\varphi_0(\ope{X}_e)) & = (e-q) \varphi_0(\ope{X}_e)
\quad
\ope{Q}(\varphi_0^\dag(\ope{X}_e)) = (e+q) \varphi_0^\dag(\ope{X}_e)
\\
\ope{Q}(\varphi_0^{\pm}(\ope{X}_e)) & = (e-q) \varphi_0^{\pm}(\ope{X}_e)
\quad
\ope{Q}(\varphi_0^{\dag\,\pm}(\ope{X}_e))
			= (e+q) \varphi_0^{\dag\,\pm}(\ope{X}_e) .
\\ \!\!\!
\ope{Q}(\varphi_0^{\pm}(\bk)(\ope{X}_e))
				& = (e-q) \varphi_0^{\pm}(\bk)(\ope{X}_e)
\quad \!\!\!
\ope{Q}(\varphi_0^{\dag\,\pm}(\bk)(\ope{X}_e))
			= (e+q) \varphi_0^{\dag\,\pm}(\bk)(\ope{X}_e) .
	\end{split}
	\end{equation}
Therefore $\varphi_0$, $\varphi_0^\pm$ and $\varphi_0^\pm(\bk)$ decrease the
field's charge by $q$, while $\varphi_0^\dag$, $\varphi_0^{\dag\,\pm}$ and
$\varphi_0^{\dag\,\pm}(\bk)$ increase it by the same quantity. So, in a
summary,
$\varphi_0^+(\bk)$ and $\varphi_0^{\dag\,+}(\bk)$ create particles with
4\ndash momentum $(\sqrt{m^2c^2+\bk^2},\bk)$ and charges $(-q)$ and $(+q)$,
respectively, while $\varphi_0^-(\bk)$ and $\varphi_0^{\dag\,-}(\bk)$
annihilate particles with 4\ndash momentum $(\sqrt{m^2c^2+\bk^2},\bk)$ and
charges $(+q)$ and $(-q)$, respectively.%
\footnote{~%
By convention, the particles created by $\varphi_0^+$ or annihilated by
$\varphi_0^{\dag\,-}(\bk)$ are called `particles', while the ones created by
$\varphi_0^{\dag\,+}(\bk)$ or annihilated by $\varphi_0^{-}(\bk)$ are called
`antiparticles'.%
}$^{,~}$%
\footnote{~%
Since~\eref{11.34} originates from the equation~\eref{11.14}, which is
external to the Lagrangian formalism, one should accept the given
interpretation of $\varphi_0^\pm(\bk)$ and $\varphi_0^{\dag\,\pm}(\bk)$ by
some reserve. However, this interpretation is confirmed in the later
development of the theory on the base of a notion of $n$\ndash particle,
$n\in\field[N]$, states (see Sect.~\ref{Sect19}). %
}


\section
[The field equations in terms of creation and annihilation operators]
{The field equations in terms of creation and\\ annihilation operators}
	\label{Sect15}

	If one wants to obtain from~\eref{11.10} a system of equations for
the momentum and field operators, an explicit expression for $\ope{P}_\mu$,
as a function (functional) of the field operators $\varphi_0$ and
$\varphi_0^\dag$, is required. To find it, we shall proceed as in
Sect.~\ref{Sect7}, when the Hermitian case, $\varphi_0^\dag=\varphi_0$, was
investigated.

	Since~\eref{11.29}--\eref{11.31} imply
	\begin{equation}	\label{11.34new}
	\begin{split}
[\varphi_0,\ope{P}_\mu]_{\_}
& =
\int \{ k_\mu(- \varphi_0^{+}(k) + \varphi_0^{-}(k)) \}
				|_{k_0=\sqrt{m^2c^2+{\bs k}^2}} \Id^3\bs k
\\
[\varphi_0^\dag,\ope{P}_\mu]_{\_}
& =
\int \{ k_\mu(- \varphi_0^{\dag\,+}(k) + \varphi_0^{\dag\,-}(k)) \}
				|_{k_0=\sqrt{m^2c^2+{\bs k}^2}} \Id^3\bs k ,
	\end{split}
	\end{equation}
the energy-momentum operators~\eref{11.9}, in view of~\eref{11.30}, can be
written as:
	\begin{subequations}	\label{11.35}
	\begin{gather}
			\label{11.35a}
	\begin{split}
{ \lindex[\mspace{-3mu}\ope{T}]{}{\prime} }_{\mu\nu}^{(3)}
=
&  \frac{1}{1+\tau(\varphi_0)} c^2 \int\Id^3\bs k\Id^3\bs k'
\{
(-k_\mu k'_\nu -k_\nu k'_\mu + \eta_{\mu\nu} k_\lambda k^{\prime\,\lambda} )
\\
& \times
( - \varphi_0^{\dag\,+}(k) + \varphi_0^{\dag\,-}(k) )  \circ
( - \varphi_0^{+}(k') + \varphi_0^{-}(k') )
\\
& +
\eta_{\mu\nu} m^2c^2
(  \varphi_0^{\dag\,+}(k) + \varphi_0^{\dag\,-}(k) )  \circ
(  \varphi_0^{+}(k') + \varphi_0^{-}(k') )
\}
	\end{split}
\displaybreak[1]\\			\label{11.35b}
	\begin{split}
{ \lindex[\mspace{-3mu}\ope{T}]{}{\prime\prime} }_{\mu\nu}^{(4)}
=
&  \frac{1}{1+\tau(\varphi_0)} c^2 \int\Id^3\bs k\Id^3\bs k'
\{
(-k_\mu k'_\nu -k_\nu k'_\mu + \eta_{\mu\nu} k_\lambda k^{\prime\,\lambda} )
\\
& \times
( - \varphi_0^{+}(k) + \varphi_0^{-}(k) )  \circ
( - \varphi_0^{\dag\,+}(k') + \varphi_0^{\dag\,-}(k') )
\\
& +
\eta_{\mu\nu} m^2c^2
(  \varphi_0^{+}(k) + \varphi_0^{-}(k) )  \circ
(  \varphi_0^{\dag\,+}(k') + \varphi_0^{\dag\,-}(k') )
\}
	\end{split}
\displaybreak[1]\\			\label{11.35c}
	\begin{split}
{ \lindex[\mspace{-3mu}\ope{T}]{}{\prime\prime\prime} }_{\mu\nu}^{(2)}
=
&  \frac{1}{2(1+\tau(\varphi_0))} c^2 \int\Id^3\bs k\Id^3\bs k'
\{
(-k_\mu k'_\nu -k_\nu k'_\mu + \eta_{\mu\nu} k_\lambda k^{\prime\,\lambda} )
\\ & \times
( - \varphi_0^{\dag\,+}(k) + \varphi_0^{\dag\,-}(k) )  \circ
( - \varphi_0^{+}(k') + \varphi_0^{-}(k') )
\\ & +
(-k_\mu k'_\nu -k_\nu k'_\mu + \eta_{\mu\nu} k_\lambda k^{\prime\,\lambda} )
\\ & \times
( - \varphi_0^{+}(k) + \varphi_0^{-}(k) )  \circ
( - \varphi_0^{\dag\,+}(k') + \varphi_0^{\dag\,-}(k') )
\\
& +
\eta_{\mu\nu} m^2c^2
(  \varphi_0^{\dag\,+}(k) + \varphi_0^{\dag\,-}(k) )  \circ
(  \varphi_0^{+}(k') + \varphi_0^{-}(k') )
\\ & +
\eta_{\mu\nu} m^2c^2
(  \varphi_0^{+}(k) + \varphi_0^{-}(k) )  \circ
(  \varphi_0^{\dag\,+}(k') + \varphi_0^{\dag\,-}(k') )
\}
	\end{split}
	\end{gather}
	\end{subequations}
where $k_0=\sqrt{m^2c^2+{\bs k}^2}$ and $k'_0=\sqrt{m^2c^2+{\bs k'}^2}$ .

	Performing with these expressions the same manipulations as the ones
leading from~\eref{3.33} to~\eref{3.36}, we derive the following expressions
for the momentum operator:%
\footnote{~%
Notice, the equalities~\eref{3.35new} remain valid in the general case.
Besides, these equations hold if in them some or all of the operators
$\varphi_0^{\pm}(k)$ and $\varphi_0^{\pm}(k')$ are replaced with
$\varphi_0^{\dag\,\pm}(k)$ and $\varphi_0^{\dag\,\pm}(k')$, respectively.%
}$^{,}$~%
\footnote{~%
For the other 8 combinations of the Lagrangians~\eref{11.2-1}--\eref{11.2-3}
and energy\ndash momentum operators~\eref{11.7a}--\eref{11.7d}, in the
integrands in the r.h.s.\ of~\eref{11.36} terms proportional to
$\varphi_0^{\pm}(k)\circ \varphi_0^{\dag\,\pm}(k_0,-\bs k)$ and
$\varphi_0^{\dag\,\pm}(k)\circ\varphi_0^{\pm}(k_0,-\bs k)$
will appear. They are responsible for the contradictions mentioned in
Sect.~\ref{Sect11} (see, in particular, footnote~\ref{Contradictions}).%
}
	\begin{subequations}	\label{11.36}
	\begin{gather}
			\label{11.36a}
	\begin{split}
{ \lindex[\mspace{-6mu}\ope{P}]{}{\prime} }_\mu^{(3)}
=
\frac{1}{1+\tau(\varphi_0)}\int
	k_\mu |_{ k_0=\sqrt{m^2c^2+{\bs k}^2} }
 \{
\varphi_0^{\dag\,+}(\bs k)\circ\varphi_0^{-}(\bs k)
 +
\varphi_0^{\dag\,-}(\bs k)\circ\varphi_0^{+}(\bs k)
\}
\Id^3\bs k
	\end{split}
\\			\label{11.36b}
	\begin{split}
{ \lindex[\mspace{-6mu}\ope{P}]{}{\prime\prime} }_\mu^{(4)}
=
\frac{1}{1+\tau(\varphi_0)}\int
	k_\mu |_{ k_0=\sqrt{m^2c^2+{\bs k}^2} }
 \{
\varphi_0^{      +}(\bs k)\circ\varphi_0^{\dag\,-}(\bs k)
 +
\varphi_0^{-}(\bs k)\circ\varphi_0^{\dag\,+}(\bs k)
\}
\Id^3\bs k
	\end{split}
\\			\label{11.36c}
	\begin{split}
{ \lindex[\mspace{-6mu}\ope{P}]{}{\prime\prime\prime} }_\mu^{(2)}
=
\frac{1}{2(1+\tau(\varphi_0))}\int
& 	k_\mu |_{ k_0=\sqrt{m^2c^2+{\bs k}^2} }
\{
\varphi_0^{\dag\,+}(\bs k)\circ\varphi_0^{-}(\bs k)
+
\varphi_0^{\dag\,-}(\bs k)\circ\varphi_0^{+}(\bs k)
\\ & +
\varphi_0^{+}(\bs k)\circ\varphi_0^{\dag\,-}(\bs k)
+
\varphi_0^{-}(\bs k)\circ\varphi_0^{\dag\,+}(\bs k)
\}
\Id^3\bs k .
	\end{split}
	\end{gather}
	\end{subequations}
Here $\tau(\varphi_0)$ is defined by~\eref{11.37} and the following shortcuts
are introduced:
	\begin{gather}
			\label{11.38}
	\begin{split}
\varphi_0^{\pm} (\bs k)
&:=
\bigl\{\bigl( 2 c (2\pi\hbar)^3 k_0 \bigr)^{1/2}
     \varphi_0^{\pm} (k) \bigr\}\big|_{ k_0=\sqrt{m^2c^2+{\bs k}^2} }
\\
\varphi_0^{\dag\,\pm} (\bs k)
&:=
\bigl\{\bigl( 2 c (2\pi\hbar)^3 k_0 \bigr)^{1/2}
     \varphi_0^{\dag\,\pm} (k) \bigr\}\big|_{ k_0=\sqrt{m^2c^2+{\bs k}^2} }
\quad .
	\end{split}
	\end{gather}
The operators $\varphi_0^{+} (\bs k)$ and $\varphi_0^{\dag\,+} (\bs k)$
(resp.\ $\varphi_0^{-} (\bs k)$ and $\varphi_0^{\dag\,-} (\bs k)$)
are called the \emph{creation} (resp.\ \emph{annihilation}) operators (of the
field (field's particles)).

	Obviously, for a Hermitian field, $\varphi_0^\dag=\varphi_0$, all of
the three expressions in~\eref{11.36} reduce to the `right Hermitian'
result~\eref{3.36}; besides, the `3\ndash dimensional' creation/annihilation
operators~\eref{11.38} reduce to~\eref{3.37}, as one should expect. In the
non\ndash Hermitian case, $\varphi_0^\dag\not=\varphi_0$, the
operator~\eref{11.36b}, as a function of the creation/annihilation
operators~\eref{11.38}, formally coincides with the momentum operator
obtained from the Lagrangian~\eref{11.2} in the
literature~\cite[eq.~(3.39)]{Bogolyubov&Shirkov}. However, it should be
remarked, our creation and annihilation operators for the
Lagrangian~\eref{11.2} with $\varphi_0^\dag\not=\varphi_0$ differ by a phase
factor from the ones in Heisenberg picture used in the literature (see below
equations~\eref{3.38new}). Generally, the three momentum
operators~\eref{11.36} are different, but, after normal ordering, they result
into one and the same momentum operator (see Sect.~\ref{Sect18},
equation~\eref{11.71}).

	For a comparison with expressions in (the momentum representation of)
Heisenberg picture, it is worth to be noticed that, due to
proposition~\ref{Prop11.2}, the creation/annihilation operators
$\tope{\varphi}^{\pm}(\bk)$ and $\tope{\varphi}^{\dag\,\pm}(\bk)$ in (the
momentum representation of) Heisenberg picture are (cf.~\eref{3.37new1})
	\begin{equation}	\label{3.38new}
\tope{\varphi}^{\pm} (\bk)
=
\e^{\pm\iih x_0^\mu k_\mu} \bigr|_{k_0=\sqrt{m^2c^2+\bk^2}}
\varphi_0^{\pm}
\quad
\tope{\varphi}^{\dag\,\pm} (\bk)
=
\e^{\pm\iih x_0^\mu k_\mu} \bigr|_{k_0=\sqrt{m^2c^2+\bk^2}}
\varphi_0^{\dag\,\pm} .
	\end{equation}
(Relations like~\eref{3.37new2} are, of course, also valid;~\eref{3.38new}
is a consequence from them for $x^\mu=x_0^\mu$.) Therefore, quadratic
expressions, like the ones in the integrands in~\eref{11.36}, look in one and
the same way in momentum and Heisenberg pictures.

	Now we are ready to obtain the field equations in terms of creation
and annihilation operators. Since~\eref{11.29}--\eref{11.31} are equivalent
to~\eref{11.10}, the equations, we want to derive, are~\eref{11.31} with
$\ope{P}_\mu$ given via~\eref{11.36}. At this stage of the development of
the theory, we will get \emph{three}, generally different, systems of
equations, corresponding to the three Lagrangians,~\eref{11.2-1},
\eref{11.2-2} and \eref{11.2-3}, we started off. But, as after normal
ordering the three momentum operators~\eref{11.36} become identical, these
systems of equations will turn to be identical after normal ordering.

	In terms of the operators~\eref{11.38}, the
equations~\eref{11.29}--\eref{11.31} can equivalently be rewritten as:
	\begin{gather}
			\label{11.39}
	\begin{split}
\varphi_0
& =	\int \bigl( 2 c (2\pi\hbar)^3 k_0 \bigr)^{-1/2}
	(\varphi_0^{+}(\bs k) + \varphi_0^{-}(\bs k)) \Id^3\bs k
\\
\varphi_0^\dag
& =	\int \bigl( 2 c (2\pi\hbar)^3 k_0 \bigr)^{-1/2}
	(\varphi_0^{\dag\,+}(\bs k) + \varphi_0^{\dag\,-}(\bs k)) \Id^3\bs k
	\end{split}
\\			\label{11.40}
	\begin{split}
[\varphi_0^\pm(\bs k),\ope{P}_\mu]_{\_}   = \mp k_\mu \varphi_0^\pm(\bs k)
\quad 
[\varphi_0^{\dag\,\pm}(\bs k),\ope{P}_\mu]_{\_}
				  = \mp k_\mu \varphi_0^{\dag\,\pm}(\bs k)
\qquad
k_0=\sqrt{m^2c^2+{\bs k}^2}~ .
	\end{split}
	\end{gather}
Inserting the equalities
(with $k_0=\sqrt{m^2c^2+{\bs k}^2}$ and $q_0=\sqrt{m^2c^2+{\bs q}^2}$)
	\begin{align*}
\mp k_\mu\varphi_0^{\pm} (\bs k)
 =
\int \mp q_\mu \varphi_0^{\pm} (\bs k) \delta^3(\bs q-\bs k) \Id^3\bs q
\qquad
\mp k_\mu\varphi_0^{\dag\,\pm} (\bs k)
 =
\int \mp q_\mu \varphi_0^{\dag\,\pm} (\bs k) \delta^3(\bs q-\bs k) \Id^3\bs q
	\end{align*}
and~\eref{11.36} into~\eref{11.40}, we, after some algebra, obtain the
next variants of the systems of field equations for an arbitrary scalar field
in terms of creation and annihilation operators:
	\begin{subequations}	\label{11.41}
	\begin{gather}
				\label{11.41a}
	\begin{split}
\int q_\mu |_{q_0=\sqrt{m^2c^2+{\bs q}^2}}
\bigl\{ &
[\varphi_0^{\pm}(\bs k),
	\varphi_0^{\dag\,+}(\bs q) \circ \varphi_0^{-}(\bs q)
      +	\varphi_0^{\dag\,-}(\bs q) \circ \varphi_0^{+}(\bs q) ]_{\_}
\\ &
\pm (1+\tau(\varphi_0)) \varphi_0^{\pm}(\bs k) \delta^3(\bs q-\bs k)
\bigr\} \Id^3\bs q = 0
	\end{split}
\\				\label{11.41b}
	\begin{split}
\int q_\mu |_{q_0=\sqrt{m^2c^2+{\bs q}^2}}
\bigl\{ &
[\varphi_0^{\dag\,\pm}(\bs k),
	\varphi_0^{\dag\,+}(\bs q) \circ \varphi_0^{-}(\bs q)
      +	\varphi_0^{\dag\,-}(\bs q) \circ \varphi_0^{+}(\bs q) ]_{\_}
\\ &
\pm (1+\tau(\varphi_0)) \varphi_0^{\dag\,\pm}(\bs k) \delta^3(\bs q-\bs k)
\bigr\} \Id^3\bs q = 0
	\end{split}
	\end{gather}
	\end{subequations}
	\begin{subequations}	\label{11.42}
	\begin{gather}
				\label{11.42a}
	\begin{split}
\int q_\mu |_{q_0=\sqrt{m^2c^2+{\bs q}^2}}
\bigl\{ &
[\varphi_0^{\pm}(\bs k),
	\varphi_0^{+}(\bs q) \circ \varphi_0^{\dag\,-}(\bs q)
      +	\varphi_0^{-}(\bs q) \circ \varphi_0^{\dag\,+}(\bs q) ]_{\_}
\\ &
\pm (1+\tau(\varphi_0)) \varphi_0^{\pm}(\bs k) \delta^3(\bs q-\bs k)
\bigr\} \Id^3\bs q = 0
	\end{split}
\\				\label{11.42b}
	\begin{split}
\int q_\mu |_{q_0=\sqrt{m^2c^2+{\bs q}^2}}
\bigl\{ &
[\varphi_0^{\dag\,\pm}(\bs k),
	\varphi_0^{+}(\bs q) \circ \varphi_0^{\dag\,-}(\bs q)
      +	\varphi_0^{-}(\bs q) \circ \varphi_0^{\dag\,+}(\bs q) ]_{\_}
\\ &
\pm (1+\tau(\varphi_0)) \varphi_0^{\dag\,\pm}(\bs k) \delta^3(\bs q-\bs k)
\bigr\} \Id^3\bs q = 0
	\end{split}
	\end{gather}
	\end{subequations}
	\begin{subequations}	\label{11.43}
	\begin{gather}
				\label{11.43a}
	\begin{split}
\int q_\mu |_{q_0=\sqrt{m^2c^2+{\bs q}^2}}
\bigl\{ &
[\varphi_0^{\pm}(\bs k),
	[ \varphi_0^{\dag\,+}(\bs q), \varphi_0^{-}(\bs q)]_{+}
      +	[ \varphi_0^{+}(\bs q) , \varphi_0^{\dag\,-}(\bs q)]_{+} ]_{\_}
\\ &
\pm 2 (1+\tau(\varphi_0)) \varphi_0^{\pm}(\bs k) \delta^3(\bs q-\bs k)
\bigr\} \Id^3\bs q = 0
	\end{split}
\\				\label{11.43b}
	\begin{split}
\int q_\mu |_{q_0=\sqrt{m^2c^2+{\bs q}^2}}
\bigl\{ &
[\varphi_0^{\dag\,\pm}(\bs k),
	[ \varphi_0^{\dag\,+}(\bs q), \varphi_0^{-}(\bs q)]_{+}
      +	[ \varphi_0^{+}(\bs q) , \varphi_0^{\dag\,-}(\bs q)]_{+} ]_{\_}
\\ &
\pm 2 (1+\tau(\varphi_0)) \varphi_0^{\dag\,\pm}(\bs k) \delta^3(\bs q-\bs k)
\bigr\} \Id^3\bs q = 0 .
	\end{split}
	\end{gather}
	\end{subequations}

	Consequently, to the Lagrangians~\eref{11.2-1}--\eref{11.2-3}
correspond respectively the following three systems of equations for
$\varphi_0^{\pm}$ and $\varphi_0^{\dag\,\pm}$:
	\begin{subequations}	\label{11.44}
	\begin{gather}
				\label{11.44a}
	\begin{split}
[\varphi_0^{\pm}(\bs k),
	\varphi_0^{\dag\,+}(\bs q) & \circ \varphi_0^{-}(\bs q)
     +	\varphi_0^{\dag\,-}(\bs q) \circ \varphi_0^{+}(\bs q) ]_{\_}
\\ &
\pm (1+\tau(\varphi_0)) \varphi_0^{\pm}(\bs k) \delta^3(\bs q-\bs k)
= {\lindex[\mspace{-3mu}f]{}{\prime}}^{\pm}(\bs k,\bs q)
	\end{split}
\\				\label{11.44b}
	\begin{split}
[\varphi_0^{\dag\,\pm}(\bs k),
	\varphi_0^{\dag\,+}(\bs q) & \circ \varphi_0^{-}(\bs q)
      +	\varphi_0^{\dag\,-}(\bs q) \circ \varphi_0^{+}(\bs q) ]_{\_}
\\ &
\pm (1+\tau(\varphi_0)) \varphi_0^{\dag\,\pm}(\bs k) \delta^3(\bs q-\bs k)
= {\lindex[\mspace{-3mu}f]{}{\prime}}^{\dag\,\pm} (\bs k,\bs q)
	\end{split}
	\end{gather}
	\end{subequations}
	\begin{subequations}	\label{11.45}
	\begin{gather}
				\label{11.45a}
	\begin{split}
[\varphi_0^{\pm}(\bs k),
	\varphi_0^{+}(\bs q) & \circ \varphi_0^{\dag\,-}(\bs q)
      +	\varphi_0^{-}(\bs q) \circ \varphi_0^{\dag\,+}(\bs q) ]_{\_}
\\ &
\pm (1+\tau(\varphi_0)) \varphi_0^{\pm}(\bs k) \delta^3(\bs q-\bs k)
= {\lindex[\mspace{-3mu}f]{}{\prime\prime}}^{\pm} (\bs k,\bs q)
	\end{split}
\\				\label{11.45b}
	\begin{split}
[\varphi_0^{\dag\,\pm}(\bs k),
	\varphi_0^{+}(\bs q) & \circ \varphi_0^{\dag\,-}(\bs q)
      +	\varphi_0^{-}(\bs q) \circ \varphi_0^{\dag\,+}(\bs q) ]_{\_}
\\ &
\pm (1+\tau(\varphi_0)) \varphi_0^{\dag\,\pm}(\bs k) \delta^3(\bs q-\bs k)
= {\lindex[\mspace{-3mu}f]{}{\prime\prime}}^{\dag\,\pm} (\bs k,\bs q)
	\end{split}
	\end{gather}
	\end{subequations}
	\begin{subequations}	\label{11.46}
	\begin{gather}
				\label{11.46a}
	\begin{split}
[\varphi_0^{\pm}(\bs k),
	[ \varphi_0^{\dag\,+}(\bs q) & , \varphi_0^{-}(\bs q)]_{+}
      +	[ \varphi_0^{+}(\bs q) , \varphi_0^{\dag\,-}(\bs q)]_{+} ]_{\_}
\\ &
\pm 2 (1+\tau(\varphi_0)) \varphi_0^{\pm}(\bs k) \delta^3(\bs q-\bs k)
= {\lindex[\mspace{-3mu}f]{}{\prime\prime\prime}}^{\pm} (\bs k,\bs q)
	\end{split}
\\				\label{11.46b}
	\begin{split}
[\varphi_0^{\dag\,\pm}(\bs k),
	[ \varphi_0^{\dag\,+}(\bs q) & , \varphi_0^{-}(\bs q)]_{+}
      +	[ \varphi_0^{+}(\bs q) , \varphi_0^{\dag\,-}(\bs q)]_{+} ]_{\_}
\\ &
\pm 2 (1+\tau(\varphi_0)) \varphi_0^{\dag\,\pm}(\bs k) \delta^3(\bs q-\bs k)
= {\lindex[\mspace{-3mu}f]{}{\prime\prime\prime}}^{\dag\,\pm} (\bs k,\bs q) ,
	\end{split}
	\end{gather}
	\end{subequations}
where the operator-valued (generalized) functions
$ {\lindex[\mspace{-3mu}f]{}{a}}^{\pm} (\bs k,\bs q)$ and
$ {\lindex[\mspace{-3mu}f]{}{a}}^{\dag\,\pm} (\bs k,\bs q)$ with
 $a=\prime,\prime\prime,\prime\prime\prime$ must be such that
	\begin{equation}	\label{11.47}
\int q_\mu |_{q_0=\sqrt{m^2c^2+{\bs q}^2}}
	{\lindex[\mspace{-3mu}f]{}{a}}^{\pm} (\bs k,\bs q) \Id^3\bs q
=
\int q_\mu |_{q_0=\sqrt{m^2c^2+{\bs q}^2}}
	{\lindex[\mspace{-3mu}f]{}{a}}^{\dag\,\pm} (\bs k,\bs q) \Id^3\bs q
= 0 .
	\end{equation}
Generally the systems of equations~\eref{11.44},~\eref{11.45}
and~\eref{11.46} are different. They will become equivalent after normal
ordering, when the equations in them will turn to be identical. From the
derivation of~\eref{11.44}--\eref{11.47}, it is clear that these systems of
equations are equivalent to the initial system of Klein\ndash Gordon
equations~\eref{11.10}. Thus, we can say that these systems represent the
field equations in terms of creation and annihilation operators.

	As a verification of the self-consistence of the theory, one can prove
the commutativity between the components of any one of the momentum
operators~\eref{11.36}, \ie
	\begin{equation}	\label{11.47-1}
[\ope{P}_\mu,\ope{P}_\nu]_{\_} = 0 ,
	\end{equation}
where
\(
\ope{P}_\mu
=
 { \lindex[\mspace{-6mu}\ope{P}]{}{\prime} }_\mu^{(3)},
 { \lindex[\mspace{-6mu}\ope{P}]{}{\prime\prime} }_\mu^{(4)},
 { \lindex[\mspace{-6mu}\ope{P}]{}{\prime\prime\prime} }_\mu^{(2)} ,
\)
is a consequence of~\eref{11.36}, \eref{11.44}--\eref{11.46} and the
identity~\eref{3.43-3} with $\varepsilon=-1$.


\section
[The charge and orbital angular momentum operators]
{The charge and orbital angular momentum operators}
	\label{Sect17}

	In Sect.~\ref{Sect11}, we introduced the charge
operator~\eref{11.11} (see also~\eref{11.11-1}) and pointed to different
possible definitions of the defining it current operator. In particular, to
the Lagrangians~\eref{11.2-1}--\eref{11.2-3} correspond respectively the
current operators~\eref{11.22a}--\eref{11.22c} in momentum picture. Below we
shall express the charge  operator $\ope{Q}$ through the
creation and annihilation operators~\eref{11.28}.

	Substituting~\eref{11.29},~\eref{11.30} and~\eref{11.34new}
into~\eref{11.22}, we get:%
\footnote{~%
Since we apply only results based on the general properties of the momentum
operator, the below obtained equalities for the current and charge operators
are independent of the concrete choice of energy\ndash momentum operator,
like~\eref{11.7a}--\eref{11.7d}, and, consequently, of the particular form of
the momentum operator, like~\eref{11.36}. However, they depend on the
Lagrangians~\eref{11.2-1}--\eref{11.2-3} from which the theory is derived.%
}
	\begin{subequations}	\label{11.68}
	\begin{align}
		\notag
{ \lindex[\mspace{-6mu}\ope{J}]{}{\prime} }_\mu^{(3)}
 =
\frac{1}{2} q c^2 \int
& \Id^3\bs k\Id^3\bs k'
\bigl\{
 k'_\mu ( \varphi_0^{\dag\,+}(\bs k) + \varphi_0^{\dag\,-}(\bs k)) \circ
       (-\varphi_0^{+}(\bs k') + \varphi_0^{-}(\bs k') )
\\		\label{11.68a}
&- k_\mu (-\varphi_0^{\dag\,+}(\bs k) + \varphi_0^{\dag\,-}(\bs k)) \circ
       ( \varphi_0^{+}(\bs k') + \varphi_0^{-}(\bs k') )
\bigr\}
\displaybreak[1]\\[0.5ex]		\notag
{ \lindex[\mspace{-6mu}\ope{J}]{}{\prime\prime} }_\mu^{(4)}
 =
\frac{1}{2} q c^2 \int
& \Id^3\bs k\Id^3\bs k'
\bigl\{
k_\mu  (-\varphi_0^{+}(\bs k) + \varphi_0^{-}(\bs k)) \circ
       ( \varphi_0^{\dag\,+}(\bs k') + \varphi_0^{\dag\,-}(\bs k') )
\\	    		\label{11.68b}
& -k'_\mu( \varphi_0^{+}(\bs k) + \varphi_0^{-}(\bs k)) \circ
       (-\varphi_0^{\dag\,+}(\bs k') + \varphi_0^{\dag\,-}(\bs k') )
\bigr\}
\displaybreak[1]\\[0.5ex]	    		\notag
{ \lindex[\mspace{-6mu}\ope{J}]{}{\prime\prime\prime} }_\mu^{(2)}
 =
\frac{1}{4} q c^2 \int
& \Id^3\bs k\Id^3\bs k'
\bigl\{
k'_\mu ( \varphi_0^{\dag\,+}(\bs k) + \varphi_0^{\dag\,-}(\bs k)) \circ
       (-\varphi_0^{+}(\bs k') + \varphi_0^{-}(\bs k') )
\\		\notag
& +k_\mu (-\varphi_0^{+}(\bs k) + \varphi_0^{-}(\bs k)) \circ
       ( \varphi_0^{\dag\,+}(\bs k') + \varphi_0^{\dag\,-}(\bs k') )
\\		\notag
& -k_\mu (-\varphi_0^{\dag\,+}(\bs k) + \varphi_0^{\dag\,-}(\bs k)) \circ
       ( \varphi_0^{+}(\bs k') + \varphi_0^{-}(\bs k') )
\\			\label{11.68c}
& -k'_\mu( \varphi_0^{+}(\bs k) + \varphi_0^{-}(\bs k)) \circ
       (-\varphi_0^{\dag\,+}(\bs k') + \varphi_0^{\dag\,-}(\bs k') )
\bigr\}~~ .
	\end{align}
	\end{subequations}

	Now we have to perform the following three steps:
	(i)~insert these expressions with $\mu=0$ into~\eref{11.11};
	(ii)~apply~\eref{3.35new}, possibly with
$\varphi_0^{\dag\,\pm}(\bs k)$ and/or $\varphi_0^{\dag\,\pm}(\bs k')$ for
$\varphi_0^{\pm}(\bs k)$ and/or $\varphi_0^{\pm}(\bs k')$;
	(iii)~integrate over $\bs x$, which gives
$(2\pi\hbar)^3\delta^3(\bs k+\bs k')$ for products of equal\ndash frequency
operators and
$(2\pi\hbar)^3\delta^3(\bs k-\bs k')$ for products of different\ndash
frequency operators.
	As a result of these steps, we obtain:%
\footnote{~%
When deriving~\eref{11.69}, one gets the equalities in it with
$\frac{q}{1+\tau(\varphi_0)}$
for $q$ with $\tau(\varphi_0)$ defined by~\eref{11.37}. Since $q=0$ for
$\varphi_0^\dag=\varphi_0$ and $\tau(\varphi_0)=0$ for
$\varphi_0^\dag\not=\varphi_0$, we have
$\frac{q}{1+\tau(\varphi_0)} \equiv q$ in all cases.%
}
	\begin{subequations}	\label{11.69}
	\begin{align}
				\label{11.69a}
{ \lindex[\mspace{-6mu}\tope{Q}]{}{\prime} }^{(3)}
 =
q  \int
& \Id^3\bs k
\bigl\{
  \varphi_0^{\dag\,+}(\bs k) \circ \varphi_0^{-}(\bs k)
- \varphi_0^{\dag\,-}(\bs k) \circ \varphi_0^{+}(\bs k)
\bigr\}
\displaybreak[1]\\[0.5ex]		\label{11.69b}
{ \lindex[\mspace{-6mu}\tope{Q}]{}{\prime\prime} }^{(4)}
 =
- q  \int
& \Id^3\bs k
\bigl\{
  \varphi_0^{+}(\bs k) \circ \varphi_0^{\dag\,-}(\bs k)
- \varphi_0^{-}(\bs k) \circ \varphi_0^{\dag\,+}(\bs k)
\bigr\}
\displaybreak[1]\\[0.5ex]	        \notag
{ \lindex[\mspace{-6mu}\tope{Q}]{}{\prime\prime\prime} }^{(2)}
 =
\frac{1}{2} q \int
& \Id^3\bs k
\bigl\{
  \varphi_0^{\dag\,+}(\bs k) \circ \varphi_0^{-}(\bs k)
- \varphi_0^{+}(\bs k) \circ \varphi_0^{\dag\,-}(\bs k)
\\					 \label{11.69c}
& + \varphi_0^{-}(\bs k) \circ \varphi_0^{\dag\,+}(\bs k)
- \varphi_0^{\dag\,-}(\bs k) \circ \varphi_0^{+}(\bs k)
\bigr\} .
	\end{align}
	\end{subequations}
Recall, here the `3-dimensional' creation/annihilation operators are defined
via~\eref{11.38}. Notice, if $\varphi_0^\dag=\varphi_0$, the
charges~\eref{11.69a} and~\eref{11.69b}, as well as the defining them
respective current operators~\eref{11.22a} and~\eref{11.22b}, vanish as $q=0$
in this case, while the charge~\eref{11.69c} and the defining it current
operator~\eref{11.22c} vanish for \emph{two} reasons, if
$\varphi_0^\dag=\varphi_0$: due to $q=0$ and  due to the vanishment of the
integrand in~\eref{11.69c} or the expression in parentheses
in~\eref{11.22c}.

	Generally, the three charge operators~\eref{11.69a}--\eref{11.69c},
originating respectively from the Lagrangians~\eref{11.2-1}--\eref{11.2-3},
are different, but they become identical after normal ordering (see
Sect.~\ref{Sect18}).

	Let us turn our attention now to the orbital angular momentum
operator of free scalar field. In Heisenberg picture, it is given
via~\eref{3.9-3} To obtain an explicit expression for the orbital angular
momentum operator $\tope{L}_{\mu\nu}$  (in Heisenberg picture) through the
creation and annihilation operators $\varphi_{0}^{\pm}(\bk)$ and
$\varphi_{0}^{\dag\,\pm}(\bk)$ (see~\eref{11.38}), we have to do
the following: substitute each of the equalities in~\eref{11.35} into the last
equality in~\eref{3.9-3}, to apply~\eref{3.35new} to all terms in the
obtained equation, then to integrate over $x$, which results in $\delta$\ndash
function terms, and, at last, to perform the integration over $\bk'$ by
means of the arising $\delta$\ndash functions.%
\footnote{~
	 The integrals, one has to calculate, are of the type
	\begin{multline}	\label{7.12-1}
J
 = \sum_{\alpha} \sum_{\varepsilon,\varepsilon'=+,-}
\int\Id^3\bs x \int\Id^3\bk\Id^3\bk'
x_a
\ope{U}^{-1}(x,x_0)
 \circ
\{ \varphi_0^{\dag\,\varepsilon}(\bk)
A_{\varepsilon\varepsilon'}^{\alpha} (\bk,\bk')
   \circ \varphi_{0}^{\varepsilon'}(\bk')
\}
\circ\ope{U}(x,x_0)
\\
 = \sum_{\alpha} \sum_{\varepsilon,\varepsilon'=+,-}
\int\!\!\Id^3\bs x \int\!\!\Id^3\bk\Id^3\bk'
\{ \varphi_0^{\dag\,\varepsilon}(\bk)
A_{\varepsilon\varepsilon'}^{\alpha} (\bk,\bk')
   \circ \varphi_{s'}^{\varepsilon'}(\bk')
\}
x_a
\e^{-\iih (x^\lambda-x_0^\lambda)
	(\varepsilon k_\lambda + \varepsilon' k'_\lambda)} ,
	\end{multline}
where $a=1,2,3$, $A_{\varepsilon\varepsilon'}^{\alpha} (\bk,\bk')$,
$\alpha=1,2,\dots$, are some functions, $k_0=\sqrt{m^2c^2+\bk^2}$, and
$k'_0=\sqrt{m^2c^2+{\bk'}^2}$. The integration over $x$ results in
\(
(2\pi\hbar)^3 \Bigl( -\ih
\frac{\pd}
{ \pd( \varepsilon k^a + \varepsilon' k^{\prime\, a} ) }
\Bigr) \delta^3 ( \varepsilon \bk + \varepsilon' \bk' ) .
\)
Simple manipulations with the remaining terms, by invoking the equality
\(
f(y)\frac{\pd\delta(y)}{\pd y} = -\frac{\pd f(y)} {\pd y} \delta(y)
\)
in the form
	\begin{multline*}
\int \Id^3\bs y \Id^3\bs z
f(\bs y,\bs z) \frac{\pd\delta^3(\bs y-\bs z)}{\pd(y^a-z^a)}
=
\int \Id^3\bs y \Id^3\bs z
\frac{1}{2}\bigl( f(\bs y,\bs z) - f(\bs z,\bs y) \bigr)
\frac{\pd\delta^3(\bs y-\bs z)}{\pd(y^a-z^a)}
\\
=
- \frac{1}{2} \int \Id^3\bs y \Id^3\bs z  \delta^3(\bs y-\bs z)
\Big( \frac{\pd}{\pd y^a} - \frac{\pd}{\pd z^a} \Bigr)
 f(\bs y,\bs z) ,
	\end{multline*}
gives the following result:
	\begin{multline}	\label{7.12-2}
J =
(2\pi\hbar)^3
  \sum_{\alpha} \sum_{\varepsilon,\varepsilon'=+,-}
  \int\Id^3\bk\Id^3\bk'
\delta^3(\varepsilon \bk + \varepsilon'\bk')
\e^{ - \iih (x^0-x_0^0) k_0 (\varepsilon 1 + \varepsilon' 1) }
\\ \times
\Bigl\{
(x^0-x_0^0) \frac{1}{2}\Bigl( \frac{k_a}{k_0} + \frac{k'_a}{k_0} \Bigr)
+ x_{0a}
+ \frac{1}{2} \ih
\Bigl( \varepsilon \frac{\pd}{\pd k^a}
     + \varepsilon'\frac{\pd}{\pd {k'}^a} \Bigr)
\Bigr\}
\bigl\{  \varphi_0^{\dag\,\varepsilon}(\bk)
A_{\varepsilon\varepsilon'}^{\alpha} (\bk,\bk')
  \circ \varphi_{0}^{\varepsilon'}(\bk') \bigr\} .
	\end{multline}
The particular form of $A_{\varepsilon\varepsilon'}^{\alpha} (\bk,\bk')$ is
clear from~\eref{3.9-3} and~\eref{11.9}. So, applying several
times~\eref{7.12-2}, calculating the appearing derivatives, and, at last,
performing the trivial integration over $\bk$ or $\bk'$ by means of
 $\delta^3(\varepsilon \bk + \varepsilon'\bk')$,
one can derive~\eref{17.1} after simple, but lengthy and tedious algebraic
manipulations.%
} 
This procedure results in:
	\begin{subequations}
				\label{17.1}
	\begin{multline}	\label{17.1a}
{ \lindex[\mspace{-6mu}\tope{L}]{}{\prime} }_{\mu\nu}^{(3)}
=
x_{0\,\mu} { \lindex[\mspace{-6mu}\ope{P}]{}{\prime} }_\nu^{(3)}
-
x_{0\,\nu} { \lindex[\mspace{-6mu}\ope{P}]{}{\prime} }_\mu^{(3)}
+
\frac{\ih}{2(1+\tau(\varphi_0))} \int\Id^3\bk
\Bigl\{
\varphi_{0}^{\dag\,+}(\bk)
\Bigl( \xlrarrow{ k_\mu \frac{\pd}{\pd k^\nu} }
     - \xlrarrow{ k_\nu \frac{\pd}{\pd k^\mu} } \Bigr)
\circ \varphi_{0}^-(\bk)
\\  -
\varphi_{0}^{\dag\,-}(\bk)
\Bigl( \xlrarrow{ k_\mu \frac{\pd}{\pd k^\nu} }
     - \xlrarrow{ k_\nu \frac{\pd}{\pd k^\mu} } \Bigr)
\circ \varphi_{0}^+(\bk)
\Bigr\}
\Big|_{	k_0=\sqrt{m^2c^2+\bk^2} }
	\end{multline}
\vspace{-3ex}
	\begin{multline}	\label{17.1b}
{ \lindex[\mspace{-6mu}\tope{L}]{}{\prime\prime} }_{\mu\nu}^{(4)}
=
x_{0\,\mu} { \lindex[\mspace{-6mu}\ope{P}]{}{\prime\prime} }_\nu^{(4)}
-
x_{0\,\nu} { \lindex[\mspace{-6mu}\ope{P}]{}{\prime\prime} }_\mu^{(4)}
+
\frac{\ih}{2(1+\tau(\varphi_0))} \int\Id^3\bk
\Bigl\{
\varphi_{0}^{+}(\bk)
\Bigl( \xlrarrow{ k_\mu \frac{\pd}{\pd k^\nu} }
     - \xlrarrow{ k_\nu \frac{\pd}{\pd k^\mu} } \Bigr)
\circ \varphi_{0}^{\dag\,-}(\bk)
\\  -
\varphi_{0}^{-}(\bk)
\Bigl( \xlrarrow{ k_\mu \frac{\pd}{\pd k^\nu} }
     - \xlrarrow{ k_\nu \frac{\pd}{\pd k^\mu} } \Bigr)
\circ \varphi_{0}^{\dag\,+}(\bk)
\Bigr\}
\Big|_{	k_0=\sqrt{m^2c^2+\bk^2} }
	\end{multline}
\vspace{-3ex}
	\begin{multline}	\label{17.1c}
{ \lindex[\mspace{-6mu}\tope{L}]{}{\prime\prime\prime} }_{\mu\nu}^{(2)}
=
x_{0\,\mu} { \lindex[\mspace{-6mu}\ope{P}]{}{\prime\prime\prime} }_\nu^{(2)}
-
x_{0\,\nu} { \lindex[\mspace{-6mu}\ope{P}]{}{\prime\prime\prime} }_\mu^{(2)}
\\ +
\frac{\ih}{4(1+\tau(\varphi_0))} \int\Id^3\bk
\Bigl\{
\varphi_{0}^{\dag\,+}(\bk)
\Bigl( \xlrarrow{ k_\mu \frac{\pd}{\pd k^\nu} }
     - \xlrarrow{ k_\nu \frac{\pd}{\pd k^\mu} } \Bigr)
\circ \varphi_{0}^-(\bk)
\\
+
\varphi_{0}^{+}(\bk)
\Bigl( \xlrarrow{ k_\mu \frac{\pd}{\pd k^\nu} }
     - \xlrarrow{ k_\nu \frac{\pd}{\pd k^\mu} } \Bigr)
\circ \varphi_{0}^{\dag\,-}(\bk)
    -
\varphi_{0}^{\dag\,-}(\bk)
\Bigl( \xlrarrow{ k_\mu \frac{\pd}{\pd k^\nu} }
     - \xlrarrow{ k_\nu \frac{\pd}{\pd k^\mu} } \Bigr)
\circ \varphi_{0}^+(\bk)
\\  -
\varphi_{0}^{-}(\bk)
\Bigl( \xlrarrow{ k_\mu \frac{\pd}{\pd k^\nu} }
     - \xlrarrow{ k_\nu \frac{\pd}{\pd k^\mu} } \Bigr)
\circ \varphi_{0}^{\dag\,+}(\bk)
\Bigr\}
\Big|_{	k_0=\sqrt{m^2c^2+\bk^2} } \ \ .
	\end{multline}
	\end{subequations}
Here the derivatives with respect to $k_0$, like
$\frac{\pd}{\pd k^0}\varphi_{0}^{\pm}(\bk)$, must be set equal to zero, and%
\footnote{~%
More generally, if $\omega\colon\{\Hil\to\Hil\}\to\{\Hil\to\Hil\}$ is a
mapping on the operator space over the system's Hilbert space, we put
 $A\xlrarrow{\omega}\circ B := -\omega(A)\circ B + A\circ \omega(B)$
for any $A,B\colon\Hil\to\Hil$. Usually~\cite{Bjorken&Drell,Itzykson&Zuber},
this notation is used for $\omega=\pd_\mu$.%
}
	\begin{multline}	\label{17.2}
A(\bk) \xlrarrow{ k_\mu\frac{\pd}{\pd k^\nu} } \circ B(\bk)
:=
-
\Bigl( k_\mu\frac{\pd A(\bk)}{\pd k^\nu} \Bigr) \circ B(\bk)
+
\Bigl( A(\bk) \circ k_\mu\frac{\pd B(\bk)}{\pd k^\nu} \Bigr)
\\ =
k_\mu \Bigl(  A(\bk) \xlrarrow{ \frac{\pd}{\pd k^\nu} } \circ B(\bk) \Bigr)
	\end{multline}
for operators $A(\bk)$ and $B(\bk)$ having $C^1$ dependence on $\bk$.
	If the operators $A(\bk)$ and $B(\bk)$ tend to zero sufficiently fast
at spacial infinity, then, by integration by parts, one can prove the
equality
	\begin{multline}	\label{17.2-1}
\int\Id^3\bk
\Bigl\{ A(\bk)
\Bigl(
  \xlrarrow{ k_\mu\frac{\pd}{\pd k^\nu} }
- \xlrarrow{ k_\nu\frac{\pd}{\pd k^\mu} }
\Bigr) \circ B(\bk)
\Bigr\}
\Big|_{	k_0=\sqrt{m^2c^2+\bk^2} }
\\
=
2 \int\Id^3\bk
\Bigl\{ A(\bk) \circ
\Bigl(
  { k_\mu\frac{\pd}{\pd k^\nu} }
- { k_\nu\frac{\pd}{\pd k^\mu} }
\Bigr) B(\bk)
\Bigr\}
\Big|_{	k_0=\sqrt{m^2c^2+\bk^2} }
\\
=
-2 \int\Id^3\bk
\Bigl\{
\Bigl( \Bigl(
  { k_\mu\frac{\pd}{\pd k^\nu} }
- { k_\nu\frac{\pd}{\pd k^\mu} }
\Bigr) A(\bk) \Bigr) \circ B(\bk)
\Bigr\}
\Big|_{	k_0=\sqrt{m^2c^2+\bk^2} } \ .
	\end{multline}
By means of these equations, one can reduce (two times) the number of terms
in~\eref{17.1}, but we prefer to retain the `more (anti)symmetric' form of
the results by invoking the operation introduced via~\eref{17.2}.

	Since for a neutral scalar field $\varphi_0^\dag=\varphi_0$ and
$\tau(\varphi_0)=1$, in this case the three operators~\eref{17.1} reduce to
	\begin{multline}	\label{17.3}
{ \lindex[\mspace{-6mu}\tope{L}]{}{0} }_{\mu\nu}
=
x_{0\,\mu} { \lindex[\mspace{-6mu}\ope{P}]{}{} }_\nu
-
x_{0\,\nu} { \lindex[\mspace{-6mu}\ope{P}]{}{} }_\mu
+
\frac{\ih}{4} \int\Id^3\bk
\Bigl\{
\varphi_{0}^{+}(\bk)
\Bigl( \xlrarrow{ k_\mu \frac{\pd}{\pd k^\nu} }
     - \xlrarrow{ k_\nu \frac{\pd}{\pd k^\mu} } \Bigr)
\circ \varphi_{0}^-(\bk)
\\  -
\varphi_{0}^{-}(\bk)
\Bigl( \xlrarrow{ k_\mu \frac{\pd}{\pd k^\nu} }
     - \xlrarrow{ k_\nu \frac{\pd}{\pd k^\mu} } \Bigr)
\circ \varphi_{0}^+(\bk)
\Bigr\}
\Big|_{	k_0=\sqrt{m^2c^2+\bk^2} }
	\end{multline}
with $\ope{P}_\mu$ given by~\eref{3.36}. The reader may verify that the
equality~\eref{17.3} holds for any one of the energy\ndash momentum
operators~\eref{3.9} (or~\eref{3.8}) for the Lagrangian~\eref{3.2}
(or~\eref{3.1}).

	It is a trivial calculation to be shown that in terms of the
operators~\eref{3.38new}, representing the creation/annihilation operators in
(momentum representation of) Heisenberg picture, in the equations~\eref{17.1}
and~\eref{17.3} the first two terms, proportional to the momentum operator,
should be deleted and tildes over all creation and annihilation operators
must be added.

	Using the explicit formulae~\eref{11.36},~\eref{11.69}
and~\eref{17.1}, by means of the identity~\eref{3.43-3}, with
$\varepsilon=-1$, and the field equations~\eref{11.44}--\eref{11.46}, one can
verify the equations:
	\begin{align}
			\label{17.4}
& [ \tope{Q},\ope{P}_\lambda ] = 0
\\			\label{17.5}
& [ \tope{L}_{\mu\nu} , \ope{P}_\lambda ]
 = - \ih ( \eta_{\lambda\mu}\ope{P}_\nu - \eta_{\lambda\nu}\ope{P}_\mu ) ,
	\end{align}
where $\tope{L}_{\mu\nu}$ (resp.\ $\tope{Q}$) denotes any one of the orbital
momentum operators in~\eref{17.1} (resp.\ charge operators in~\eref{11.65}).
The first of these equalities confirms the validity of the (external to the
Lagrangian formalism) equation~\eref{11.16}.  We emphasize on the sign before
$\ih$ in the r.h.s.\ of~\eref{17.5}, which is opposite to the one usually
assumed in the literature, for instance
in~\cite[p.~77, eq.~(2-87)]{Roman-QFT} or in~\cite[eq.~(2.187)]{Ryder-QFT}.%
\footnote{~%
Equation~\eref{17.5} with $+\ih$ for $-\ih$ in its r.h.s.\ is part of the
commutation relations for the Lie algebra of the Poincar\'e group -- see,
e.g.,~\cite[pp.143--147]{Bogolyubov&et_al.-AxQFT}
or~\cite[sec.~7.1]{Bogolyubov&et_al.-QFT}. However, such a change of the sign
in the r.h.s.\ of~\eref{17.5} contradicts to the results and the physical
interpretation of the creation and annihilation operators -- \emph{vide
infra}.%
}

	From~\eref{17.4} and~\eref{12.112}, we get
	\begin{gather}	\label{17.4-1}
[\tope{Q},\ope{U}(x,x_0)]_{\_} = 0
\\\intertext{and, hence,}
			\label{17.4-2}
\ope{Q} = \tope{Q},
	\end{gather}
due to~\eref{12.114}. Therefore the \emph{charge operator is one and the same
in momentum and Heisenberg pictures.}

	Applying~\eref{17.5}, we get the orbital angular momentum of a scalar
field in \emph{momentum picture} as
	\begin{align}  \notag
\ope{L}_{\mu\nu}
& = \ope{U}(x,x_0) \circ \tope{L}_{\mu\nu} \circ \ope{U}^{-1}(x,x_0)
  = \tope{L}_{\mu\nu} + [\ope{U}(x,x_0),\tope{L}_{\mu\nu}]_{\_} \circ
  						\ope{U}^{-1}(x,x_0)
\\ 	\label{17.6}
& =
\tope{L}_{\mu\nu}
+ (x_\mu-x_{0\,\mu}) \ope{P}_\nu - (x_\nu-x_{0\,\nu}) \ope{P}_\mu ,
	\end{align}
due to the equality
	\begin{equation}	\label{17.7}
[\tope{L}_{\mu\nu}, \ope{U}(x,x_0)]_{\_}
=
- \{ (x_\mu-x_{0\,\mu}) \tope{P}_\nu
   - (x_\nu-x_{0\,\nu}) \tope{P}_\mu \} \circ \ope{U}(x,x_0) .
	\end{equation}
which is a consequence of~\eref{17.5} and~\eref{12.112}.%
\footnote{~\label{CommutativityWithMomentum}%
To derive equation~\eref{17.7}, notice that~\eref{17.5} implies
\(
[\tope{L}_{\mu\nu}, \tope{P}_{\mu_1}\circ\dots\circ\tope{P}_{\mu_n} ]_{\_}
=
- \sum_{i=1}^{n}
\bigl( \eta_{\mu\mu_i} \tope{P}_{\nu} - \eta_{\nu\mu_i} \tope{P}_{\mu} \bigr)
\tope{P}_{\mu_1}\circ\dots\circ
\tope{P}_{\mu_{i-1}}\circ\tope{P}_{\mu_{i+1}} \circ \dots \circ
\tope{P}_{\mu_n},
\)
due to $[A,B\circ C]_{\_} = [A,B]_{\_}\circ C+ B\circ [A,C]_{\_}$,
and expand the exponent in~\eref{12.112} into a power series.
More generally, if $[A(x),\tope{P}_\mu]_{\_}=B_\mu(x)$ with
$[B_\mu(x),\tope{P}_\nu]_{\_}=0$, then
 $[A(x),\ope{U}(x,x_0)]_{\_} = \iih (x^\mu-x_0^\mu)B_\mu(x) \ope{U}(x,x_0)$;
in particular, $[A(x),\tope{P}_\mu]_{\_}=0$ implies
 $[A(x),\ope{U}(x,x_0)]_{\_}=0$. Notice, we consider $(x^\mu-x_0^\mu)$ as a
real parameter by which the corresponding operators are multiplied and which
operators are supposed to be linear in it.%
}
Thus, in momentum picture, the equations~\eref{17.1} read:
	\begin{subequations}
				\label{17.8}
	\begin{multline}	\label{17.8a}
{ \lindex[\mspace{-6mu}\ope{L}]{}{\prime} }_{\mu\nu}^{(3)}
=
x_{\mu} { \lindex[\mspace{-6mu}\ope{P}]{}{\prime} }_\nu^{(3)}
-
x_{\nu} { \lindex[\mspace{-6mu}\ope{P}]{}{\prime} }_\mu^{(3)}
+
\frac{\ih}{2(1+\tau(\varphi_0))} \int\Id^3\bk
\Bigl\{
\varphi_{0}^{\dag\,+}(\bk)
\Bigl( \xlrarrow{ k_\mu \frac{\pd}{\pd k^\nu} }
     - \xlrarrow{ k_\nu \frac{\pd}{\pd k^\mu} } \Bigr)
\circ \varphi_{0}^-(\bk)
\\  -
\varphi_{0}^{\dag\,-}(\bk)
\Bigl( \xlrarrow{ k_\mu \frac{\pd}{\pd k^\nu} }
     - \xlrarrow{ k_\nu \frac{\pd}{\pd k^\mu} } \Bigr)
\circ \varphi_{0}^+(\bk)
\Bigr\}
\Big|_{	k_0=\sqrt{m^2c^2+\bk^2} }
	\end{multline}
\vspace{-3ex}
	\begin{multline}	\label{17.8b}
{ \lindex[\mspace{-6mu}\ope{L}]{}{\prime\prime} }_{\mu\nu}^{(4)}
=
x_{\mu} { \lindex[\mspace{-6mu}\ope{P}]{}{\prime\prime} }_\nu^{(4)}
-
x_{\nu} { \lindex[\mspace{-6mu}\ope{P}]{}{\prime\prime} }_\mu^{(4)}
+
\frac{\ih}{2(1+\tau(\varphi_0))} \int\Id^3\bk
\Bigl\{
\varphi_{0}^{+}(\bk)
\Bigl( \xlrarrow{ k_\mu \frac{\pd}{\pd k^\nu} }
     - \xlrarrow{ k_\nu \frac{\pd}{\pd k^\mu} } \Bigr)
\circ \varphi_{0}^{\dag\,-}(\bk)
\\  -
\varphi_{0}^{-}(\bk)
\Bigl( \xlrarrow{ k_\mu \frac{\pd}{\pd k^\nu} }
     - \xlrarrow{ k_\nu \frac{\pd}{\pd k^\mu} } \Bigr)
\circ \varphi_{0}^{\dag\,+}(\bk)
\Bigr\}
\Big|_{	k_0=\sqrt{m^2c^2+\bk^2} }
	\end{multline}
\vspace{-3ex}
	\begin{multline}	\label{17.8c}
{ \lindex[\mspace{-6mu}\ope{L}]{}{\prime\prime\prime} }_{\mu\nu}^{(2)}
=
x_{\mu} { \lindex[\mspace{-6mu}\ope{P}]{}{\prime\prime\prime} }_\nu^{(2)}
-
x_{\nu} { \lindex[\mspace{-6mu}\ope{P}]{}{\prime\prime\prime} }_\mu^{(2)}
+
\frac{\ih}{4(1+\tau(\varphi_0))} \int\Id^3\bk
\Bigl\{
\varphi_{0}^{\dag\,+}(\bk)
\Bigl( \xlrarrow{ k_\mu \frac{\pd}{\pd k^\nu} }
     - \xlrarrow{ k_\nu \frac{\pd}{\pd k^\mu} } \Bigr)
\circ \varphi_{0}^-(\bk)
\\
+
\varphi_{0}^{+}(\bk)
\Bigl( \xlrarrow{ k_\mu \frac{\pd}{\pd k^\nu} }
     - \xlrarrow{ k_\nu \frac{\pd}{\pd k^\mu} } \Bigr)
\circ \varphi_{0}^{\dag\,-}(\bk)
    -
\varphi_{0}^{\dag\,-}(\bk)
\Bigl( \xlrarrow{ k_\mu \frac{\pd}{\pd k^\nu} }
     - \xlrarrow{ k_\nu \frac{\pd}{\pd k^\mu} } \Bigr)
\circ \varphi_{0}^+(\bk)
\\  -
\varphi_{0}^{-}(\bk)
\Bigl( \xlrarrow{ k_\mu \frac{\pd}{\pd k^\nu} }
     - \xlrarrow{ k_\nu \frac{\pd}{\pd k^\mu} } \Bigr)
\circ \varphi_{0}^{\dag\,+}(\bk)
\Bigr\}
\Big|_{	k_0=\sqrt{m^2c^2+\bk^2} } \ \ .
	\end{multline}
	\end{subequations}
These three angular momentum operators are different but, after normal
ordering, they will be mapped into one and the same operator (see
Sect.~\ref{Sect18}).


\section{The commutation relations}
	\label{Sect16}

	The trilinear systems of equations~\eref{11.44}--\eref{11.46} are
similar to the (system of) Klein\ndash Gordon equation(s)~\eref{3.41} and,
correspondingly, will be treated in an analogous way.

	Since the equations~\eref{11.44b},~\eref{11.45b}, and~\eref{11.46b}
can be obtained from~\eref{11.44a}, \eref{11.45a}, and~\eref{11.46a} by
replacing $\varphi_0^{\pm}(\bs k)$ with $\varphi_0^{\dag\,\pm}(\bs k)$ and
${\lindex[\mspace{-3mu}f]{}{a}}^{\pm} (\bs k,\bs q)$ with
${\lindex[\mspace{-3mu}f]{}{a}}^{\dag\,\pm} (\bs k,\bs q)$, respectively, all
of the next `intermediate' considerations will be done only for the former set
of equations; only the more essential and final results will be doubled, \ie
written for
 $\varphi_0^{\pm}(\bs k)$ and $\varphi_0^{\dag\,\pm}(\bs k)$.

	First of all, applying the identity~\eref{3.43-3} several times, we
rewrite~\eref{11.44a}, \eref{11.45a}, and~\eref{11.46a} respectively as
(recall, $\varepsilon=\pm1$)
	\begin{gather}
				\label{11.48}
	\begin{split}
[\varphi_0^{\pm}(\bs k), \varphi_0^{\dag\,+}(\bs q)]_{\varepsilon}
					\circ \varphi_0^{-}(\bs q)
& - \varepsilon \varphi_0^{\dag\,+}(\bs q) \circ
	[\varphi_0^{\pm}(\bs k), \varphi_0^{-}(\bs q) ]_{\varepsilon}
\\
+ [\varphi_0^{\pm}(\bs k), \varphi_0^{\dag\,-}(\bs q)]_{\varepsilon}
					\circ \varphi_0^{+}(\bs q)
& - \varepsilon \varphi_0^{\dag\,-}(\bs q) \circ
	[\varphi_0^{\pm}(\bs k), \varphi_0^{+}(\bs q) ]_{\varepsilon}
\\
\pm (1 & + \tau(\varphi_0)) \varphi_0^{\pm}(\bs k) \delta^3(\bs k-\bs q)
= {\lindex[\mspace{-3mu}f]{}{\prime}}^{\pm}(\bs k,\bs q)
	\end{split}
\displaybreak[1]\\[0.5ex]		\label{11.49}
	\begin{split}
[\varphi_0^{\pm}(\bs k), \varphi_0^{      +}(\bs q)]_{\varepsilon}
					\circ \varphi_0^{\dag\,-}(\bs q)
& - \varepsilon \varphi_0^{      +}(\bs q) \circ
	[\varphi_0^{\pm}(\bs k), \varphi_0^{\dag\,-}(\bs q) ]_{\varepsilon}
\\
+ [\varphi_0^{\pm}(\bs k), \varphi_0^{      -}(\bs q)]_{\varepsilon}
					\circ \varphi_0^{\dag\,+}(\bs q)
& - \varepsilon \varphi_0^{      -}(\bs q) \circ
	[\varphi_0^{\pm}(\bs k), \varphi_0^{\dag\,+}(\bs q) ]_{\varepsilon}
\\
\pm (1 & +\tau(\varphi_0)) \varphi_0^{\pm}(\bs k) \delta^3(\bs k-\bs q)
= {\lindex[\mspace{-3mu}f]{}{\prime\prime}}^{\pm}(\bs k,\bs q)
	\end{split}
\displaybreak[1]\\[0.5ex]		\label{11.50}
	\begin{split}
[\varphi_0^{\pm}(\bs k), \varphi_0^{\dag\,+}(\bs q)]_{\varepsilon}
					\circ \varphi_0^{-}(\bs q)
& - \varepsilon \varphi_0^{\dag\,+}(\bs q) \circ
	[\varphi_0^{\pm}(\bs k), \varphi_0^{-}(\bs q) ]_{\varepsilon}
\\
+ [\varphi_0^{\pm}(\bs k), \varphi_0^{      -}(\bs q)]_{\varepsilon}
					\circ \varphi_0^{\dag\,+}(\bs q)
& - \varepsilon \varphi_0^{      -}(\bs q) \circ
	[\varphi_0^{\pm}(\bs k), \varphi_0^{\dag\,+}(\bs q) ]_{\varepsilon}
\\
+ [\varphi_0^{\pm}(\bs k), \varphi_0^{      +}(\bs q)]_{\varepsilon}
					\circ \varphi_0^{\dag\,-}(\bs q)
& - \varepsilon \varphi_0^{      +}(\bs q) \circ
	[\varphi_0^{\pm}(\bs k), \varphi_0^{\dag\,-}(\bs q) ]_{\varepsilon}
\\
+ [\varphi_0^{\pm}(\bs k), \varphi_0^{\dag\,-}(\bs q)]_{\varepsilon}
					\circ \varphi_0^{+}(\bs q)
&- \varepsilon \varphi_0^{\dag\,-}(\bs q) \circ
	[\varphi_0^{\pm}(\bs k), \varphi_0^{+}(\bs q) ]_{\varepsilon}
\\
\pm 2 (1 & +\tau(\varphi_0)) \varphi_0^{\pm}(\bs k) \delta^3(\bs k-\bs q)
= {\lindex[\mspace{-3mu}f]{}{\prime\prime\prime}}^{\pm} (\bs k,\bs q)  .
	\end{split}
	\end{gather}

	Now, following the know
argumentation~\cite{Bogolyubov&Shirkov,Roman-QFT,Bjorken&Drell}, we shall
impose an \emph{additional condition} saying that the commutators,
$\varepsilon=-1$, or anticommutators, $\varepsilon=+1$, of all combinations
of creation and/or annihilation operators should be proportional to the
identity operator $\id_\Hil$ of the Hilbert space $\Hil$ of the considered
free arbitrary scalar field.

	It is easily seen, the equations~\eref{11.48} and~\eref{11.49} (and
similar ones obtained from them with
 $\varphi_0^{\dag\,\pm}(\bs k)$ for
 $\varphi_0^{\pm}(\bs k)$)
do not make any difference between the choices $\varepsilon=-1$ and
$\varepsilon=+1$. But, for equation~\eref{11.50}, the situation is
completely different. Indeed, for $\varepsilon=+1$, which corresponds to
quantization of a scalar field by anticommutators, equation~\eref{11.50}
reduces to
\(
\pm2(1+\tau(\varphi_0))\varphi_0^\pm\delta^3(\bs k -\bs q)
= {\lindex[\mspace{-3mu}f]{}{\prime\prime\prime}}^{\pm} (\bs k,\bs q)
\)
which, when inserted in~\eref{11.47}, entails
 $k_\mu|_{k_0=\sqrt{m^2c^2+{\bs k}^2}} \varphi^\pm(\bs k) = 0$
for any $\bs k$; a similar result follows from~\eref{11.46b}, \ie we have
	\begin{gather}
			\label{11.51}
k_\mu|_{k_0=\sqrt{m^2c^2+{\bs k}^2}} \varphi_0^\pm(\bs k) = 0
\quad
k_\mu|_{k_0=\sqrt{m^2c^2+{\bs k}^2}} \varphi_0^{\dag\,\pm}(\bs k) = 0
\qquad
\text{for }\varepsilon =+1
\intertext{which, by~\eref{11.36c}, implies}
			\label{11.52}
{ \lindex[\mspace{-6mu}\ope{P}]{}{\prime\prime\prime} }_\mu^{(2)} = 0.
\intertext{Consequently, since~\eref{11.52} and the Klein-Gordon
equations~\eref{11.10} imply}
			\label{11.53}
m^2c^2\varphi_0 = 0 \qquad m^2c^2\varphi_0^\dag = 0,
\intertext{the choice $\varepsilon=+1$ for~\eref{11.50} is possible only for
the degenerate (unphysical) solutions~\eref{11.26} (or~\eref{11.26'} in
Heisenberg picture) and for the solution}
			\label{11.54}
\varphi_0=0 \quad \varphi_0^\dag=0 \qquad\text{for $m\not=0$},
\intertext{the last of which is equivalent to}
			\label{11.55}
\varphi_0^\pm(\bs k)=0 \quad
\varphi_0^{\dag\,\pm}(\bs k)=0 \qquad\text{for $m\not=0$}.
	\end{gather}
According to equations~\eref{11.20} and~\eref{11.52}, the degenerate
solutions~\eref{11.26} and~\eref{11.54} carry no 4\ndash momentum and charge
and, hence, cannot be detected. So, these solutions should be interpreted as
an absence of the scalar field and, if one starts from the
Lagrangian~\eref{11.2-3}, they are the only ones that can be quantized by
anticommutators.%
\footnote{~%
In fact, the last assertion completes the proof of spin-statistics theorem
for free arbitrary scalar field. Notice, in this proof we have not used any
additional hypotheses, like charge conjugation/symmetry or positivity of the
Hilbert space metric (cf.~\cite[sec.~10.2]{Bogolyubov&Shirkov}).%
}

	Let us return to the consideration of
equations~\eref{11.48}--\eref{11.50} and similar ones with
$\varphi_0^{\dag\,\pm}(\bs k)$ for $\varphi_0^{\pm}(\bs k)$ having in mind
that $\varepsilon=-1$ for~\eref{11.50}. Writing explicitly them for the upper,
``$+$'', and lower, ``$-$'', signs, we see that they can equivalently be
represented respectively in the forms:
	\begin{gather}
				\label{11.56}
	\begin{split}
    [\varphi_0^{\pm}(\bs k), \varphi_0^{\dag\,\pm}(\bs q) ]_{\varepsilon}
    \circ \varphi_0^{\mp}(\bs q)
&- \varepsilon \varphi_0^{\dag\,\mp}(\bs q) \circ
    [\varphi_0^{\pm}(\bs k), \varphi_0^{\pm}(\bs q)]_{\varepsilon}
\\
+  [\varphi_0^{\pm}(\bs k), \varphi_0^{\dag\,\mp}(\bs q) ]_{\varepsilon}
  \circ \varphi_0^{\pm}(\bs q)
&- \varepsilon \varphi_0^{\dag\,\pm}(\bs q) \circ
   [\varphi_0^{\pm}(\bs k), \varphi_0^{\mp}(\bs q)]_{\varepsilon}
\\
\pm (1 & + \tau(\varphi_0)) \varphi_0^{\pm}(\bs k) \delta^3(\bs k-\bs q)
= {\lindex[\mspace{-3mu}f]{}{\prime}}^{\pm}(\bs k,\bs q)
	\end{split}
\\[0.5ex]				\label{11.57}
	\begin{split}
[\varphi_0^{\pm}(\bs k), \varphi_0^{\pm}(\bs q)]_{\varepsilon}
\circ \varphi_0^{\dag\,\mp}(\bs q)
& - \varepsilon \varphi_0^{\mp}(\bs q) \circ
	[\varphi_0^{\pm}(\bs k), \varphi_0^{\dag\,\pm}(\bs q) ]_{\varepsilon}
\\
+ [\varphi_0^{\pm}(\bs k), \varphi_0^{\mp}(\bs q)]_{\varepsilon}
  \circ \varphi_0^{\dag\,\pm}(\bs q)
& -\varepsilon \varphi_0^{\pm}(\bs q) \circ
	[\varphi_0^{\pm}(\bs k), \varphi_0^{\dag\,\mp}(\bs q) ]_{\varepsilon}
\\
\pm (1 & + \tau(\varphi_0)) \varphi_0^{\pm}(\bs k) \delta^3(\bs k-\bs q)
= {\lindex[\mspace{-3mu}f]{}{\prime\prime}}^{\pm}(\bs k,\bs q)
	\end{split}
\\				\label{11.58}
	\begin{split}
 \varphi_0^{\dag\,\mp}(\bs q) \circ
[\varphi_0^{\pm}(\bs k), \varphi_0^{\pm}(\bs q)]_{\varepsilon}
& + \varphi_0^{\mp}(\bs q) \circ
	[\varphi_0^{\pm}(\bs k), \varphi_0^{\dag\,\pm}(\bs q) ]_{\varepsilon}
\\
 \varphi_0^{\dag\,\pm}(\bs q) \circ
  [\varphi_0^{\pm}(\bs k), \varphi_0^{\mp}(\bs q)]_{\varepsilon}
& + \varphi_0^{\pm}(\bs q) \circ
	[\varphi_0^{\pm}(\bs k), \varphi_0^{\dag\,\mp}(\bs q) ]_{\varepsilon}
\\
\pm (1 & + \tau(\varphi_0)) \varphi_0^{\pm}(\bs k) \delta^3(\bs k-\bs q)
= \frac{1}{2}
	{\lindex[\mspace{-3mu}f]{}{\prime\prime\prime}}^{\pm}(\bs k,\bs q)  .
	\end{split}
	\end{gather}

	Let us write explicitly the above-stated \emph{additional
condition} concerning the (anti)commu\-ta\-tors of creation and annihilation
operators. We have (\cf equations~\eref{3.43}):
	\begin{equation}	\label{11.59}
	\begin{split}
[\varphi_0^{\pm}(\bs k), \varphi_0^{\pm}(\bs q) ]_{\varepsilon}
	= a_\varepsilon^{\pm}(\bs k,\bs q) \id_\Hil
\qquad
[\varphi_0^{\dag\,\pm}(\bs k), \varphi_0^{\dag\,\pm}(\bs q) ]_{\varepsilon}
	= a_\varepsilon^{\dag\,\pm}(\bs k,\bs q) \id_\Hil
\\
[\varphi_0^{\mp}(\bs k), \varphi_0^{\pm}(\bs q) ]_{\varepsilon}
	= b_\varepsilon^{\pm}(\bs k,\bs q) \id_\Hil
\qquad
[\varphi_0^{\dag\,\mp}(\bs k), \varphi_0^{\dag\,\pm}(\bs q) ]_{\varepsilon}
	= b_\varepsilon^{\dag\,\pm}(\bs k,\bs q) \id_\Hil
\\
[\varphi_0^{\pm}(\bs k), \varphi_0^{\dag\,\pm}(\bs q) ]_{\varepsilon}
	= d_\varepsilon^{\pm}(\bs k,\bs q) \id_\Hil
\qquad
[\varphi_0^{\dag\,\pm}(\bs k), \varphi_0^{\pm}(\bs q) ]_{\varepsilon}
	= \varepsilon  d_\varepsilon^{\pm}(\bs q,\bs k) \id_\Hil
\\
[\varphi_0^{\mp}(\bs k), \varphi_0^{\dag\,\pm}(\bs q) ]_{\varepsilon}
	= e_\varepsilon^{\pm}(\bs k,\bs q) \id_\Hil
\qquad
[\varphi_0^{\dag\,\mp}(\bs k), \varphi_0^{\pm}(\bs q) ]_{\varepsilon}
	= \varepsilon  e_\varepsilon^{\mp}(\bs q,\bs k) \id_\Hil
	\end{split}
	\end{equation}
where $\varepsilon=\pm1$ for~\eref{11.48} and~\eref{11.49}, $\varepsilon=-1$
for~\eref{11.50}, and $a_\varepsilon^{\pm}$, $a_\varepsilon^{\dag\,\pm}$,
\dots, $e_\varepsilon^{\pm}$ are some complex\ndash valued (generalized)
functions, which we have to determine. These last functions are subjected to
a number of restrictions which can be derived in the same way as ~\eref{3.53}
and~\eref{3.53-1} in the Hermitian case.%
\footnote{~%
Notice,~\eref{3.50}--\eref{3.52} remain valid if we replace in
them $\varphi_0^{\pm}(\bs k)$ and/or $\varphi_0^{\pm}(\bs q)$
with $\varphi_0^{\dag\,\pm}(\bs k)$ and/or $\varphi_0^{\dag\,\pm}(\bs q)$
respectively; see~\eref{11.32}.%
}
One can easily verify that these restrictions are:
	\begin{subequations}	\label{11.60}
	\begin{gather}	\label{11.60a}
(\bs k+\bs q) a_\varepsilon^{\pm} (\bs k,\bs q) = 0 \quad
(\bs k+\bs q) a_\varepsilon^{\dag\,\pm} (\bs k,\bs q) = 0 \quad
(\bs k+\bs q) d_\varepsilon^{\pm} (\bs k,\bs q) = 0
\\
			\label{11.60b}
(\bs k-\bs q) b_\varepsilon^{\pm} (\bs k,\bs q) = 0 \quad
(\bs k-\bs q) b_\varepsilon^{\dag\,\pm} (\bs k,\bs q) = 0 \quad
(\bs k-\bs q) e_\varepsilon^{\pm} (\bs k,\bs q) = 0
	\end{gather}
	\end{subequations}
\vspace{-5ex}
	\begin{subequations}	\label{11.60-1}
	\begin{gather}	\label{11.60-1a}
\bigl( \sqrt{m^2c^2+{\bs k}^2} + \sqrt{m^2c^2+{\bs q}^2} \bigr)
	\alpha(\bs k,\bs q) = 0
\qquad\text{for }
\alpha=a_\varepsilon^{\pm}, a_\varepsilon^{\dag\,\pm}, d_\varepsilon^{\pm}
\\
			\label{11.60-1b}
\bigl( \sqrt{m^2c^2+{\bs k}^2} - \sqrt{m^2c^2+{\bs q}^2} \bigr)
	\beta(\bs k,\bs q)  = 0
\qquad\text{for }
\beta=b_\varepsilon^{\pm}, b_\varepsilon^{\dag\,\pm}, e_\varepsilon^{\pm} .
	\end{gather}
	\end{subequations}

	Regarding $a_\varepsilon^{\pm}$, $a_\varepsilon^{\dag\,\pm}$,
\dots,$e_\varepsilon^{\pm}$ as distributions, from~\eref{11.60}, we derive
(cf.~\eref{3.54}):
	\begin{subequations}	\label{11.61}
	\begin{gather}	\label{11.61a}
f(\bs q) \alpha  (\bs k,\bs q) =  f(-\bs k) \alpha  (\bs k,\bs q)
\qquad\text{for }
\alpha=a_\varepsilon^{\pm}, a_\varepsilon^{\dag\,\pm}, d_\varepsilon^{\pm}
\\
			\label{11.61b}
f(\bs q) \beta (\bs k,\bs q) =  f(+\bs k) \beta (\bs k,\bs q)
\qquad\text{for }
\beta=b_\varepsilon^{\pm}, b_\varepsilon^{\dag\,\pm}, e_\varepsilon^{\pm}
	\end{gather}
	\end{subequations}
for any function $f$ which is polynomial or convergent power series. In view
of~\eref{11.61}, the equalities~\eref{11.60-1b} are identically satisfied,
while~\eref{11.60-1a} are equivalent to the equations
	\begin{equation}	\label{11.61-1}
\sqrt{m^2c^2+{\bs k}^2} \alpha(\bs k,\bs q) = 0
\qquad\text{for }
\alpha= a_\varepsilon^{\pm}, a_\varepsilon^{\dag\,\pm}, d_\varepsilon^{\pm} .
	\end{equation}

	Substituting the equalities~\eref{11.59} into
equations~\eref{11.56}--\eref{11.58} and similar ones with
$\varphi_0^{\dag\,\pm}(\bs k)$ for $\varphi_0^{\pm}(\bs k)$, we see that the
restrictions~\eref{11.47}, in view of~\eref{11.61}, give the next systems of
equations for the unknown (generalized) functions $a_\varepsilon^{\pm}$,
$a_\varepsilon^{\dag\,\pm}$, \dots, $e_\varepsilon^{\pm}$:
	\begin{subequations}	\label{11.62}
	\begin{align}
	\begin{split}
				\label{11.62a}
k_a \int\Id^3\bs q \bigl\{
  \varepsilon \varphi_0^{\dag\,\mp}(\bs q) a_\varepsilon^{\pm}(\bs k,\bs q)
- \varphi_0^{\mp}(\bs q) d_\varepsilon^{\pm}(\bs k,\bs q)
- \varepsilon \varphi_0^{\dag\,\pm}(\bs q) b_\varepsilon^{\mp}(\bs k,\bs q)
\\
+ \varphi_0^{\pm}(\bs q)
	\bigl( e_\varepsilon^{\mp} (\bs k,\bs q) \pm\sigma
	(1+\tau(\varphi_0)) \delta^3(\bs k-\bs q) \bigr)
\bigr\}
= 0
	\end{split}
\\[0.75ex]				\label{11.62b}
	\begin{split}
& \sqrt{m^2c^2+{\bs k}^2} \int\Id^3\bs q \bigl\{
- \varepsilon \varphi_0^{\dag\,\mp}(\bs q) a_\varepsilon^{\pm}(\bs k,\bs q)
+ \varphi_0^{\mp}(\bs q) d_\varepsilon^{\pm}(\bs k,\bs q)
\\
& - \varepsilon \varphi_0^{\pm}(\bs q) b_\varepsilon^{\mp}(\bs k,\bs q)
+ \varphi_0^{\dag\,\pm}(\bs q)
	\bigl( e_\varepsilon^{\mp} (\bs k,\bs q) \pm\sigma
	(1+\tau(\varphi_0)) \delta^3(\bs k-\bs q) \bigr)
\bigr\}
= 0
	\end{split}
\displaybreak[1]\\		\label{11.62c}
	\begin{split}
k_a \int\Id^3\bs q \bigl\{
  \varphi_0^{\dag\,\mp}(\bs q) d_\varepsilon^{\pm}(\bs q,\bs k)
- \varphi_0^{\mp}(\bs q) a_\varepsilon^{\dag\,\pm}(\bs k,\bs q)
+ \varphi_0^{\pm}(\bs q) b_\varepsilon^{\dag\,\mp}(\bs k,\bs q)
\\
+ \varphi_0^{\dag\,\pm}(\bs q)
	\bigl( - e_\varepsilon^{\pm} (\bs q,\bs k) \pm\sigma
	(1+\tau(\varphi_0)) \delta^3(\bs k-\bs q) \bigr)
\bigr\}
= 0
	\end{split}
\displaybreak[1]\\[0.75ex]	\label{11.62d}
	\begin{split}
& \sqrt{m^2c^2+{\bs k}^2}  \int\Id^3\bs q \bigl\{
- \varphi_0^{\dag\,\mp}(\bs q) d_\varepsilon^{\pm}(\bs q,\bs k)
+ \varphi_0^{\mp}(\bs q) a_\varepsilon^{\dag\,\pm}(\bs k,\bs q)
\\
& + \varphi_0^{\pm}(\bs q) b_\varepsilon^{\dag\,\mp}(\bs k,\bs q)
 + \varphi_0^{\dag\,\pm}(\bs q)
	\bigl( - e_\varepsilon^{\pm} (\bs q,\bs k) \pm\sigma
	(1+\tau(\varphi_0)) \delta^3(\bs k-\bs q) \bigr)
\bigr\}
= 0 .
	\end{split}
	\end{align}
	\end{subequations}
Here: $a=1,2,3$, $\sigma=-\varepsilon$ for~\eref{11.57} and $\sigma=1$
for~\eref{11.56} and~\eref{11.58}, $\varepsilon=\pm1$ for~\eref{11.56}
and~\eref{11.57}, and $\varepsilon=-1$ for~\eref{11.58} (\emph{vide supra}).
Notice,~\eref{11.62a} and~\eref{11.62b} correspond
to~\eref{11.56}--\eref{11.58}, while~\eref{11.62c} and~\eref{11.62d}
correspond to the same equations with $\varphi_0^{\dag\,\pm}(\bs k)$ for
$\varphi_0^{\pm}(\bs k)$.

	If we impose a \emph{second}, after~\eref{11.59}, \emph{additional
condition}, namely that equation~\eref{11.62} to be valid for arbitrary
$\varphi_0^{\pm}(\bs q)$ and $\varphi_0^{\dag\,\pm}(\bs q)$, we see that, if
$(m,\bs k)\not=(0,\bs0)$, the only solution of~\eref{11.62} relative to
$a_\varepsilon^{\pm}$, $a_\varepsilon^{\dag\,\pm}$, \dots,
$e_\varepsilon^{\pm}$ is:
	\begin{subequations}\label{11.63}
	\begin{align}
				\label{11.63a}
  a_{\varepsilon(\varphi_0)}^{\pm} (\bs k,\bs q)
= a_{\varepsilon(\varphi_0)}^{\dag\,\pm} (\bs k,\bs q)
= d_{\varepsilon(\varphi_0)}^{\pm} (\bs k,\bs q)
= 0
\\				\label{11.63b}
  b_{\varepsilon(\varphi_0)}^{\pm} (\bs k,\bs q)
= b_{\varepsilon(\varphi_0)}^{\dag\,\pm} (\bs k,\bs q)
= \pm \tau(\varphi_0) \delta^3(\bs k-\bs q)
\\				\label{11.63c}
e_{\varepsilon(\varphi_0)}^{\pm} (\bs k,\bs q)
= \pm \sigma(\varphi_0) \delta^3(\bs k-\bs q) ,
	\end{align}
	\end{subequations}
where $\tau(\varphi_0)$ is defined via~\eref{11.37} and:
	\begin{equation}	\label{11.64}
	\begin{split}
\varepsilon(\varphi_0)
&=
	\begin{cases}
-1	& \text{for equation~\eref{11.58} with any $\varphi_0$}\\
	& \text{and for equations~\eref{11.56} and~\eref{11.57}
		with $\varphi_0^\dag=\varphi_0$}\\
\pm 1	&\text{for equations~\eref{11.56} and~\eref{11.57}
		with $\varphi_0^\dag\not=\varphi_0$}
	\end{cases}
\\
\sigma(\varphi_0)
&=
	\begin{cases}
-\varepsilon(\varphi_0)	& \text{for equation~\eref{11.57}  }\\
+ 1	&\text{for equations~\eref{11.56} and~\eref{11.58} }
	\end{cases}
\quad .
	\end{split}
	\end{equation}
Evidently,~\eref{11.63a} converts~\eref{11.61-1} into identity and,
consequently, under the hypotheses made, \eref{11.63} is the general
solution of our problem.

	It should be emphasized on the fact that the function
$\tau(\varphi_0)$ in~\eref{11.63} takes care of what is the field
$\varphi_0$, Hermitian or non\ndash Hermitian, while the functions
$\varepsilon(\varphi_0)$ and $\sigma(\varphi_0)$ take care of from what
Lagrangian,~\eref{11.2-1}--\eref{11.2-3}, we have started off.

	Before commenting on the solutions~\eref{11.63}, we want to say some
words on the case $m=0$ and $\bs k= \bs0$ for which the
equations~\eref{11.62} and~\eref{11.61-1} take the form of the identity $0=0$
and, consequently, no information can be extracted from them. The above
analysis reveals that, under the additional conditions~\eref{11.59}, the
field equations do not impose some restrictions on the operators

	\begin{equation}	\label{11.64new}
\varphi_0^{\pm}(\bs k) \text{ and } \varphi_0^{\dag\,\pm}(\bs k)
\qquad\text{for $m=0$ and $\bs k= \bs0$,}
	\end{equation}
\ie these operators must satisfy ~\eref{11.59} with  $m=0$ and $\bs k= \bs0$
and arbitrary $a_\varepsilon^{\pm}(\bs 0,\bs q)$,
$a_\varepsilon^{\dag\,\pm}(\bs 0,\bs q)$, \dots,
$e_\varepsilon^{\pm}(\bs 0,\bs q)$ (with $\varepsilon=\pm1$ for~\eref{11.48}
and~\eref{11.49} and $\varepsilon=-1$ for~\eref{11.50}). To ensure a
continuous limit $(m,\bs k)\to(0,\bs 0)$, we shall assume \emph{by
convention} that $a_\varepsilon^{\pm}(\bs 0,\bs q)$,
$a_\varepsilon^{\dag\,\pm}(\bs 0,\bs q)$, \dots,
$e_\varepsilon^{\pm}(\bs 0,\bs q)$ are given via~\eref{11.63} with
$\bs k=\bs0$ (and $m=0$). From physical point of view (see
Sect.~\ref{Sect14}), the operators~\eref{11.64new} describe
creation/annihilation of massless particles with vanishing 4\ndash momentum
and charge $\pm q$, which is zero for a Hermitian (neutral) filed and
non\ndash zero for a non\ndash Hermitian (charged) one. Consequently,
\emph{in the non\ndash Hermitian case, the theory admits existence of free,
charged, massless scalar particles with vanishing 4\ndash momentum},
which are quanta of free, charged, massless scalar field. As far as the
author of these lines knows, such particles/fields have not been observed
until now. In the Hermitian case, as we pointed in Sect.~\ref{Sect8}, the
operators~\eref{11.64new} reduce to~\eref{3.56-1} and describe unphysical
particles/fields which are experimentally unobservable.

	Since the (anti)commutation relations~\eref{11.59} are extremely
important for quantum field theory, we shall write them explicitly for the
obtained solutions~\eref{11.63}. As the three versions of some equations,
like~\eref{11.62}, \eref{11.48}--\eref{11.50}, \eref{11.44}--\eref{11.46},
\eref{11.36}, etc., originate from the
Lagrangians~\eref{11.2-1}--\eref{11.2-3} we have started off, we shall
associate the found (anti)commutation relations with the initial Lagrangians
rather than with the particular equations utilized in their derivation.

	Since equations~\eref{11.58}, for which $\varepsilon(\varphi_0)=-1$
and $\sigma(\varphi_0)=+1$, originate from the Lagrangian~\eref{11.2-3}, we
can assert that the \emph{Lagrangian~\eref{11.2-3} implies the following
commutation relations}:
	\begin{align}	\notag
&[\varphi_0^{\pm}(\bs k), \varphi_0^{\pm}(\bs q) ]_{\_}
	= 0
&&
[\varphi_0^{\dag\,\pm}(\bs k), \varphi_0^{\dag\,\pm}(\bs q) ]_{\_}
	= 0
\\	\notag
&[\varphi_0^{\mp}(\bs k), \varphi_0^{\pm}(\bs q) ]_{\_}
	= \pm \tau(\varphi_0) \delta^3(\bs k-\bs q) \id_\Hil
&&
[\varphi_0^{\dag\,\mp}(\bs k), \varphi_0^{\dag\,\pm}(\bs q) ]_{\_}
	= \pm \tau(\varphi_0) \delta^3(\bs k-\bs q) \id_\Hil
\\	\notag
&[\varphi_0^{\pm}(\bs k), \varphi_0^{\dag\,\pm}(\bs q) ]_{\_}
	= 0
&&
[\varphi_0^{\dag\,\pm}(\bs k), \varphi_0^{\pm}(\bs q) ]_{\_}
	= 0
\\	\label{11.65}
&[\varphi_0^{\mp}(\bs k), \varphi_0^{\dag\,\pm}(\bs q) ]_{\_}
	= \pm \delta^3(\bs k-\bs q) \id_\Hil
&&
[\varphi_0^{\dag\,\mp}(\bs k), \varphi_0^{\pm}(\bs q) ]_{\_}
	= \pm \delta^3(\bs k-\bs q) \id_\Hil
	\end{align}
where $0$ denotes the zero operator on $\Hil$ and $\tau(\varphi_0)$ takes
care of is the field neutral ($\varphi_0^\dag=\varphi_0$,
$\tau(\varphi_0)=1$) or charged ($\varphi_0^\dag\not=\varphi_0$,
$\tau(\varphi_0)=0$) and ensures a correct commutation relations in the
Hermitian case (see~\eref{3.57}).

	Since the equations~\eref{11.56} are consequences of the
Lagrangian~\eref{11.2-1}, we can assert that the next \emph{(anti)commutation
relations follow from the Lagrangian~\eref{11.2-1}}:
	\begin{align}	\notag
&[\varphi_0^{\pm}(\bs k), \varphi_0^{\pm}(\bs q) ]_{\varepsilon}
	= 0
&&
[\varphi_0^{\dag\,\pm}(\bs k), \varphi_0^{\dag\,\pm}(\bs q) ]_{\varepsilon}
	= 0
\\	\notag
&[\varphi_0^{\mp}(\bs k), \varphi_0^{\pm}(\bs q) ]_{\varepsilon}
	= \pm \tau(\varphi_0) \delta^3(\bs k-\bs q) \id_\Hil
&&
[\varphi_0^{\dag\,\mp}(\bs k), \varphi_0^{\dag\,\pm}(\bs q) ]_{\varepsilon}
	= \pm \tau(\varphi_0) \delta^3(\bs k-\bs q) \id_\Hil
\\	\notag
&[\varphi_0^{\pm}(\bs k), \varphi_0^{\dag\,\pm}(\bs q) ]_{\varepsilon}
	= 0
&&
[\varphi_0^{\dag\,\pm}(\bs k), \varphi_0^{\pm}(\bs q) ]_{\varepsilon}
	= 0
\\	\label{11.66}
&[\varphi_0^{\mp}(\bs k), \varphi_0^{\dag\,\pm}(\bs q) ]_{\varepsilon}
	= \pm \delta^3(\bs k-\bs q) \id_\Hil
&&
[\varphi_0^{\dag\,\mp}(\bs k), \varphi_0^{\pm}(\bs q) ]_{\varepsilon}
	= \pm \delta^3(\bs k-\bs q) \id_\Hil
	\end{align}
where
$\varepsilon=-1$ (commutation relations) for a Hermitian filed,
$\varphi_0^\dag=\varphi_0$, and
$\varepsilon=\pm1$ (commutation or anticommutation relations) for a
non\ndash Hermitian filed, $\varphi_0^\dag\not=\varphi_0$.

	At last, since the Lagrangian~\eref{11.2-2} entails~\eref{11.57}, the
\emph{Lagrangian~\eref{11.2-2} implies the following (anti)commutation
relations}:
	\begin{align}	\notag
&[\varphi_0^{\pm}(\bs k), \varphi_0^{\pm}(\bs q) ]_{\varepsilon}
	= 0
&&
[\varphi_0^{\dag\,\pm}(\bs k), \varphi_0^{\dag\,\pm}(\bs q) ]_{\varepsilon}
	= 0
\\	\notag
&[\varphi_0^{\mp}(\bs k), \varphi_0^{\pm}(\bs q) ]_{\varepsilon}
	= \pm \tau(\varphi_0) \delta^3(\bs k-\bs q) \id_\Hil
&&
[\varphi_0^{\dag\,\mp}(\bs k), \varphi_0^{\dag\,\pm}(\bs q) ]_{\varepsilon}
	= \pm \tau(\varphi_0) \delta^3(\bs k-\bs q) \id_\Hil
\\	\notag
&[\varphi_0^{\pm}(\bs k), \varphi_0^{\dag\,\pm}(\bs q) ]_{\varepsilon}
	= 0
&&
[\varphi_0^{\dag\,\pm}(\bs k), \varphi_0^{\pm}(\bs q) ]_{\varepsilon}
	= 0
\\	\label{11.67}
&[\varphi_0^{\mp}(\bs k), \varphi_0^{\dag\,\pm}(\bs q) ]_{\varepsilon}
	= \mp\varepsilon \delta^3(\bs k-\bs q) \id_\Hil
&&
[\varphi_0^{\dag\,\mp}(\bs k), \varphi_0^{\pm}(\bs q) ]_{\varepsilon}
	= \mp\varepsilon \delta^3(\bs k-\bs q) \id_\Hil
	\end{align}
where
$\varepsilon=-1$ (commutation relations) for a Hermitian filed,
$\varphi_0^\dag=\varphi_0$, and
$\varepsilon=\pm1$ (commutation or anticommutation relations) for a
non\ndash Hermitian filed, $\varphi_0^\dag\not=\varphi_0$.

	It should be emphasized, for a Hermitian (neutral, real) field, when
$\varepsilon=-1$ in~\eref{11.66} and~\eref{11.67}, the \emph{commutation}
relations~\eref{11.65}, \eref{11.66}, and~\eref{11.67} coincide and, due to
$\tau(\varphi_0)=1$ in this case, are identical with~\eref{3.57}; thus, they
correctly reproduce the already established results in Sect.~\ref{Sect8}.
However, for a \emph{non\ndash Hermitian} (charged) field, for which
$\tau(\varphi_0)=0$, we have \emph{three} independent sets of
(anti)commutation relations:
	\begin{description}
\item{(i)}
	the commutation relations~\eref{11.65} correspond to the
Lagrangians~\eref{11.2-1} and~\eref{11.2-2}, with the \emph{choice}
$\varepsilon=-1$ for the both ones, and the Lagrangian~\eref{11.2-3};
\item{(ii)}
	the \emph{anticommutation} relations~\eref{11.66} with
$\varepsilon=+1$ correspond to the Lagrangian~\eref{11.2-1} with the
\emph{choice} $\varepsilon=+1$;
\item{(iii)}
	the \emph{anticommutation} relations~\eref{11.67} with
$\varepsilon=+1$ correspond to the Lagrangian~\eref{11.2-2} with the
\emph{choice} $\varepsilon=+1$.
	\end{description}

	The relations~\eref{11.67} with $\varepsilon=+1$ differ
from~\eref{11.66} with $\varepsilon=+1$ only in the sign before the
$\delta$\ndash function in the last row. This is quite understandable as the
Lagrangian~\eref{11.2-2} can be obtained from~\eref{11.2-1} by replacing
$\tope{\varphi}$ with $\tope{\varphi}^\dag$ and
$\tope{\varphi}^\dag$ with $\tope{\varphi}$. If we make the same change
in~\eref{11.66}, \ie $\varphi_0^{\pm}\leftrightarrow\varphi_0^{\dag\,\pm}$,
we see that~\eref{11.66} transforms into~\eref{11.67}. Since~\eref{11.66}
and~\eref{11.67} are identical (resp.\ different) for $\varepsilon=-1$
(resp.\ $\varepsilon=+1$), we conclude that the theory is invariant (resp.\
non\ndash invariant) under the change
$\varphi_0\leftrightarrow\varphi_0^{\dag}$ or, equivalently,
$\varphi_0^{\pm}\leftrightarrow\varphi_0^{\dag\,\pm}$, called \emph{charge
conjugation}~\cite{Bogolyubov&Shirkov,Bjorken&Drell-2,Roman-QFT}, if and only
if it is quantized via commutators (resp.\ anticommutators) if one starts
from any one of the Lagrangians~\eref{11.2-1} and~\eref{11.2-2}. The theory
is always charge symmetric, \ie invariant under charge conjugation, if one
starts from the Lagrangian~\eref{11.2-3}.

	Thus, for a free non-Hermitian scalar field, we see a principal
difference between the Lagrangian~\eref{11.2-3}, on one hand, and the
Lagrangians~\eref{11.2-1} and~\eref{11.2-2}, on the other hand: the first
Lagrangian entails quantization with commutators, while the other two imply
quantization either with commutators (identical with the one of the previous
case) or with anticommutators and one needs a \emph{new additional
condition/hypothesis} to make a distinction between these two cases.  As it is
well known, the correct quantization of a free scalar field is via
commutators, not by
anticommutators~\cite{Bogolyubov&Shirkov,Roman-QFT,Bjorken&Drell-2}. Usually
(see \emph{loc.\ cit.}) this result is derived, for a charged field, from the
Lagrangian~\eref{11.2} by invoking a \emph{new additional condition}, like
charge symmetry, or positivity of the Hilbert space metric, or spin\ndash
statistics theorem.%
\footnote{~%
The particular additional conditions mentioned above are, in fact, equivalent
to \emph{postulating} quantization via commutators in the case of free scalar
field if one starts from someone of the Lagrangians~\eref{11.2}, \eref{11.2-1},
and~\eref{11.2-2}.%
}
The above considerations show that these additional conditions are not
required if one starts from the Lagrangian~\eref{11.2-3}; in fact, these
conditions are corollaries from it, as we saw with the charge symmetry and
spin\ndash statistics theorem (saying that a scalar field, which is a spin
zero field, must be quantized via commutators). This is not a surprising
result, if we recall that the Lagrangian~\eref{11.2-3} is a sum of the
Lagrangians of two (independent) Hermitian scalar fields (see
Sect.~\ref{Sect11}) for which a similar result was established in
Sect.~\ref{Sect8}.  In conclusion, the Lagrangian~\eref{11.2-3} is richer in
consequences than the Lagrangians~\eref{11.2-1} and~\eref{11.2-2};%
\footnote{~%
And also the Lagrangian~\eref{11.2} which is two times~\eref{11.2-1} and is,
usually, used in the literature.%
}
the cause for this is that $\tope{\varphi}$ and $\tope{\varphi}^\dag$ enter
in~\eref{11.2-3} on equal footing, i.e.~\eref{11.2-3} is invariant under the
change $\tope{\varphi}\leftrightarrow\tope{\varphi}^\dag$, which cannot be
said relative to~\eref{11.2-1} and~\eref{11.2-2}.

	Relying on the above discussion, the commutation
relations~\eref{11.65} will be accepted from now on in this paper. As we
proved, under the hypotheses made, they are equivalent to the initial
system~\eref{11.10} of Klein\ndash Gordon equations. If, by some reason one
rejects these hypotheses, the system~\eref{11.10} of Klein\ndash Gordon
equations will be equivalent to the trilinear
relations~\eref{11.44}--\eref{11.46} corresponding to the
Lagrangians~\eref{11.2-1}--\eref{11.2-3}. But, at present, it seems that
correct description of the real physical world is given by~\eref{11.65}, not
by the more general trilinear equations mentioned, \ie there are indications
that the so\ndash called parafields, satisfying trilinear equations similar
to~\eref{11.46}, do not exist in the Nature~\cite{Greenberg&Messiah-1965}.

	Similarly to the said at the end of Sect.~\ref{Sect8}, the
considerations in the present section naturally lead to the operator\ndash
valued distribution character of the field variables and hence of the
creation/annihilation operators. However, such a rigorous treatment is out of
the range of this paper, in which it will be incorporated in the appearance
of Dirac's delta function in some formulae.

	Ending this section, we want to say that, due to the
above considerations, the Lagrangian~\eref{11.2-3} is the `best' one for the
correct description of arbitrary, neutral or charged, free scalar field.


\section{The vacuum and normal ordering}
	\label{Sect18}

	The arguments, leading to a correct definition of a vacuum and the
need of normal ordering of compositions (products) of creation and/or
annihilation operators, are practically the same as in the Hermitian case,
studied in Sect.~\ref{Sect9}. Without repeating them \emph{mutatis mutandis},
we shall point only to the difference when the field $\varphi_0$ is non\ndash
Hermitian (charged). There are two of them: (i)~since in this case we have
two types of annihilation operators, \viz $\varphi_0^{-}(\bs k)$ and
$\varphi_0^{\dag\,-}(\bs k)$, the condition~\eref{8.3} should be doubled,
\ie to it one must add the equality $\varphi_0^{\dag\,-}(\bs
k)(\ope{X}_0)=0$, $\ope{X}_0$ being the vacuum (state vector); (ii)~as now
the field possesses a non\ndash vanishing charge operator, the combinations
of~\eref{11.69} with the commutation relations~\eref{11.65} leads to
infinities, like~\eref{8.4} for the momentum operator of a Hermitian field.%
\footnote{~%
Applying~\eref{11.36} and~\eref{11.65}, the reader can easily obtain the
versions of~\eref{8.4} for a non\ndash Hermitian field. The results will be
senseless infinities, like the ones in~\eref{8.4}, which are removed via
normal ordering (\emph{vide infra}).%
}

	Thus, arguments, similar to the ones in Sect.~\ref{Sect9}, lead to
the following definition of a vacuum for a free arbitrary scalar field.

	\begin{Defn}	\label{Defn18.1}
The vacuum of a free arbitrary scalar field $\varphi_0$ is its physical state
that contains no particles and possesses vanishing 4\ndash momentum and
(total) charge. It is described by a state vector, denoted by $\ope{X}_0$ (in
momentum picture) and called also the vacuum (of the field), such that:
	\begin{subequations}	\label{11.70}
	\begin{align}
				\label{11.70a}
& \ope{X}_0 \not= 0
\\				\label{11.70b}
& \ope{X}_0 = \tope{X}_0
\\				\label{11.70c}
& \varphi_0^-(\bs k) (\ope{X}_0)
= \varphi_0^{\dag\,-}(\bs k) (\ope{X}_0) = 0
\\				\label{11.70d}
& \langle\ope{X}_0|\ope{X}_0\rangle = 1 .
	\end{align}
	\end{subequations}
	\end{Defn}

	As we said above, the formulae~\eref{11.36} and~\eref{11.69},
together with the commutation relations~\eref{11.65}, imply senseless
(infinity) values for the 4\ndash momentum and charge of the vacuum. They are
removed by redefining the dynamical variables, like the Lagrangian, momentum
operator and charge operator, by writing the compositions (products) of the
field, and/or creation and/or annihilation operators in normal order, exactly
in the same way as described in Sect.~\ref{Sect9}. Besides, the definition of
the normal ordering operator (mapping) $\ope{N}$ is also retained the same as
in Sect.~\ref{Sect9}, with the only remark that now it concerns all
creation/annihilation operators, \ie $\varphi_0^{\pm}(\bs k)$ and
$\varphi_0^{\dag\,\pm}(\bs k)$.

	Since the evident equalities
\(
\ope{N}\bigl( \varphi_0^{+}(\bs k) \circ\varphi_0^{\dag\,-}(\bs k) \bigr)
=
\ope{N}\bigl( \varphi_0^{\dag\,-}(\bs k) \circ\varphi_0^{+}(\bs k) \bigr)
=
 \varphi_0^{+}(\bs k) \circ\varphi_0^{\dag\,-}(\bs k)
\) and
\(
\ope{N}\bigl( \varphi_0^{-}(\bs k) \circ\varphi_0^{\dag\,+}(\bs k) \bigr)
=
\ope{N}\bigl( \varphi_0^{\dag\,+}(\bs k) \circ\varphi_0^{-}(\bs k) \bigr)
=
 \varphi_0^{\dag\,+}(\bs k) \circ\varphi_0^{-}(\bs k)
\)
hold, the \emph{three} momentum operators~\eref{11.36} transform, after
normal ordering, into a \emph{single momentum operator}, \viz into the
operator
	\begin{gather}
			\label{11.71}
	\begin{split}
\ope{P}_\mu
=
\frac{1}{1+\tau(\varphi_0)}\int
	k_\mu |_{ k_0=\sqrt{m^2c^2+{\bs k}^2} }
 \{
\varphi_0^{\dag\,+}(\bs k)\circ\varphi_0^{-}(\bs k)
 +
\varphi_0^{+}(\bs k) \circ \varphi_0^{\dag\,-}(\bs k)
\}
\Id^3\bs k ~.
	\end{split}
\\\intertext{Similarly, the \emph{three} charge operators~\eref{11.69}
transform, after normal ordering, into a \emph{single charge operator}, \viz
the operator}
				\label{11.72}
\ope{Q}
 =
q  \int \bigl\{
  \varphi_0^{\dag\,+}(\bs k) \circ \varphi_0^{-}(\bs k)
- \varphi_0^{+}(\bs k) \circ   \varphi_0^{\dag\,-}(\bs k)
\bigr\} \Id^3\bs k  ~.
	\end{gather}

	One can verify the equations
	\begin{equation*}
	\begin{split}
    \ope{N} \bigl( \varphi_0^{-}(\bk)
       \xlrarrow{A} \circ \varphi_0^{\dag\,+}(\bk) \bigr)
= - \ope{N} \bigl( \varphi_0^{\dag\,+}(\bk)
	\xlrarrow{A} \circ \varphi_0^{-}(\bk) \bigr)
= - \varphi_0^{\dag\,+}(\bk) \xlrarrow{A} \circ a_s^{-}(\bk)
\\
    \ope{N} \bigl( \varphi_0^{\dag\,-}(\bk)
	\xlrarrow{A} \circ \varphi_0^{+}(\bk) \bigr)
= - \ope{N} \bigl( \varphi_0^{+}(\bk)
	\xlrarrow{A} \circ \varphi_0^{\dag\,-}(\bk) \bigr)
= - a_s^{+}(\bk) \xlrarrow{A} \circ a_s^{\dag\,-}(\bk) ,
	\end{split}
	\end{equation*}
with $A=k_\mu\frac{\pd}{\pd k^\mu}$, as a result of which the \emph{three}
angular momentum operators~\eref{17.8} transform, after normal ordering, into
a \emph{single} orbital angular momentum operator given by
	\begin{multline}	\label{11.72-1}
 \ope{L}_{\mu\nu}
=
x_{\mu} \ope{P}_\nu - x_{\nu} \ope{P}_\mu
+
\frac{\ih}{2(1+\tau(\varphi_0))} \int\Id^3\bk
\Bigl\{
\varphi_{0}^{\dag\,+}(\bk)
\Bigl( \xlrarrow{ k_\mu \frac{\pd}{\pd k^\nu} }
     - \xlrarrow{ k_\nu \frac{\pd}{\pd k^\mu} } \Bigr)
\circ \varphi_{0}^-(\bk)
\\  +
\varphi_{0}^+(\bk)
\Bigl( \xlrarrow{ k_\mu \frac{\pd}{\pd k^\nu} }
     - \xlrarrow{ k_\nu \frac{\pd}{\pd k^\mu} } \Bigr)
\circ \varphi_{0}^{\dag\,-}(\bk)
\Bigr\}
\Big|_{	k_0=\sqrt{m^2c^2+\bk^2} }
	\end{multline}
where $\ope{P}_\mu$ is given by~\eref{11.71}. This equation, in Heisenberg
picture and expressed via the Heisenberg creation and annihilation
operators~\eref{3.38new}, reads
	\begin{multline}	\label{11.72-2}
 \tope{L}_{\mu\nu}
=
\frac{\ih}{2(1+\tau(\tope{\varphi}_0))} \int\Id^3\bk
\Bigl\{
\tope{\varphi}_{0}^{\dag\,+}(\bk)
\Bigl( \xlrarrow{ k_\mu \frac{\pd}{\pd k^\nu} }
     - \xlrarrow{ k_\nu \frac{\pd}{\pd k^\mu} } \Bigr)
\circ \tope{\varphi}_{0}^-(\bk)
\\  +
\tope{\varphi}_{0}^+(\bk)
\Bigl( \xlrarrow{ k_\mu \frac{\pd}{\pd k^\nu} }
     - \xlrarrow{ k_\nu \frac{\pd}{\pd k^\mu} } \Bigr)
\circ \tope{\varphi}_{0}^{\dag\,-}(\bk)
\Bigr\}
\Big|_{	k_0=\sqrt{m^2c^2+\bk^2} } \ .
	\end{multline}
In a case of neutral (Hermitian) scalar field, when
$\varphi_0^\dag=\varphi_0$ and $\tau(\varphi_0)=1$, the last expression for
the orbital angular momentum operator reproduces the one presented
in~\cite[eq.~(3.54)]{Itzykson&Zuber}, due to the first equality
in~\eref{17.2-1}.

	Applying~\eref{11.72-1},~\eref{17.2-1},~\eref{11.39}
and~\eref{11.65}, one can verify the equations
	\begin{equation}	\label{11.72-2new}
[\varphi_0 , \ope{L}_{\mu\nu} ]_{\_}
=
   x_\mu [\varphi_0 , \ope{P}_{\nu} ]_{\_}
-  x_\nu [\varphi_0 , \ope{P}_{\mu} ]_{\_}
\quad
[\varphi_0^\dag , \ope{L}_{\mu\nu} ]_{\_}
=
   x_\mu [\varphi_0^\dag , \ope{P}_{\nu} ]_{\_}
-  x_\nu [\varphi_0^\dag , \ope{P}_{\mu} ]_{\_} ,
	\end{equation}
which in Heisenberg picture respectively read
	\begin{equation}	\label{11.72-3new}
[\tope{\varphi}(x) , \tope{L}_{\mu\nu} ]_{\_}
= \ih (x_\mu \pd_\nu -  x_\nu \pd_\mu) \tope{\varphi}(x)
\quad
[\tope{\varphi}^\dag(x) , \tope{L}_{\mu\nu} ]_{\_}
= \ih (x_\mu \pd_\nu -  x_\nu \pd_\mu) \tope{\varphi}^\dag(x) .
	\end{equation}
These equations, together with~\eref{2.28}, express the relativistic
covariance of the theory considered~\cite{Bjorken&Drell-2}.

	Similarly, applying~\eref{11.72},~\eref{11.39} and~\eref{11.65}, we
obtain the equations
	\begin{equation}	\label{11.72-4new}
[\varphi_0,\ope{Q}]_{\_} = q \varphi_0
\quad
[\varphi_0^\dag,\ope{Q}]_{\_} = - q \varphi_0^\dag ,
	\end{equation}
which in Heisenberg picture take the form~\eref{11.14}. Analogously, we have
	\begin{equation}	\label{11.72-5}
[\varphi_0^{\pm}(\bk),\ope{Q}]_{\_} = q \varphi_0^{\pm}(\bk)
\quad
[\varphi_0^{\dag\,\pm}(\bk),\ope{Q}]_{\_} = - q \varphi_0^{\dag\,\pm}(\bk) ,
	\end{equation}
which, evidently, entail
	\begin{equation}	\label{11.72-6}
[\ope{P}_\mu,\ope{Q}]_{\_} = 0
\quad
[\ope{Q},\ope{Q}]_{\_} = 0
\quad
[\ope{L}_{\mu\nu},\ope{Q}]_{\_} = 0.
	\end{equation}

	At last, we shall derive the commutation relations between the
components of the orbital angular momentum operator~\eref{11.72-1}, which
coincides with the total angular momentum operator. To simplify the proof and
to safe some space, we shall work in Heisenberg picture and employ the
Heisenberg creation and annihilation operators~\eref{3.38new}, which satisfy
the same commutation relations as their momentum picture counterparts. At
first, we notice that~\eref{11.72-2} and~\eref{11.65} imply the equations
($p_0:=\sqrt{m^2c^2+{\bs p}^2}$)
	\begin{equation}	\label{11.72-7}
	\begin{split}
[ \tope{\varphi}_0^{\pm}(\bs p) , \tope{L}_{\mu\nu} ]_{\_}
& =
\ih \Bigl\{ p_\mu\frac{\pd}{\pd p^\nu} - p_\nu\frac{\pd}{\pd p^\mu} \Bigr\}
\tope{\varphi}_0^{\pm}(\bs p)
\\
[ \tope{\varphi}_0^{\dag\,\pm}(\bs p) , \tope{L}_{\mu\nu} ]_{\_}
& =
\ih \Bigl\{ p_\mu\frac{\pd}{\pd p^\nu} - p_\nu\frac{\pd}{\pd p^\mu} \Bigr\}
\tope{\varphi}_0^{\dag\,\pm}(\bs p) .
	\end{split}
	\end{equation}
If $-(\mu\leftrightarrow\nu)$ denotes antisymmetrization with respect to the
indices $\mu$ and $\nu$, \ie a subtraction of the previous terms combined with
the change $\mu\leftrightarrow\nu$ and
\(
+ ( \tope{\varphi}_0^{\pm}(\bs p) \leftrightarrow
\tope{\varphi}_0^{\dag\,\pm}(\bs p) )
\)
means that one has to add the previous terms by making the change
\(
 \tope{\varphi}_0^{\pm}(\bs p) \leftrightarrow
\tope{\varphi}_0^{\dag\,\pm}(\bs p)  ,
\)
then we get by applying~\eref{11.72-2}, \eref{11.72-7} and~\eref{11.72-1}:
	\begin{multline*}
[ \tope{L}_{\varkappa\lambda} , \tope{L}_{\mu\nu} ]_{\_}
=
\frac{\ih}{2(1+\tau(\tope{\varphi})} 2 \ih \int \Id^3\bk
\Bigl\{
k_\mu \frac{\pd \tope{\varphi}_0^{\dag\,+}(\bk)} {\pd k^\nu}
k_\varkappa \frac{\pd \tope{\varphi}_0^{-}(\bk)} {\pd k^\lambda}
-
k_\varkappa \frac{\pd \tope{\varphi}_0^{\dag\,+}(\bk)} {\pd k^\lambda}
k_\mu \frac{\pd \tope{\varphi}_0^{-}(\bk)} {\pd k^\nu}
\\
+ ( \tope{\varphi}_0^{\pm}(\bk) \leftrightarrow
				\tope{\varphi}_0^{\dag\,\pm}(\bk) )
- (\varkappa\leftrightarrow\lambda) - (\mu\leftrightarrow\nu)
\Bigr\}
\\
=
\ih \frac{\ih}{(1+\tau(\tope{\varphi})} \int \Id^3\bk
\Bigl\{
\eta_{\mu\lambda} \tope{\varphi}_0^{\dag\,+}(\bk)
k_\varkappa \frac{\pd \tope{\varphi}_0^{-}(\bk)} {\pd k^\nu}
-
\eta_{\varkappa\nu} \tope{\varphi}_0^{\dag\,+}(\bk)
k_\mu \frac{\pd \tope{\varphi}_0^{-}(\bk)} {\pd k^\lambda}
\\
+ ( \tope{\varphi}_0^{\pm}(\bk) \leftrightarrow
				\tope{\varphi}_0^{\dag\,\pm}(\bk) )
- (\varkappa\leftrightarrow\lambda) - (\mu\leftrightarrow\nu)
\Bigr\}
\\
=
\ih \bigl\{
\eta_{\varkappa\nu} \tope{L}_{\lambda\mu}
- (\varkappa\leftrightarrow\lambda) - (\mu\leftrightarrow\nu)
\bigr\}
=
- \ih \bigl\{
\eta_{\varkappa\mu} \tope{L}_{\lambda\nu}
- (\varkappa\leftrightarrow\lambda) - (\mu\leftrightarrow\nu)
\bigr\} ,
	\end{multline*}
where $k_0:=\sqrt{m^2c^2+\bk^2}$ in the integrands, the terms containing
derivatives with respect to $k^\nu$ were integrated by parts and the
antisymmetries relative to $\varkappa$ and $\lambda$ and $\mu$ and $\nu$
were taken into account. The explicit form of the result obtained is:
	\begin{equation}	\label{11.72-8}
[ \tope{L}_{\varkappa\lambda} , \tope{L}_{\mu\nu} ]_{\_}
=
- \ih \bigl\{
\eta_{\varkappa\mu}	\tope{L}_{\lambda\nu} -
\eta_{\lambda\mu}	\tope{L}_{\varkappa\nu} -
\eta_{\varkappa\nu}	\tope{L}_{\lambda\mu} +
\eta_{\lambda\nu}	\tope{L}_{\varkappa\mu}
\bigr\} ,
	\end{equation}
which in momentum picture reads
	\begin{equation}	\label{11.72-9}
[ \ope{L}_{\varkappa\lambda} , \ope{L}_{\mu\nu} ]_{\_}
=
- \ih \bigl\{
\eta_{\varkappa\mu}	\ope{L}_{\lambda\nu} -
\eta_{\lambda\mu}	\ope{L}_{\varkappa\nu} -
\eta_{\varkappa\nu}	\ope{L}_{\lambda\mu} +
\eta_{\lambda\nu}	\ope{L}_{\varkappa\mu}
\bigr\} .
	\end{equation}
It should be noted the minus sign in the multiplier $-\ih$ in the r.h.s.\
of~\eref{11.72-8} relative to a similar one in the last equation
in~\cite[eqs.~(3.51)]{Itzykson&Zuber} for a neutral scalar field.

	The equations~\eref{11.72-2new}--\eref{11.72-9} are valid also before
the normal ordering is performed, \ie if the orbital angular momentum,
momentum and charge operators are replaced with any one of the corresponding
operators in~\eref{17.8}, \eref{11.36} and~\eref{11.69}, respectively.

	We would like to emphasize, equation~\eref{11.72-8}
(or~\eref{11.72-9}) is a consequence of~\eref{11.72-2} and~\eref{11.72-7},
which is equivalent to~\eref{11.72-3new}, and this conclusion is independent
of the validity of the commutation relations~\eref{11.65} and/or the normal
ordering (before normal ordering equation~\eref{11.72-8} follows
from~\eref{11.72-7} and~\eref{17.1}). Similar result concerns
equations~\eref{11.72-4new} and~\eref{11.72-6}.

	So, we see that, at the very end of building of the theory of free
scalar fields, all of the three Lagrangians~\eref{11.2-1}--\eref{11.2-3} lead
to one and the same final theory.%
\footnote{~%
Recall (see Sect.~\ref{Sect16}), the Lagrangians~\eref{11.2-1}
and~\eref{11.2-2} require an additional hypothesis for the establishment of
the commutation relations~\eref{11.65} and, in this sense the arising from
them theory is not equivalent to the one build from the
Lagrangian~\eref{11.2-3}.%
}
	This is a remarkable fact which is far from evident at the beginning
and all intermediate stages of the theory.

	Acting with the operators~\eref{11.71} and~\eref{11.72} on the vacuum
$\ope{X}_0$, we get
	\begin{equation}	\label{11.73}
\ope{P}_\mu     (\ope{X}_0) = 0 \quad
\ope{Q}         (\ope{X}_0) = 0 \quad
\ope{L}_{\mu\nu}(\ope{X}_0) = 0
	\end{equation}
which agrees with definition~\ref{Defn18.1} and takes off the problem with the
senseless expressions for the 4\ndash momentum, charge and orbital angular
momentum of the vacuum before redefining the dynamical variables via normal
ordering.

	As a result of the above uniqueness of the momentum operator after
normal ordering, the \emph{three} systems of field
equations~\eref{11.44}--\eref{11.46}, together with the
conditions~\eref{11.47}, transform after normal ordering into the next
\emph{unique} system of equations:%
\footnote{~%
The normal ordering must be applied only to the anticommutators
in~\eref{11.44}--\eref{11.46} as these terms originate from the corresponding
momentum operators~\eref{11.36} before normal ordering.%
}
	\begin{subequations}	\label{11.72-4}
	\begin{gather}
				\label{11.72-4a}
	\begin{split}
[\varphi_0^{\pm}(\bs k),
	\varphi_0^{\dag\,+}(\bs q) & \circ \varphi_0^{-}(\bs q)
     +	\varphi_0^{+}(\bs q)\circ  \varphi_0^{\dag\,-}(\bs q) ]_{\_}
\\ &
\pm (1+\tau(\varphi_0)) \varphi_0^{\pm}(\bs k) \delta^3(\bs q-\bs k)
= f^{\pm}(\bs k,\bs q)
	\end{split}
\\				\label{11.72-4b}
	\begin{split}
[\varphi_0^{\dag\,\pm}(\bs k),
	\varphi_0^{\dag\,+}(\bs q) & \circ \varphi_0^{-}(\bs q)
      +	\varphi_0^{+}(\bs q) \circ \varphi_0^{\dag\,-}(\bs q) ]_{\_}
\\ &
\pm (1+\tau(\varphi_0)) \varphi_0^{\dag\,\pm}(\bs k) \delta^3(\bs q-\bs k)
= f^{\dag\,\pm} (\bs k,\bs q)
	\end{split}
\\		\label{11.72-4c}
\int q_\mu |_{q_0=\sqrt{m^2c^2+{\bs q}^2}}
	f^{\pm} (\bs k,\bs q) \Id^3\bs q
=
\int q_\mu |_{q_0=\sqrt{m^2c^2+{\bs q}^2}}
	f^{\dag\,\pm} (\bs k,\bs q) \Id^3\bs q
= 0 .
	\end{gather}
	\end{subequations}
Applying~\eref{3.43-3} with $\varepsilon=-1$, one can verify
that~\eref{11.72-4} are identically valid due to the commutation
relations~\eref{11.65}. In this sense, we can say that the commutation
relations~\eref{11.65} play a role of field equations with respect to the
creation and annihilation operators (under the hypotheses made in their
derivation).


\section{State vectors}
	\label{Sect19}

	A state vector of a free arbitrary scalar field is, of course, given
via the general formula~\eref{9.1} in which, now, the momentum operator
$\ope{P}_\mu$ is given by~\eref{11.71}. This means that the evolution
operator $\ope{U}(x,x_0)$ is
	\begin{equation}	\label{11.74}
\ope{U}(x,x_0)
=
\exp
\Bigl\{
\iih\frac{x^\mu-x_0^\mu}{1+\tau(\varphi_0)}
\int
k_\mu |_{ k_0=\sqrt{m^2c^2+{\bs k}^2} }
\{
\varphi_0^{\dag\,+}(\bs k)\circ\varphi_0^{-}(\bs k)
 +
\varphi_0^{+}(\bs k) \circ \varphi_0^{\dag\,-}(\bs k)
\}
\Id^3\bs k
\Bigr\} ~ .
	\end{equation}

	A state vector of a state with fixed 4-momentum is, of course,
described by~\eref{9.4}.

	Similarly to the neutral field case, the amplitude, describing a
transition from an initial state $\ope{X}_i(x_i)$ to final state
$\ope{X}_f(x_f)$, is~\eref{9.5} and admits the representation~\eref{9.6}
through the `$S$\ndash matrix' $\ope{U}(x_i,x_f)$.  The  expansion of the
exponent in~\eref{11.74} into a power series results in the following series
for $\ope{U}(x_i,x_f)$ (cf.~\eref{9.7} and~\eref{9.8})
	\begin{equation}	\label{11.75}
\ope{U}(x_i,x_f) = \id_\Hil + \sum_{n=1}^{\infty} \ope{U}^{(n)}(x_i,x_f)
	\end{equation}
\vspace{-2.4ex}
	\begin{multline}	\label{11.76}
\ope{U}^{(n)}(x_i,x_f)
:=
\frac{1}{n!} \frac{1}{\bigl( \ih(1+\tau(\varphi_0)) \bigr)^n}
(x_i^{\mu_1}-x_f^{\mu_1}) \dots (x_i^{\mu_n}-x_f^{\mu_n})
\\
\times \int \mspace{-7mu} \Id^3\bs k^{(1)}\dots\Id^3\bs k^{(n)}
k_{\mu_1}^{(1)}\dotsb k_{\mu_n}^{(n)}
\bigl\{ \varphi_0^{\dag\,+}(\bs k^{(1)})\circ\varphi_0^-(\bs k^{(1)})
+
\varphi_0^+(\bs k^{(1)})\circ\varphi_0^{\dag\,-}(\bs k^{(1)}) \bigr\}
\\
\circ\dotsb\circ
\bigl\{ \varphi_0^{\dag\,+}(\bs k^{(n)})\circ\varphi_0^-(\bs k^{(n)})
+
\varphi_0^+(\bs k^{(n)})\circ\varphi_0^{\dag\,-}(\bs k^{(n)}) \bigr\}
	\end{multline}
where $k_0^{(a)}=\sqrt{m^2c^2+ ({\bs k}^{(a)})^2}$, $a=1,\dots,n$.

	According to~\eref{9.4} and the considerations in
Sect.~\ref{Sect14}, a state vector of a state containing $n'$
particles and $n^{\prime\prime}$ antiparticles,
$n^{\prime},n^{\prime\prime}\ge0$, such that
the $i^{\prime\,\text{th}}$ particle has 4\ndash momentum $p'_{i'}$ and
the $i^{\prime\prime\,\text{th}}$ antiparticle has 4\ndash momentum
$p^{\prime\prime}_{i^{\prime\prime}}$, where $i'=0,1,\dots,n'$ and
$i^{\prime\prime}=0,1,\dots,n^{\prime\prime}$, is given by the equality
	\begin{multline}	\label{11.77}
\ope{X}(
x;p'_1;\ldots;p'_{n'};
  p^{\prime\prime}_1;\ldots;
  p^{\prime\prime}_{n^{\prime\prime}}
)
\\
=
\frac{1}{\sqrt{n^{\prime}! n^{\prime\prime}!}}
\exp\Bigl\{
\iih (x^{\mu} - x_0^\mu) \sum_{i'=1}^{n'} (p'_{i'})_\mu
+
\iih (x^{\mu} - x_0^\mu) \sum_{i^{\prime\prime}=1}^{n^{\prime\prime}}
			  (p^{\prime\prime}_{i^{\prime\prime}})_\mu
\Bigr\}
\\ \times
\bigl(
\varphi_0^+(\bs p'_1)\circ\dots\circ\varphi_0^+(\bs p'_{n'})
\circ
\varphi_0^{\dag\,+}(\bs p^{\prime\prime}_1)\circ\dots\circ
	\varphi_0^{\dag\,+}(\bs p^{\prime\prime}_{n^{\prime\prime}})
\bigr)
(\ope{X}_0) ,
	\end{multline}
where, in view of the commutation relations~\eref{11.65}, the order  of the
creation operators is inessential. If $n'=0$ (resp.\ $n^{\prime\prime}=0$),
the particle (resp.\ antiparticle) creation operators and the first (resp.\
second) sum in the exponent should be absent. In particular, the vacuum
corresponds to~\eref{11.77} with $n'=n^{\prime\prime}=0$. The state
vector~\eref{11.77} is
	an eigenvector of the momentum operator~\eref{11.71}  with
eigenvalue (4\ndash momentum)
\(
\sum_{i'=1}^{n'} p'_{i'}
+
\sum_{i^{\prime\prime}=1}^{n^{\prime\prime}}
	p^{\prime\prime}_{i^{\prime\prime}}
\)
	and is also an eigenvector of the charge operator~\eref{11.72} with
eigenvalue  $(-q)(n'-n^{\prime\prime})$.%
\footnote{~%
Recall (see Sect.~\ref{Sect14}), the operator $\varphi_0^{+}(\bs k)$
creates a particle with 4\ndash momentum $k_\mu$ and charge $-q$, while
$\varphi_0^{\dag\,+}(\bs k)$ creates a particle with 4\ndash momentum $k_\mu$
and charge $+q$, where, in the both cases, $k_0=\sqrt{m^2c^2+{\bs k}^2}$.%
}

	The reader may verify, using~\eref{11.65} and~\eref{11.28new}, that
the transition amplitude between two states of a charged field,
like~\eref{11.77}, is:
	\begin{multline}	\label{11.78}
\langle
\ope{X}(
y;q'_1;\ldots;q'_{n'};
  q^{\prime\prime}_1;\ldots;
  q^{\prime\prime}_{n^{\prime\prime}}
)
|
\ope{X}(
x;p'_1;\ldots;p'_{m'};
  p^{\prime\prime}_1;\ldots;
  p^{\prime\prime}_{m^{\prime\prime}}
)
\rangle
\\
=
\frac{1}{n^{\prime}! n^{\prime\prime}!}
\delta_{m'n'} \delta_{m^{\prime\prime}n^{\prime\prime}}
\exp\Bigl\{
\iih (x^{\mu} - y^{\mu}) \sum_{i'=1}^{n'} (p'_{i'})_\mu
+
\iih (x^{\mu} - y^{\mu}) \sum_{i^{\prime\prime}=1}^{n^{\prime\prime}}
			 ( p^{\prime\prime}_{i^{\prime\prime}} )_\mu
\Bigr\}
\\ \times
\sum_{(i'_1,\dots,i'_{n'})}
\delta^3(\bs p'_{n'} - \bs q'_{i'_1}) \delta^3(\bs p'_{n'-1} - \bs q'_{i'_2})
\dots
\delta^3(\bs p'_{1} - \bs q'_{i'_{n'}})
\\ \times
\sum_{(i^{\prime\prime}_1,\dots,i_{n^{\prime\prime}}^{\prime\prime})}
	\delta^3(\bs p^{\prime\prime}_{n^{\prime\prime}} -
		\bs q^{\prime\prime}_{i^{\prime\prime}_1})
	\delta^3(\bs p^{\prime\prime}_{n^{\prime\prime}-1} -
		\bs q^{\prime\prime}_{i^{\prime\prime}_2})
\dots
\delta^3(\bs p^{\prime\prime}_{1} -
\bs q^{\prime\prime}_{i^{\prime\prime}_{n^{\prime\prime}}})
	\end{multline}
where the summations are over all
permutations $(i'_1,\dots,i'_{n'})$ of $(1,\dots,n')$ and
$(i^{\prime\prime}_1,\dots,i_{n^{\prime\prime}}^{\prime\prime})$ of
$(1,\dots,n^{\prime\prime})$.
	The conclusions from this formula are similar to the ones
from~\eref{9.12} in Sect.~\ref{Sect10}. For instance, the only non\ndash
forbidden transition from an
$n'$\ndash particle~+~$n^{\prime\prime}$\ndash antiparticle state is into
$n'$\ndash particle~+~$n^{\prime\prime}$\ndash antiparticle state; the both
states may differ only in the spacetime positions of the (anti)particles in
them. This result is quite natural as we are dealing with free
particles/fields.

	In particular, if $\ope{X}_n$ denotes any state containing $n$
particles and/or antiparticles, $n=0,1,\dots$, then~\eref{11.78} says that
	\begin{equation}	\label{11.79}
\langle \ope{X}_n|\ope{X}_0 \rangle = \delta_{n0} ,
	\end{equation}
which expresses the stability of the vacuum.

	We shall end the present section with a simple example. Consider the
one (anti)particle states
$\varphi_0^{+}(\bs p)(\ope{X}_0)$ and
$\varphi_0^{\dag\,+}(\bs p)(\ope{X}_0)$.
Applying~\eref{11.71}, \eref{11.72}, \eref{11.72-1} and~\eref{11.65}, we find
($p_0:=\sqrt{m^2c^2+{\bs p}^2}$):%
\footnote{~%
In Heisenberg picture and in terms of the Heisenberg creation/annihilation
operators, in equations~\eref{11.79-2} the terms proportional to
$( x_\mu p_\nu - x_\nu p_\mu )$ are absent and tildes over all operators
should be added. Equations~\eref{11.79-1} remain unchanged in Heisenberg
picture (in terms of the corresponding Heisenberg operators).%
}
	\begin{gather}
			\label{11.79-1}
	\begin{split}
\ope{P}_\mu\bigl( \varphi_0^{+}(\bs p)(\ope{X}_0) \bigr)
& =
p_\mu \varphi_0^{+}(\bs p)(\ope{X}_0)
\quad \hphantom{+}
\ope{Q}\bigl( \varphi_0^{+}(\bs p)(\ope{X}_0) \bigr)
=
 - q \varphi_0^{+}(\bs p)(\ope{X}_0)
\\
\ope{P}_\mu\bigl( \varphi_0^{\dag\,+}(\bs p)(\ope{X}_0) \bigr)
& =
p_\mu \varphi_0^{\dag\,+}(\bs p)(\ope{X}_0)
\quad
\ope{Q}\bigl( \varphi_0^{\dag\,+}(\bs p)(\ope{X}_0) \bigr)
=
 + q \varphi_0^{\dag\,+}(\bs p)(\ope{X}_0)
	\end{split}
\displaybreak[1]\\		\label{11.79-2}
	\begin{split}
\ope{L}_{\mu\nu}(x) \bigl( \varphi_0^{+}(\bs p)(\ope{X}_0) \bigr)
& =
\Bigl\{
( x_\mu p_\nu - x_\nu p_\mu )
- \ih \Big( p_\mu\frac{\pd}{\pd p_\nu} -  p_\nu\frac{\pd}{\pd p_\mu} \Big)
\Bigr\}
  \bigl( \varphi_0^{+}(\bs p)(\ope{X}_0) \bigr)
\\
\ope{L}_{\mu\nu}(x) \bigl( \varphi_0^{\dag\,+}(\bs p)(\ope{X}_0) \bigr)
& =
\Bigl\{
( x_\mu p_\nu - x_\nu p_\mu )
- \ih \Big( p_\mu\frac{\pd}{\pd p_\nu} -  p_\nu\frac{\pd}{\pd p_\mu} \Big)
\Bigr\}
    \bigl( \varphi_0^{\dag\,+}(\bs p)(\ope{X}_0) \bigr) .
	\end{split}
	\end{gather}
These results agree completely with the interpretation of
$\varphi_0^{+}(\bk)$ and $\varphi_0^{\dag\,+}(\bk)$ as operators creating one
(anti)particle states.


\section {Conclusion}
\label{Conclusion}

	The main results of this paper, dealing with a study of free
Hermitian or non\ndash Hermitian  scalar fields, may be formulated as follows:

	\begin{description}
\item
	The creation and annihilation operators in momentum representation
in momentum picture are introduced without an explicit appeal to the Fourier
transform of the field operator(s). However, they are (up to constant phase
factor and, possibly, normalization) identical with the known ones introduced
in momentum representation in Heisenberg picture.

\item
	The quantization with commutators, not by anticommutators, is derived
from the field equations (in momentum picture) without involving the
spin\ndash statistics theorem (or other additional condition), if one stars
from a suitable Lagrangian.

\item
	The (system of) field equation(s) in terms of creation and
annihilation operators is derived. It happens to be similar to a kind of
paracommutation relations.%
\footnote{~%
More precisely, elsewhere we shall show how the parabose-commutation relations
for free arbitrary scalar field can be derived from the Klein\ndash Gordon
equations in terms of creation and annihilation operators.%
}

\item
	An analysis of the derivation of the standard commutation relations
is given. It is shown that, under some explicitly presented conditions, they
are equivalent to the (system of) field equation(s) and are not additional to
it conditions in the theory.

	\end{description}

	In forthcoming paper(s), we intend to investigate other free fields,
like vector and spinor ones, in momentum picture.




\addcontentsline{toc}{section}{References}
\bibliography{bozhopub,bozhoref}
\bibliographystyle{unsrt}
\addcontentsline{toc}{subsubsection}{This article ends at page}

\end{document}